\documentclass[10pt]{article} 
\usepackage[preprint]{tmlr}

\usepackage{amsmath,amsfonts,bm}

\def\eqref#1{equation~\ref{#1}}

\def\1{\bm{1}}

\DeclareMathAlphabet{\mathsfit}{\encodingdefault}{\sfdefault}{m}{sl}
\SetMathAlphabet{\mathsfit}{bold}{\encodingdefault}{\sfdefault}{bx}{n}

\usepackage{hyperref}
\usepackage{url}
\usepackage{cleveref} 
\usepackage{bm}
\usepackage{graphicx}
\usepackage{algorithm}
\usepackage{algpseudocode}
\usepackage{caption}
\usepackage{subfigure}
\usepackage{subcaption}
\usepackage{enumitem}\usepackage{float}
\usepackage{tikz}
\usepackage{amsmath}
\usepackage{xcolor}
\usepackage{bm}
\usepackage{multicol}
\usepackage{pythonhighlight}
\usepackage{amssymb} 
\usepackage{xspace}

\let\oldbm\bm
\renewcommand{\bm}[1]{\oldbm{#1}}
\definecolor{gray}{RGB}{200,200,200}

\newcommand{\ind}{\perp\!\!\!\perp}
\newcommand{\nind}{\not\!\perp\!\!\!\perp}
\newcommand{\cS}{{\mathcal{S}}}
\newcommand{\cG}{{\mathcal{G}}}
\newcommand{\cI}{{\mathcal{I}}}
\newcommand{\cP}{{\mathcal{P}}}
\newcommand{\cA}{{\mathcal{A}}}

\newcommand{\cC}{{\mathcal{C}}}
\newcommand{\cD}{{\mathcal{D}}}
\newcommand{\cE}{{\mathcal{E}}}
\newcommand{\cL}{{\mathcal{L}}}
\newcommand{\cU}{{\mathcal{U}}}
\newcommand{\cT}{{\mathcal{T}}}
\newcommand{\cZ}{{\mathcal{Z}}}

\newcommand{\indmod}{{\mathcal{I}}}
\newcommand{\boldeta}{\boldsymbol{\eta}}
\newcommand{\doublemarked}{\circ\!\!-\!\!\circ}
\newcommand{\confictmarked}{\times\!\!-\!\!\times}
\newcommand{\stov}{\texttt{S2V}\xspace}
\newcommand{\stovarg}{\texttt{S2V}($\cA, \cE$)\xspace}
\newcommand{\stovtwo}{\texttt{S2V-2}\xspace}
\newcommand{\adag}{\texttt{Adag}\xspace}
\newcommand{\adj}{\mathbf{Adj}}
\newcommand{\tested}{\cI_\cA^{test}}
\newcommand{\coll}{\mathcal{U}_c}
\newcommand{\aggscore}{\mathcal{Q}}
\newcommand{\poss}{\operatorname{\mathbf{Possible-D-sep}}} 
\newcommand{\pluseq}{\mathrel{+}=}
\makeatletter
\newcommand{\paranew}[1]{%
  \par 
  \noindent
  \textbf{#1\@addpunct{:}}\enspace\ignorespaces
}
\makeatother

\makeatletter
\newcommand{\paranewspace}[1]{%
  \par 
  \addvspace{\medskipamount}
  \noindent
  \textbf{#1\@addpunct{:}}\enspace\ignorespaces
}
\makeatother

\def\*#1{\mathbf{#1}}

\newtheorem{lemma}{Lemma}
\newtheorem{theorem}{Theorem}
\newtheorem{corollary}{Corollary}
\newtheorem{assumption}{Assumption}
\newtheorem{definition}{Definition}
\newtheorem{property}{Property}

\newtheorem{remark}{Remark}

\newtheorem{example}{Example}
\newtheorem{proof}{Proof}

\title{Causal discovery on vector-valued variables and consistency-guided aggregation}

\author{\name Urmi Ninad \email urmi.ninad@tu-berlin.de \\
      \addr Technische Universit\"at Berlin, Institute of Computer Engineering and Microelectronics, \\
      Berlin, Germany
      \AND
      \name Jonas Wahl$^{*}$ \email jonas.wahl@dfki.de \\
      \addr German Research Centre for Artificial Intelligence (DFKI), Berlin, Germany
      \AND
      \name Andreas Gerhardus$^{*}$ \email andreas.gerhardus@dlr.de\\
      \addr German Aerospace Center (DLR), Institute of Data Science, Jena, Germany
      \AND
      \name Jakob Runge \email jakob.runge@uni-potsdam.de\\
      \addr Department of Computer Science, University of Potsdam, Potsdam, Germany \\ 
      Center for Scalable Data Analytics and Artificial Intelligence (ScaDS.AI) Dresden/Leipzig, Germany \\ Technische Universit\"at Berlin, Institute of Computer Engineering and Microelectronics, \\ Berlin, Germany 
      \\
     \newline $^*$equal contribution
      }

\def\month{MM}  
\def\year{YYYY} 
\def\openreview{\url{https://openreview.net/forum?id=XXXX}} 

\begin{document}

\maketitle

\begin{abstract}
Causal discovery (CD) aims to discover the causal graph underlying the data generation mechanism of observed variables. In many real-world applications, the observed variables are \emph{vector-valued}, such as in climate science where variables are defined over a spatial grid and the task is called spatio-temporal causal discovery.
We motivate CD in vector-valued variable setting while considering different possibilities for the underlying model, and highlight the pitfalls of commonly-used approaches when compared to a fully vectorized approach.
Furthermore, often the vector-valued variables are high-dimensional, and \emph{aggregations} of the variables, such as averages, are considered in interest of efficiency and robustness. In the absence of interventional data, testing for the soundness of aggregate variables as consistent abstractions that map a low-level to a high-level structural causal model (SCM) is hard, and recent works have illustrated the stringency of conditions required for testing consistency. 
In this work, we take a careful look at the task of vector-valued CD via constraint-based methods, focusing on the problem of consistency of aggregation for this task. 
We derive three \emph{aggregation consistency scores}, based on compatibility of independence models and (partial) aggregation, that quantify different aspects of the soundness of an aggregation map for the CD problem. We present the argument that the consistency of causal abstractions must be separated from the task-dependent consistency of aggregation maps. As an actionable conclusion of our findings, we propose a wrapper \adag to optimize a chosen aggregation consistency score for aggregate-CD, to make the output of CD over aggregate variables more reliable. We supplement all our findings with experimental evaluations on synthetic non-time series and spatio-temporal data.
\end{abstract}

\section{Introduction}

Causal discovery (CD) is the task of inferring causal relationships from data, which may be passively observed or (partially) intervened on, in the form of a causal graph \citep{spirtes_causation_1993, PearlCausality, PetJanSch17}. Constraint-based CD is the branch of CD that aims to learn the causal graph via conditional independence (CI) tests on the data distribution through relevant assumptions that establish a connection between the distribution and the underlying causal model. 
Some of these assumptions, such as faithfulness, have repeatedly been called into question \citep{Uhler_faithfulness, boeken_faithfulness}. Nevertheless, the strength of this approach lies in being able to make statements about the underlying causal graph, when interventional data is categorically unavailable, either due to the impossibility or questionable ethical conditions of performing interventions. Other approaches to CD besides the constraint-based approach include score-based CD, asymmetry-based CD and multiple environments-based CD, see \cite{runge2023causal,Camps_Valls_phys_reps} for an overview of the various approaches. 

In many fields, ranging from the natural sciences, such as climate science or space weather research, to non-natural sciences, such as economics, the observed variables of interest are vector-valued, that is, they consist of a collection of observed scalar variables. 
As an example, Earth system scientists may wish to query whether there exists a causal relationship between a temperature field defined over an area consisting of a finite number of grid boxes and the sea surface pressure field over another area \citep{ENSO_NAO}. In governmental policy, such a query could involve determining nationwide causal drivers of voter turnout, which varies with location \citep{voter_turnout}. In \cref{fig:motivation} we present a few examples of domains where vector-valued data is prevalent. In the following, we use the terms `vector' (scalar) and `vector-valued' (scalar-valued) interchangeably.

The ubiquity of vector variables in systems of causal interest has motivated two broad approaches for CD in this setting, termed vector CD for the remaining of this work. (1) The first approach is the \emph{network} or the \emph{component-wise} approach, where the scalar components comprising the vector variables are treated individually as causal variables in their own right. The CD algorithm is then applied over the collection of scalar variables stemming from different vector variables of interest, for instance, see \cite{Uphoff_Deng} for applications to spatiotemporal data. (2) The second approach is the \emph{dimension reduction}  or \emph{aggregation} approach, which first reduces the dimension of the vector variables respectively and then performs CD over the dimensionally reduced counterparts, e.g.~averaging or using variants of principal component analysis is common when analyzing spatiotemporal teleconnections \citep{enso_teleconnections, runge2015identifying, Giannakis_VSA}. 
Both of the above-stated approaches suffer from drawbacks: the component-wise approach is unable to distinguish between the varying dynamics within and without the vector variable; for instance, the scalar components corresponding to one vector variable may have highly-dependent noises due to fast dynamics that lead to cyclic relationships inside a vector variable (but not across vector variables); the aggregation approach suffers from an inability to consistently capture the independence relationships that hold for the vector data. 

In this work, our \textbf{first contribution} is to introduce a deceptively simple but, as we will show, more robust \emph{third} approach to vector CD that is versatile in its soundness to different settings that give rise to causal vector variables. This approach, which we term the \emph{vectorized} approach, treats the vector variables as causal variables in themselves to perform CD on. The strength of this third approach lies in (1) its ability to treat the varying dynamics inside and across the vector variables consistently, and (2) its ability to faithfully capture the causally relevant components within each vector variable, thereby countering the central weakness of the first two approaches to vector CD. The efficacy of this approach is limited by the efficacy of conditional independence tests, especially when the vector variables are high-dimensional. 
Since conditional independence is a provably hard task, and has been discussed elsewhere in great depth \citep{Shah_Peters}, we do not dwell on this issue in this work, and focus on how, within the realm of constraint-based CD for vector variables with a meaningful causal interpretation, our approach proves conceptually superior. In \cref{app:mult_CI}, we cast a brief glance at conditional independence testing over vectors, and evaluate different aggregation techniques for univariate statistics within the pairwise approach to vector conditional independence testing \citep{hochsprung_pairwise}. Furthermore, the given vector variables are assumed to be causally meaningful, i.e., we do not focus on a causal representation learning \citep{towards_CRL_schölkopf} or a causal clustering task \citep{causal_clustering_Tikka}. 

Despite the strengths of vector CD, its applicability is limited by the trustworthiness of high-dimensional conditional independence tests. In practice, the second approach, i.e., the aggregation of vector variables into lower dimensional variables before performing CD is a popular choice among practitioners, to avoid the curse of dimensionality. 
The field of causal abstractions (sometimes called aggregations) has emerged in response to the need of understanding causal phenomena at coarser level of granularity among abstract or macro variables than at the fine-grained level at which measurements are observed, and formalizing a connection between the two \citep{Rubenstein, ChaEbPer17}. This task requires, as a first step, defining a consistent transformation between the causal models at the two levels that preserves interventional implications. Such transformations have been termed \emph{exact} transformations, and several other stricter as well as looser definitions of abstraction consistency have emerged since then \citep{rischel_uai, Beckers_abstracting, beckers_approximate}. 
Application of causal abstraction theory has grown around learning a consistent abstraction mapping between SCMs at the low and high level in a way that preserves intervention alignment.
In many settings, at least a partial knowledge of the SCM, such as the causal graph, is assumed to be known \citep{Chalupka_ENSO}. Furthermore, the abstraction mapping is also assumed to be partially known, either in the form of the mapping between SCMs or sets of interventions in the two SCMs \citep{felekis_optimal_transport}. These are undoubtedly remarkable developments in formalizing the micro-macro SCM connection and learning suitable causal abstractions of low-level variables. However, for the task of causal discovery which is the focus of the current work, such supporting pre-requisite knowledge in the form of the causal graph, or interventional data, or an intervention mapping, is typically absent. 
This gap in the literature, namely the need to aggregate before performing CD given insufficient background knowledge on the underlying SCM, yet an absence of any consistency checks on the aggregation for CD, leads to the \textbf{second contribution} of our work. To this end, first, we define the so-called \emph{valid aggregation map} for the task of CD and study the conditions required for this definition of validity that we present. In the absence of ground-truth knowledge, the validity of aggregation can be checked by means of consistency of the aggregation (high-level) with partial knowledge from vector variables (low-level). 
Consequentially, we define different \emph{aggregation consistency} scores to quantify the degree of consistency of aggregation, given no knowledge of the ground truth SCM. Given such scores, we propose an adaptive approach to optimize a given score to a desired level, which we name the \adag wrapper. The wrapper can be combined with any constraint-based CD method, a tunable aggregation map (defined in \cref{sec:agg_consistency_adag}), and an aggregation consistency score. While we present three aggregation consistency scores in this paper, we remark that the \adag wrapper may be used in conjunction with other user-defined scores better fitting to the problem.  

In \cref{sec:motivation} we motivate the problem of vector CD and testing aggregation consistency for the task of CD over vector variables in the absence of ground truth knowledge. In \cref{sec:data_gen} we present vector-valued data generation processes amenable to a causal interpretation and overview the diverse dynamics internal to the vector variables that are compatible with such a process. In \cref{sec:scalar_approach}, we present the component-wise approach to vector CD and study its conceptual soundness in the context of diverse internal dynamics. In \cref{sec:agg_approach}, we present the aggregation approach to vector CD, define a valid aggregation map by first defining aggregation faithfulness and aggregation sufficiency properties, and discuss the intuition behind these properties for vector CD. In \cref{sec:experiments} we present experiments on simulated data comparing all three approaches, namely, component-wise, aggregate, and vectorized, under the assumption of a causal model over vectors, to illustrate the failure modes of different approaches. In \cref{sec:agg_consistency_adag} we introduce the concept of aggregation consistency and present the \adag wrapper for consistency-guided aggregation for vector CD. Finally, in \cref{sec:savar} we present an application of the \adag wrapper to a stochastic climate model for spatiotemporal data (SAVAR) \citep{tibau_savar} that simulates teleconnections between climatic processes. We end with conclusions and outlook in \cref{sec:conclusions}.


\section{Motivation and related work}\label{sec:motivation}

\subsection{Why vector-valued causal discovery?}
\begin{figure}
    \centering
    \includegraphics[width=0.8\linewidth]{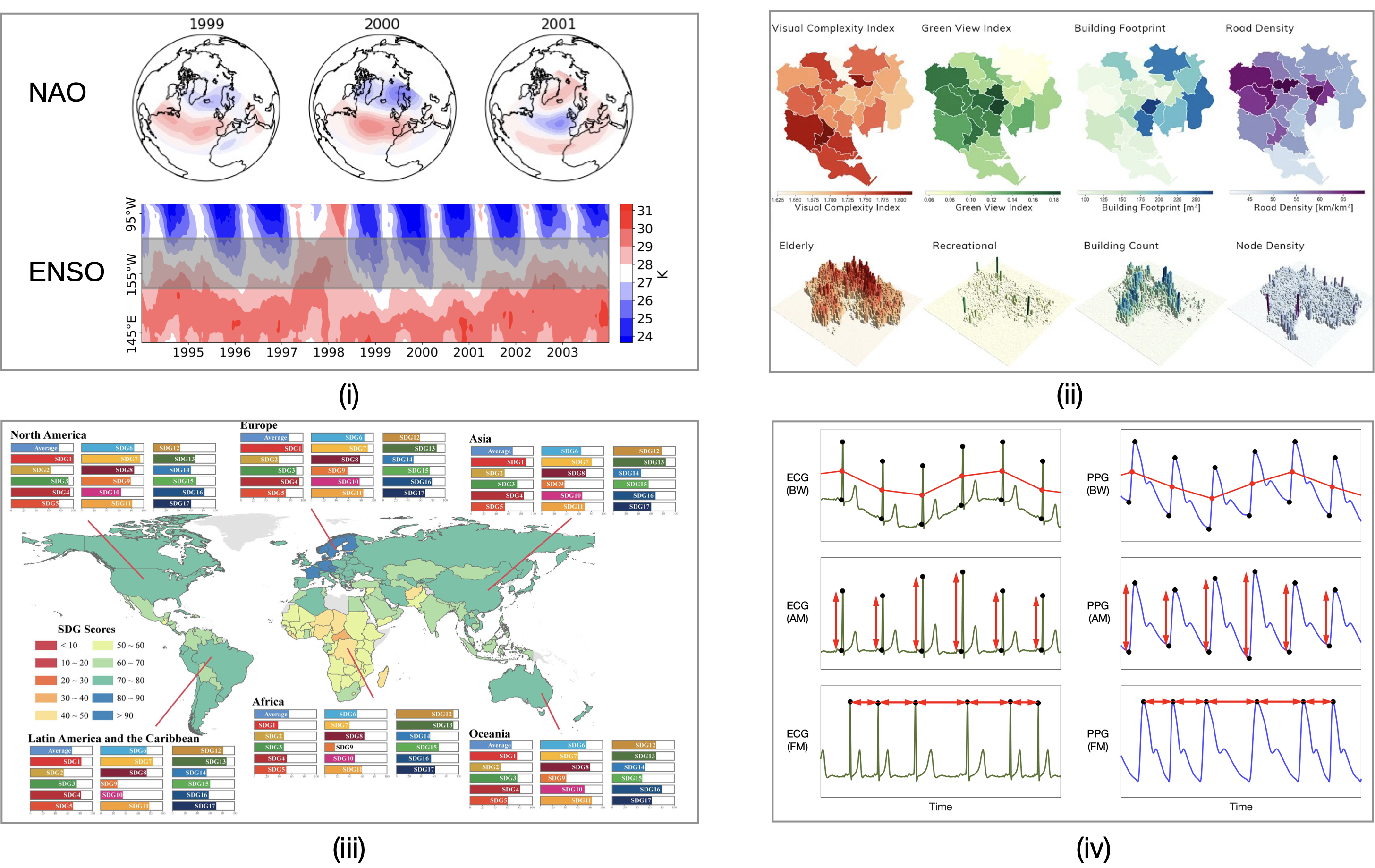}
    \caption{Examples of scientific domains exhibiting vector-valued data: (i) Earth science: spatiotemporal systems such as NAO and ENSO (see main text),
    (ii) urban systems where multiple variables represent aspects of cities \citep{urban_cities_example},
    (iii) in economics where multiple indices describe notions of sustainable development \citep{sustainable_dev_example},
(iv) oscillatory time series of the human body which are represented by spike interval length and amplitude (image from wikimedia commons). All cited images are licensed under the CC BY 4.0 license: \url{https://creativecommons.org/licenses/by/4.0/}. 
}
    \label{fig:motivation}
\end{figure}

CD over vectors is of interest to many domains where the variables over which one wishes to infer the causal graph are not scalar indices, but rather each variable corresponds to several scalar variables grouped into a vector. This grouping has its basis in the domain knowledge that the vector variable acts as a non-decomposable agent in the causal dynamics of the system of interest. For instance, the El Nino Southern Oscillation (ENSO) is defined as a coupled climate phenomenon involving a rise in the sea surface temperatures (SST) over the Pacific Ocean and a change in the surface wind patterns (from easterly to westerly). ENSO has been proven to exhibit highly complex midlatitude `teleconnections', i.e. relationships among climate or weather phenomena at widely separated spatial locations \citep{enso_teleconnections}. Among these teleconnections is a relationship with the North Atlantic oscillation  (NAO), which refers to a phenomenon that marks changes in sea-level pressure (SLP) over the Atlantic ocean \citep{ENSO_NAO, enso_nao_Toniazzo}. Therefore, a valid causal query to make in this setup is whether the vector of SSTs over the Pacific causally influences the vector of SLPs over the Atlantic, which would help analyzing an ENSO-NAO teleconnection. Similar queries can arise in the fields of neuroscience, economics, and urban sciences, to name a few (see examples in \cref{fig:motivation}).

\subsubsection{Component-wise approach}\label{subsec:motivation_scalar}
A direct approach to deal with the type of query mentioned above (SST$\xrightarrow[]{?} $ SLP) could be to perform standard CD over all scalar components of the vector variables, and study the pattern of edges that emerges between the scalar variables that belong to SST and those that belong to SLP. We name this approach the \emph{component-wise approach} to vector CD and provide details in \cref{sec:scalar_approach}. The component-wise approach poses the following problems:
\begin{itemize}
    \item[(i)] the time delay at which the SST variables causally influence one another may be much faster than the delay at which SST influences SLP, 
    \item[(ii)] the aggregating principle for multiple edges between scalar components variables corresponding to SST and SLP is not clear if the orientations of the edges differ, 
    \item[(iii)] individual components of a vector cannot be given an appropriate causal interpretation, and,
    \item[(iv)] a high number of CI tests are needed to infer causal relationships among a few variables.
\end{itemize}

The first problem of \emph{varying dynamics} within and among the vector variables is a fundamental problem that needs to be dealt with on a case-by-case basis. If the measuring instrument is precise enough to capture the fast dynamics within, then the problem is (a) a computational problem: how long of a maximum time lag should one consider to capture the slower dynamics among SST and SLP, without inflating the time window too much to avoid too many CI tests, and (b) a statistical problem: if the relative strengths of the effects within and among the variables are widely different, it may not be possible to detect the weaker effects. However, if the measurement instrument is not precise enough and leads to cycles within SST or SLP, or if the dynamics within cannot be modeled as a DAG due to dependent noises, the problem becomes a theoretical one: can CD methods that assume no latent confounding or cyclic relations be soundly applied to this problem? 
If not, one is of course free to resort to methods that do relax the assumptions of causal sufficiency and acyclicity, but this seems to be an inefficient solution if the assumption violations are restricted to within the groups. 

The second problem of \emph{ambiguity of aggregating edges} is also ultimately a statistical problem, i.e., assuming that the causal edges do only point in one direction between scalar variables corresponding to SST and SLP, a sound CD algorithm under the assumption of correct independence tests, will not detect edge orientation contradictions. However, this cannot be assumed for CD over finite samples in general.  

The third problem concerns aptness of \emph{causal meaning}: If the internal structure of a vector variable does not easily lend itself to a causal interpretation, but the structure \emph{across} vector variables does, how can one interpret the results of CD over all scalar variables causally? The answer to this problem is ambiguous and raises caution when treating a vector causal variable merely as a collection of scalar causal variables. Taking the example of ENSO, the individual grid points of the tropical Pacific cannot be regarded as autonomous assignment functions in a Pearlian SCM, because the ENSO phenomenon emerges from the spatio-temporal dynamics of the whole region. In addition, causal graphs at the grid level would depend strongly on the spatial resolution. We will deal with the question of diverse internal dynamics in \cref{subsec:internal_dynamics} and the soundness of the component-wise approach to these dynamics in \cref{sec:scalar_approach}.

The fourth problem is a statistical as well as computational problem because CI testing is provably hard \citep{Shah_Peters}, and CD algorithms require sequential CI testing, where errors at an earlier stage can multiply and affect results at later stages of the algorithm.

All of the problems mentioned above, such as varying dynamics, edge-aggregation ambiguity, and that of causal meaning that arise in the approach of `scalarizing' what is truly a vector problem, are circumvented when one handles the vector variables as non-decomposable causal variables and performs CD over the vector variables themselves. The fourth problem is not entirely circumvented if the vector variables are high-dimensional, but rather one needs to carefully study the tradeoff between too many CI tests versus high-dimensional CI tests. In \cref{fig:scalar_agg_pitfalls}(i) we illustrate these problems. 

\subsubsection{Aggregation approach}\label{subsub:agg}
\begin{figure}
    \centering
    \includegraphics[width=0.8\linewidth]{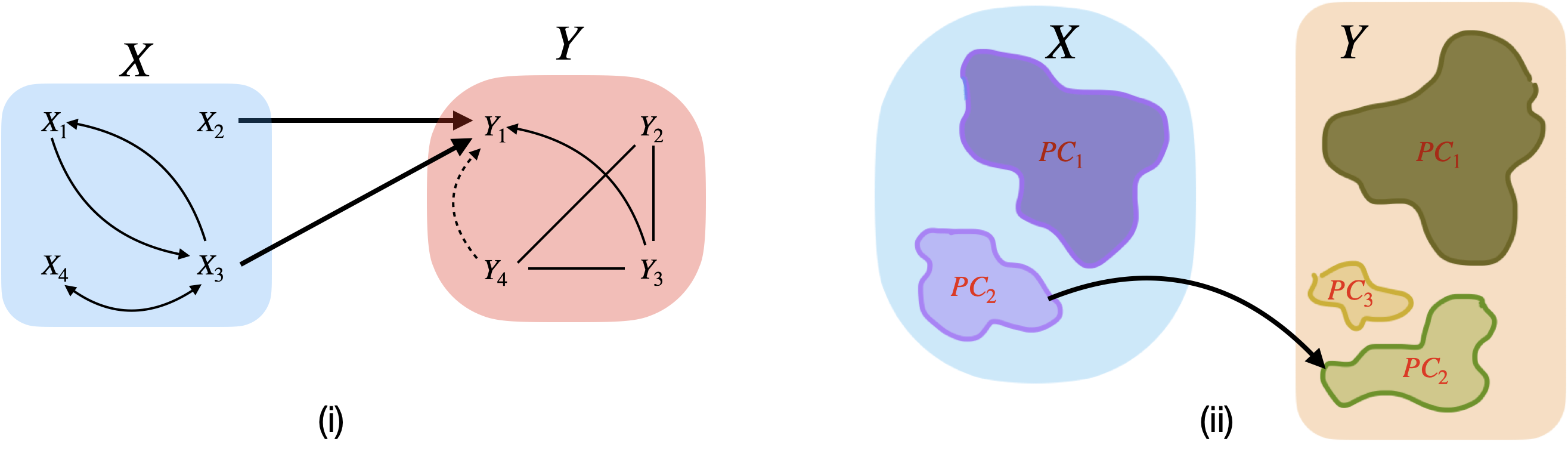}
    \caption{Illustrating the pitfalls of `scalarizing' and `aggregating' the task of CD over vector-valued variables $X$ and $Y$. (i) The dynamics internal to the vector variables may be modeled differently than as a DAG, such as with latent confounding (bidirected edge), cycles, equilibrium states (undirected edges), or (near-)deterministic relationships (dashed edges). (ii) If only the first principal components (`$PC_1$') of $X$ and $Y$, respectively, are retained in the aggregation for CD, then a dominant causal relationship, that may exist between the $PC_2$'s, will not be found.}
    \label{fig:scalar_agg_pitfalls}
\end{figure}

Oftentimes, the approach that is adopted to answer the causal query (SST$ \xrightarrow[]{?} $ SLP) is to:
\begin{itemize}
    \item[] Step 1: Aggregate each causal variable via some surjective function, such as an average or its weighted version, that reduces the dimensions of the variables, respectively.  
    \item[] Step 2: Perform CD over the aggregated variables that serve as proxies for the underlying vector variables, thereby drastically reducing the complexity of the problem.
\end{itemize}
This approach and its drawbacks were discussed briefly in \cite{Wahl_foundations}. 
Given the inadequate theoretical basis of relying on causal relationships over the aggregate variables to reflect the causal relationships of the vector variables themselves, this approach suffers from the central weakness that there is no theoretical correspondence between the causal graph over the original vector variables, and the resultant graph from CD between the aggregate variables. We schematically illustrate a special case for which the aggregation approach to vector CD is unsound in \cref{fig:scalar_agg_pitfalls}(ii). The field of causal abstractions has emerged in response to formalizing the consistency of transformation of the low-level variables from a causal perspective, see \cite{Rubenstein, Beckers_abstracting, rischel_uai}. In the next subsection, we will discuss the connection of our work with causal abstractions and provide a motivation for investigating the soundness of aggregation in a task-dependent manner. 

\subsection{Aggregation consistency for vector CD and connection to causal abstractions}\label{subsub:motivation_abstraction}
In the literature, the task of vector CD (sometimes called multivariate CD) has been handled in a number of ways, and we summarized the approaches above as (i) component-wise CD over all components of the vectors, (ii) CD over aggregations of vector variables, and (iii) vectorized CD. 
The second approach, that is opting for an aggregation, such as averaging, of the vector variables to capture relevant causal relationships while circumventing statistical challenges of high-dimensional data is widely applied due to its efficiency. The soundness of such an approach from the perspective of graph discovery remains unclear, as we briefly introduced in \cref{subsub:agg}. Nevertheless, we motivate the usefulness of aggregations for CD in the absence of the type of background knowledge that testing a consistent high-level causal abstraction would require. 

The increased interest in causal abstractions is partly responding to the observation that variables among which one wishes to learn the causal structure or quantify causal effects of interventions are often abstract variables, whose relationship to the actual observed variables is not always clear. Assuming that the low-level observed variables arise from an SCM, the aim then is to learn either a consistent higher-level SCM or, given a higher-level SCM, test whether the low and high-level SCMs are consistent. 
In several sciences, while an assumption that the low-level variables arise from some SCM can be justified, this SCM is rarely known in practice. Furthermore, an intuition about what the macro variables underlying the high-level SCM are may be completely lacking. 
However, if the central query that a scientist wishes to answer is ``What is the causal graph among the observed vector variables?'', given the knowledge that there exists a causal model among the vector variables, learning a causal abstraction from the micro to macro variables becomes a means to an end, a means that would simplify and robustify the task of CD over (possibly high-dimensional) vector-valued variables. This is the setting we wish to focus on. We further emphasize that this task of learning the causal graph should not be confused with what has been called \emph{causal sub-model selection} \citep{malinsky_cautious}, that is, learning a causal graph as a stepping stone to performing causal effect estimation. 
If one is in the causal submodel selection setting, the effect estimation must only be done over the vector variables, if one has no justification for believing the abstraction to be consistent in the sense of \cite{Rubenstein}. 

Recent works that introduce graphs over groups of variables are c-DAGs of \cite{anand_causal_2023}, coarse graphs of \cite{Wahl_foundations}, and partition graphs of \cite{shiragur_fewer}. These works offer a different perspective compared to causally consistent abstractions, in that they focus on having a coarser \emph{graphical} understanding of the model, and don't clearly connect the coarsening operation to a consistent abstraction from an interventional perspective. This connection was recently studied in \cite{schooltink_zennaro}. 

In the case that the abstraction serves as a tool for causal discovery (among low-level causal variables) as an end in itself, the insistence on intervention alignment as a criteria for consistency of the abstraction mapping seems excessive and unrealistic for those domain sciences where interventions are categorically unavailable. Consider the example of total cholesterol's ($TC$) influence on heart disease ($HD$) popularized by \cite{ChaEbPer17, Rubenstein} in the context of causal modeling. When the micro variables low-density lipoprotein ($LDL$) and high-density lipoprotein ($HDL$) are aggregated into one macro variable $TC := LDL + HDL$, intervening on $TC$ by intervening heterogeneously on $LDL$ and $HDL$ leads to a discrepancy, in that $TC$ can be found to affect HD positively or negatively depending on the intervention. However, let us consider the vector variable cholesterol $C = (LDL, HDL)$, and the task of learning whether $C$ causes $HD$. 
Now, arguing from the perspective of this task, 
aggregating $LDL$ and $HDL$ into $TC$ is allowed, and even desired to reduce the complexity of the task, as long as one infers the result to hold for the causal variable $C$, not the aggregate variable $TC$. To quote \cite{Rubenstein} ``In particular, this means that we can view the (macro variables) Y variables as causal entities, rather than only functions of underlying `truly' causal entities. Only if this is the case, causal statements such as `raising temperature increases pressure' or `$LDL$ causes heart disease' are meaningful.'' However, we emphasize that aggregation maps can still be used to make meaningful statements about the lower-level variables, without needing to ascribe a corresponding SCM at the higher level. This pragmatism is motivated by the absence of any structural knowledge of the underlying SCM and the impossibility of ever attaining interventional data, that afflicts many real-world application problems. 

Therefore, for disambiguation, in the rest of this work we refer to \emph{aggregations} when we imply the mapping from the vector-level 
to the aggregate-level 
while only assuming an SCM at the vector-level (we reserve the terms `macro' and `micro' variables to refer to the vector-valued and underlying scalar-valued variables, respectively). Aggregations, in this sense, are causally-agnostic dimension reductions, whose fittingness for the task of sound CD over vector variables remains to be tested. Testing this soundness in the absence of any ground truth knowledge and the presence of only observational data is a central contribution of our work.

\subsection{Related work}\label{subsec:related}

    For a broad overview of the foundations of CD and algorithms, see \cite{Camps_Valls_phys_reps, runge2023causal}, whereas for an overview of applications cases and their challenges see \cite{brouillard2024landscape, runge_inferring_2019}.
    Theoretical foundations for causal discovery over variable groups is a topic that has been explored in \cite{ParKas17}, \cite{wahl_ninad_vector}, \cite{Wahl_foundations}, \cite{shiragur_fewer}, while
    causal effect estimation over variable groups has been discussed in \cite{anand_causal_2023, nabi_causal_dim_red, ferreira_assaad_macro}.
    Foundations of causal abstractions and related consistency were laid down and discussed further in \cite{Rubenstein, Beckers_abstracting, beckers_approximate, rischel_uai, schooltink_zennaro, zhu_janzing}. 
    Self-benchmarking and self-correction of CD algorithms in absence of ground truth knowledge was studied in \cite{eulig_janzing_falsifying, faller_self-compatibility,  ramsey_markov_checker, faltenbacher_incoherency, faller_redundancy, schkoda_Janzing_leave_one_variable-out}.

\section{Vector-valued data generation processes} \label{sec:data_gen}

We will now model the data generation process for the case that the causal variables of interest are vectors. Assume that there are a total of $N$ vector variables $\*X^i \ , i \in \{1,\ldots N\}$, with dimension $d_i$ respectively, that is $\*X^i = \{X^i_1, \ldots X^i_{d_i}\}$, where $X^i_j$ are scalar components of the vector $\*X^i$. The variables $\*X^i$ are denoted collectively by $\*X$ and are defined by the structural equations,
\begin{equation}\label{eq:SCM} 
    \*X^i := f^i(Pa(\*X^i), \boldeta^i) 
\end{equation}
where vector noises $\boldeta^i$ are jointly independent for all $i$ but need not have independent components themselves. $Pa(\*X^i)$ denotes the parent set of the variable $\*X^i$ and $f^i(\cdot)$ denotes the mechanism that maps the parent set and noises to the variables $\*X^i$. Formally, the tuple comprising the domains of $\*X^i$ and $\boldeta^i$, the joint distribution of $P(\boldeta)$ of $\boldeta^i$, together with the mechanism functions $f^i(\cdot), \ \forall i$ together define the SCM, see \cite{PetJanSch17} for a formal definition of an SCM. 
The SCM \cref{eq:SCM} entails a graph $\cG$ over nodes $\*X^i$ that is Markovian to the joint probability distribution $P(\*X)$ over the causal variables $\*X$ that is entailed by the SCM. $\cG$ is assumed to be acyclic. 
The Markov property implies that d-separation on the graph $\cG$ implies conditional independence on the probability distribution $P(\*X)$. Specifically, for $\*X^i$, $\*X^j$ and $\cS \subset \*X$, $\*X^i \ind_d \*X^j | \cS \Rightarrow \*X^i \ind \*X^j | \cS$. Here $\ind_d$ represents d-separation. See \cite{PearlCausality, spirtes_causation_1993} for an overview of d-separation and the Markov property. We make two assumptions for the SCM \cref{eq:SCM} and graph $\cG$ to aid causal discovery:


\begin{assumption}[Causal Discovery Assumptions]\label{ass:CD}
    The faithfulness condition, namely if $\exists i, j, \text{ and } \cS \subseteq \*X$ such that $\*X^i \ind \*X^j | \cS$ then this implies $ \*X^i \ind_d \*X^j | \cS$, holds. The graph entailed by \cref{eq:SCM} is acyclic. There are no hidden confounders between $\*X^i$ and $\*X^j$, $\forall i \neq j$. 
\end{assumption}

\subsection{Diverse possibilities for internal dynamics}\label{subsec:internal_dynamics}

\begin{figure}
    \centering
    \includegraphics[width=0.8\linewidth]{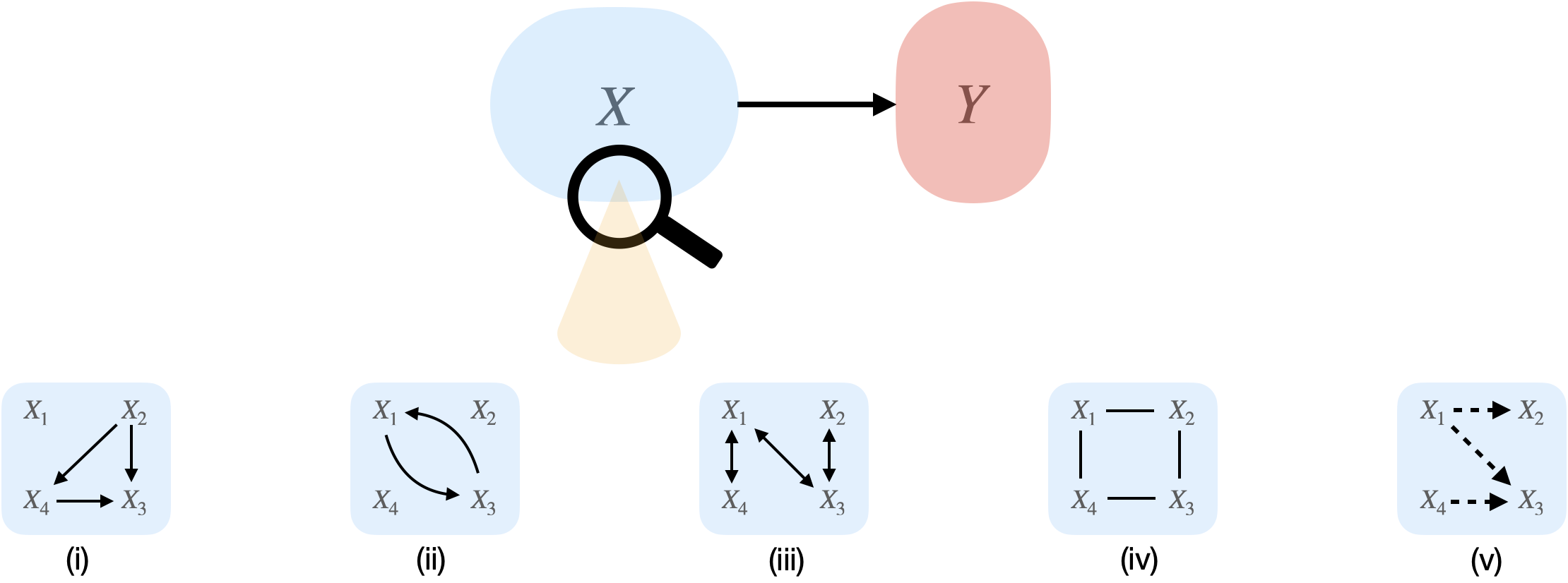}
    \caption{Figure to illustrate selected internal dynamics of vector-valued variable $\*X$ generated by an SCM which entails the causal graph $\cG: \*X \to \*Y$. The scalar components $X_i \in \*X$ could be generated by: (i) a causal DAG, (ii) a causal graph with cycles, (iii) a causal graph with latent confounding denoted by bidirected edges, (iv) an undirected graph (eg.~a Markov random field) without a causal interpretation, or (v) a causal graph with faithfulness violating relationships denoted by dashed edges. A combination of these settings, as well as an extension to time series graphs, are allowed. Note that potential internal dynamics of $\*Y$ are not displayed.}
    \label{fig:int_dynamics}
\end{figure}

The vector SCM \cref{eq:SCM} places no constraints, a priori, on the noises $\boldeta^i$ other than joint independence, i.e., $\boldeta^i \ind \boldeta^j, \  \forall i,j,\ i \neq j$. Relatedly, it is also not specified whether the relationship between the scalar components $X^i_1, X^i_2, \ldots$ corresponding to $\*X^i$ can be given a causal interpretation at all \citep{Bareinboim_Hierarchy}.
This constraint permits a number of interpretations of the internal dynamics of the vector variables and we will now discuss a few of these. We summarize the interpretations discussed below in \cref{fig:int_dynamics}.

\subsubsection{Internal DAG}
The relationships between the scalar components $X^i_1, X^i_2, \ldots, \forall i$, may permit a causal interpretation as nodes of a \emph{micro or fine-grained} causal graph $\cG_{micro}$, which specifically may take the graphical form of a DAG. 
The structural equations corresponding to the scalar variables would be given by:
\begin{equation}\label{eq:scalarSCM}
     X^i_j := f^i_j(Pa(X^i_j), \eta^i_j) \ , \forall i,j \ , 
\end{equation} 
where scalar noises $\eta^i_j, \  \forall j,$ are jointly independent  and $Pa(X^i_j) \subseteq \*X$. The graph $\cG_{micro}$ entailed by the SCM \cref{eq:scalarSCM} over the vertices $X^i_j, \ \forall i,j $ is a DAG. Then, the original SCM \cref{eq:SCM} corresponds to a partition $\cP$  of the scalar variables $X^i_j, \forall i,j$, into the vector variables $\*X^i$, i.e., $\cP = \{\*X^1, \ldots \*X^N\}$, such that coarsening $\cG_{micro}$ w.r.t.~partition $\cP$ yields $\cG$, see \cite{Wahl_foundations} for a formal definition of the coarsening operation. We call this partition the \emph{canonical partition}. The mechanisms are related by the simple relation $f^i = (f^i_1, f^i_2, \ldots f^i_{d_i})$.

Such internal dynamics are to be expected when the micro or fine-grained level admits an SCM, and where the causal relationships within a vector variable are at a similar time scale as those across the vector variables. \cite{anand_causal_2023} define so-called \emph{cluster DAGs} to represent partially understood causal relationships, relevant for fields such as medicine where the researcher is agnostic to relationships within a cluster, and only interested in relationships across clusters. 

\subsubsection{Internal latent confounding and cyclic relationships}
The scalar components $X^i_1, X^i_2, \ldots$ may be given a causal interpretation with a graph $\cG_{micro}$, while allowing for latent confounding and cyclic relationships between scalar variables within a vector variable. In this case, the SCM \cref{eq:scalarSCM} will be such that:
\begin{itemize}
    \item[(i)] the component noises $\eta^i_j, \forall j,$ comprising the vector noise $\boldeta^i$ need not be jointly independent,
    \item[(ii)] the graph $\cG_{micro}$ over all $X^i_j$ need not be acyclic, while still coarsening w.r.t.~the canonical partition $\cP$ to the DAG $\cG$.
\end{itemize}
This setting reflects those real-world scenarios where (i) either scalar components may be causally driven by unobserved variables, or, (ii) the scalar components corresponding to a vector variable interact causally at a time delay that is faster compared to the time delay at which two vectors interact (in a time series setting). Case (ii) could lead to cyclic relationships within a vector variable when the measurement device is not precise enough.  Under the assumption that the SCM \cref{eq:scalarSCM} is simple, the graph entailed by the stationary distribution of the SCM can be obtained by the acyclification operation \citep{forre_markov_2017}. The acyclified SCM contains bidirected edges and is structurally indistinguishable from an SCM with latent confounders. See \citep{Bongers_foundations} for details on SCMs with cycles and latent confounders. 

\subsubsection{Internal violations of faithfulness}
In the previous interpretations of the internal model as an SCM, we presumed that the causal faithfulness condition holds, thereby allowing us to discover $\cG_{micro}$ using the standard constraint-based algorithms. However the faithfulness condition may be violated, e.g.~due deterministic relations, XOR-type relations or fine-tuning \emph{within} a vector variable, while still preserving the faithfulness condition \emph{among} the vector variables. See \cite{zhang_faithfulness, marx_faithfulness} for a discussion on violations of faithfulness. 

Near-deterministic relationships among scalar variables $X^i_{j_1}, X^i_{j_2}$ are to be expected in those cases where the vector variable $\*X^i$ is composed of scalar variables that are nearly indistinguishable from one other, for instance, due to spatial proximity on a grid, e.g.~spatial fields in Earth system science. 
Deterministic nonlinear dynamical systems such as the Lorenz system \citep{lorenz1963deterministic} can be framed under this umbrella too. Here, the micro variables correspond to the deterministically related components or to a delay-embedding vector \citep{takens2006detecting}.

\subsubsection{Internal dynamics are `acausal'}
Finally, there is the case where the internal dynamics cannot be given a suitable causal interpretation. 
A few examples of this setting include:
\begin{itemize}
    \item[(i)] The internal structure of the vector variable is best modeled as an undirected graph, where the undirected edges only signify associations, without a straightforward causal interpretation. This may happen, for instance, for a random field in an equilibrium state. A particular example of such an undirected graph is a Markov random field, which satisfies the Markov property for undirected graphs. Such associations may be denoted by undirected edges within the vector variable. \cite{Lauritzen_chain_graphs} discuss the causal semantics of graphs that contain a mixture of directed and undirected edges, called chain graphs.  
    \item[(ii)] The internal structure can be modeled as a Bayesian network, which is a DAG that satisfies the Markov property for directed graphs. This is to say that the probabilistic structure of each vector variable is such that it admits a graphical interpretation together with a Markov property, but the graphical structure does not necessarily permit a causal interpretation \citep{Bareinboim_Hierarchy}. 
    \item[(iii)] The internal structure may not be amenable to a suitable or useful graphical interpretation.
\end{itemize}

\subsection{Agnosticism of data generation model to internal dynamics}
Naturally, all the internal dynamics that can be given a causal interpretation may be seen as special cases of the acausal setting, where the internal structure is graphical and the edges on the graph represent direct causal effects. Importantly, for the problem we wish to focus on, namely CD over vectors, we may remain entirely agnostic to the qualitative nature of the internal dynamics as long as the assumptions required for CD at the vector level are satisfied.
\cite{Wahl_foundations} discuss certain criteria under which an internal dynamics that corresponds to a causal DAG or a causal directed mixed graph (DMG) yields the faithfulness property at the vector level when an appropriate faithfulness property at the scalar level is assumed. They also prove that the Markov property at the micro level implies the Markov property at the macro level (i.e.~among vectors). In this sense, they start with a causal model, together with Markov and faithfulness properties w.r.t.~$\cG_{micro}$, at the micro-level and derive the conditions under which Markov and faithfulness hold at macro-level w.r.t.~macro graph $\cG$. 
\cite{Peña_faithfulenss_chain} discusses the faithfulness condition for chain graphs, and \cite{sadeghi_faithfulness, boeken_faithfulness} discuss the plausibility of faithfulness for Bayesian networks. It is important to note that these works consider the faithfulness condition on the micro-graph whose nodes correspond to scalar variables $X^i_j, \, \forall i,j$, from which it is not obvious how or whether faithfulness w.r.t.~the vector graph $\cG$ will follow.

An exhaustive discussion on when the various internal dynamics yield the Markov and faithfulness property at the vector level is outside the scope of the current work. In the following, we assume that the SCM over the vector variables $\*X^i$ is given by \cref{eq:SCM}, and the causal discovery~\cref{ass:CD} is satisfied. From these assumptions, the following remark follows naturally.

\begin{remark}[Correctness of Vector CD]
    Any causal discovery algorithm that is sound and complete (i.e.~discovers the Markov equivalence class of the ground truth graph in the infinite sample limit) given the Markov property, causal faithfulness, and causal sufficiency for scalar variables, is also sound and complete given \cref{eq:SCM} and \cref{ass:CD} for vector variables $\*X$. Note that the causal sufficiency assumption can be relaxed if the vectorized approach uses a CD algorithm that is sound and complete under latent confounding.
\end{remark}

\section{Component-wise approach to vector-valued causal discovery}\label{sec:scalar_approach}
In this section, we take a careful look at the component-wise approach to vector CD motivated in \cref{subsec:motivation_scalar}. 
Recall, that the component-wise approach (i) performs CD over all components $X^i_j, \ \forall i, j$, corresponding to the vector variables $\*X$, and, (ii) adopts a strategy to infer the graph $\widehat{\cG}$ over the vector variables, from the learned micro graph $\widehat{\cG}_{micro}$. There are two broad strategies to infer the macro graph from the micro graph:
\begin{enumerate}
    \item Define an \emph{edge-aggregation map} $\cE$ relative to the canonical partition $\cP = \{\*X^1, \ldots. \*X^N\}$. That is, $\cE$ maps a graph $\widehat{\cG}_{micro}$ over nodes $X^i_j, \ \forall i,j$ to a graph $\widehat{\cG}$ over nodes $\*X^i, \forall i$, or:
    \item Infer the skeleton of the macro graph $|\cG|$ and set of colliders $\coll^{\widehat{\cG}}$ on $\widehat{\cG}$ from the skeleton of the micro graph $|\widehat{\cG}_{micro}|$ and the set of colliders $\coll^{\widehat{\cG}_{micro}}$ on $\widehat{\cG}_{micro}$, potentially using additional CI tests. Then, apply orientation rules on the macro graph to obtain (the Markov equivalence class of) $\widehat{\cG}$. 
\end{enumerate}

We name the meta-algorithm corresponding to strategy (1) above \stovarg (for `Scalar-to-Vector'), given a CD algorithm $\cA$ and edge-aggregation map $\cE$, and state it in \cref{algo:S2V}. We will suppress the arguments of \stov whenever possible. 
The meta-algorithm corresponding to strategy (2) above, named \stovtwo (\cref{algo:S2Vtwo}), is relegated to the \cref{app:FCI}. 


\begin{algorithm}
\caption{Meta-algorithm \stovarg for vector CD using the component-wise approach}\label{algo:S2V}
\begin{algorithmic}[1]
\State \textbf{Input:} CD algorithm $\cA$, Edge-aggregation map $\cE$, data samples for variables $X^i_j, \ \forall i,j$.
\State \textbf{Output:} Causal Graph $\widehat{\cG}$ over vector variables $\*X^i, \forall i$
\State Step 1: Run $\cA$ on samples of $X^i_j, \ \forall i,j$, to obtain $\widehat{\cG}_{micro}$ over vertices $X^i_j , \ \forall i,j$.
\State Step 2: Use edge-aggregation map $\cE$ to obtain $\widehat{\cG}$ over vertices $\*X^i,  \ \forall i$, from $\widehat{\cG}_{micro}$
\State \Return $\widehat{\cG}$
\end{algorithmic}
\end{algorithm}


We now define the so-called \emph{majority edge-aggregation map} that requires a CPDAG as an input, a CPDAG is the markov equivalence class graph of a DAG, see \cite{spirtes_causation_1993} for details. This map counts different types of edge orientations between the scalar components corresponding to a pair of vector variables $\*X^{i}$ and $\*X^{j}$, and marks the edge on the macro graph $\widehat{\cG}$ using a majority vote. When considering the component-wise approach to vector CD, we will implicitly assume the majority edge-aggregation strategy is employed, unless stated otherwise. 
\begin{definition}[Majority edge-aggregation map]
    Let $\widehat{\cG}_{micro}$ be a CPDAG output by CD algorithm $\cA$. Further, let $\mathbf{dir}(i,j)$ denote the number of directed edges ($\to$) from $X^i_m$ to $X^j_n, \forall m,n$ and $\mathbf{undir}(i,j)$ denote the number of undirected edges ($\doublemarked$) from $X^i_m$ to $X^j_n, \forall m,n$. Note $\mathbf{undir}(i,j) = \mathbf{undir}(j,i)$. \\     
    Then an edge-aggregation map $\cE$ that yields a partially directed (but not necessarily acyclic) graph $\widehat{\cG}$ with conflicting edge orientations ($\confictmarked$) is such that: 
    
    (1) $\mathbf{X}^i \doublemarked\mathbf{X}^j $ in $\widehat{\cG}$ iff $\mathbf{undir}(i,j)>\mathbf{dir}(i,j)$ and $\mathbf{undir}(i,j)>\mathbf{dir}(j,i)$,
    
    (2) $\mathbf{X}^i \to\mathbf{X}^j $ in $\widehat{\cG}$ iff $\mathbf{dir}(i,j)>\mathbf{dir}(j,i)$ and $\mathbf{dir}(i,j) \geq \mathbf{undir}(i,j)$ (vice-versa for $\mathbf{X}^j \to\mathbf{X}^i $),
    
    (3)  $\mathbf{X}^i \confictmarked\mathbf{X}^j $ in $\widehat{\cG}$ iff $\mathbf{dir}(i,j)=\mathbf{dir}(j,i) \geq \mathbf{undir}(i,j)$.

\end{definition}

Correspondingly, we can also define a \emph{conservative edge-aggregation map} where edges in $\widehat{\cG}$ are marked as conflicts whenever $\mathbf{dir}(i,j)$ and $\mathbf{dir}(j,i)$ are both non-zero, and as directed or undirected depending on whether $\mathbf{dir}(i,j) \geq \mathbf{undir}(i,j)$ or not.
The ambiguity in choosing an appropriate edge-aggregation strategy points to a general ambiguity with the component-wise approach where edge orientations between scalar nodes corresponding to distinct vector variables conflict, as they could (and very likely would) in the finite sample case. In this case, there is no consistent way to aggregate edges except counting. A natural question emerges, if the there are $n$ edges of type "$\to$" and $n+m$ edges of type "$\leftarrow$" between vector variables $\*X$ and $\*Y$, how large should $m$ be for the edge between $\*X$,$\*Y$ to be assigned as a conflict? The majority and conservative edge-aggregation maps we propose present two ways to resolve this ambiguity, but it is unclear which map should be preferred in which setting. 

\subsection{Investigating the soundness of algorithm \ref{algo:S2V}}\label{subsec:latent_conf_soundness}

As discussed in \cref{subsec:internal_dynamics}, given our setting that satisfies~\cref{ass:CD}, a few choices for the internal dynamics include: (i) internal causal graph is a DAG, (ii) contains latent confounding (and/or cycles), (iii) faithfulness violations or (iv) the underlying scalar variables may not be well-defined causal entities at all.
Therefore, the soundness of \stov (or \stovtwo) is not automatically guaranteed if the CD algorithm $\cA$ is only sound for a restricted setting compared to the setting of the internal dynamics, for instance, if the internal dynamics contain latent confounding and the algorithm $\cA$ assumes causal sufficiency. 
However, if $\cA$ is sound for the particular internal dynamics setting, the soundness of \stov follows naturally. 
Furthermore, \stov has the advantage of additional edge orientations, as we illustrate with an example in \cref{fig:add_orient}. Naturally, the component-wise approach is not guaranteed to be sound when the internal dynamics are acausal. 
We now investigate the soundness of the \stov approach for the settings in \cref{subsec:internal_dynamics} where the internal dynamics are causal, but where algorithm $\cA$ is inappropriately chosen. 
\paranewspace{General Setting} Data for vectors $\*X^i \in \*X$ are generated by the SCM \cref{eq:SCM} which entails the causal graph $\cG$ and the probability distribution $P(\*X)$ over the vectors such that \cref{ass:CD} is satisfied.

\subsubsection{Latent confounding}\label{subsub:Scalar_latent}

\paranew{Specific Setting} 
The causal graph $\cG_{micro}$ among the scalar variables $X^i_j, \forall i,j$, that coarsens to the DAG $\cG$, is an acyclic directed mixed graph (ADMG), i.e.~it allows for latent confounders among $X^i_j$ and $X^i_k$, for any $i,j,k$. $P(\*X)$ is faithful to $\cG_{micro}$ with respect to the m-separation property \citep{Richardson_Ancestral}, i.e., no violations of faithfulness occur in the causal relationships among the scalar variables within the vector variable. 
Note, that we restrict our attention to \emph{acyclic} DMGs, and will deal with the issue of cyclic relationships in \cref{subsub:Scalar_cycles}, to clearly disambiguate these two settings and their soundness under \stov. Similarly, in the current setting we assume faithfulness to hold w.r.t.~$\cG_{micro}$, another condition that we will relax later, c.f.~\cref{subsub:Scalar_faithless}.

Here we focus on the prime representative of constraint-based CD algorithms requiring causal sufficiency, namely the PC algorithm \citep{spirtes_causation_1993}. 
The pitfalls of applying PC to the latent confounding setting are two-fold: (i) In the \textbf{skeleton discovery} phase, the PC algorithm would not delete all possible edges, 
due to the presence of inducing paths or possible d-separators that are not a subset of the adjacency set~\citep{spirtes_causation_1993}. 
(ii) The \textbf{edge orientation} phase might yield conflicting orientations due to causal sufficiency violations. E.g.~let $X \to Y \leftrightarrow Z \leftarrow W$ and no other edges between $X,Y,Z,W$ be the true graph. Then the PC skeleton phase soundly deletes the spurious edges, but the collider orientation phase detects a conflict in the orientation of $Y\doublemarked Z$ edge due to the colliders at $Y$ and $Z$. See Section 4.3 in \cite{colombo_stable} for a complete discussion on extensions of the PC algorithm that mark orientation conflicts.

\begin{figure}
    \centering
    \includegraphics[width=0.9\linewidth]{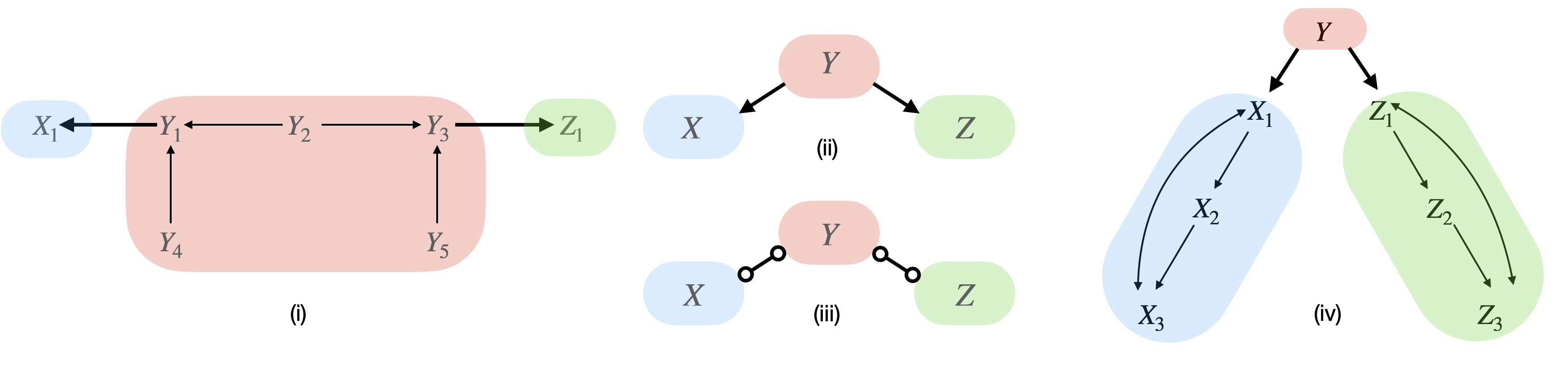}
    \caption{\stov (\cref{algo:S2V}) may result in more edge orientations across vector variables compared to 
    the vectorized approach: (i) Ground truth graph among vector variables $X, Y, Z$, (ii) The graph discovered by \stov together with the PC algorithm and majority edge aggregation strategy, (iii) The graph discovered by vectorized-PC, i.e.~PC over vector variables. (iv) Example to illustrate the accidental soundness of the PC algorithm when internal dynamics contain latent confounding. Here:  $X_3 \nind Z_3 | \{X_1, X_2, Z_1, Z_2\} $, but $X_3 \ind Z_3 | \{Y\} $. The edge $X_3 - Z_3$ is deleted even though $Y$ is not adjacent to $X_3$ or $Z_3$ in the true graph, but due to the induced adjacency resulting from the inducing path $X_3 (Z_3) \leftrightarrow X_1 (Z_1)\leftarrow Y$.}
    \label{fig:add_orient}
\end{figure}

In general, the PC algorithm cannot be guaranteed to be sound for the setting when latent confounding exists within a vector variable, even when the result of PC over all component vertices is edge-aggregated as in Step 2 of \stov (\cref{algo:S2V}).
In \cref{fig:add_orient} we present an example that illustrates how the PC algorithm may, nevertheless, learn the correct results but for the wrong reasons. In the example, note that  $X_3 \nind Z_3 | \{X_1, X_2, Z_1, Z_2\} $, namely $\adj(X_3, \cG_{micro}) \cup \adj(Z_3, \cG_{micro})$ is insufficient to separate $X_3$ and $Z_3$. However, the algorithm will not be able to d-separate $X_3$ and $Y$ or $Z_3$ and $Y$ due to the presence of inducing paths between both pairs. Thus, given that $Y$ is effectively adjacent to $X_3$ and $Z_3$, these vertices can be d-separated by $Y$. However, from such considerations it cannot be extrapolated whether all issues that internal latent confounding poses for the PC algorithm will go away.

We briefly remark that another candidate constraint-based algorithm that is sound under latent confounding and searches non-locally for separating sets, such as FCI \citep{spirtes_causation_1993}, may be applied at the micro level to yield a sound skeleton. Combined with an appropriate edge-aggregation strategy or the approach of \cref{algo:S2Vtwo}, the component-wise approach is sound in this setting. However, FCI requires many more CI tests compared to the PC algorithm, which already performs many unnecessary CI tests in the component-wise setting. We also note that if the internal dynamics contain no latent confounding, and $\cG_{micro}$ is a DAG, then \stov discovers more orientations compared to \stovtwo, that is to say, more orientations than in the Markov equivalence class of the ground truth DAG $\cG$, due to the reason illustrated in \cref{fig:add_orient}. 

\subsubsection{Cycles}\label{subsub:Scalar_cycles}
\paranew{Specific Setting} 
The causal graph $\cG_{micro}$ among the scalar variables $X^i_j, \forall i,j$, that coarsens to the DAG $\cG$, is a directed graph, i.e.~it allows for cycles among $X^i_j$ corresponding to a vector variable $\*X^i$ but no latent confounding. 
$P(\*X)$ is $\sigma$-Markovian and $\sigma$-faithful w.r.t.~$\cG_{micro}$. 

For a definition of the $\sigma$-Markov and faithfulness properties for cyclic graphs see \cite{Bongers_foundations}.
In this setting, we focus on the repercussions of cyclic relationships within the vector variable, without the violations of other standard CD assumptions. In \cite{mooij_constraint-based_2020} the consistency of constraint-based algorithms such as FCI and PC was extended to the $\sigma$-separation setting. In particular, 
they interpret causal sufficiency as a type of background knowledge that is compatible with the \emph{acyclification} operation (see Assumption 1 in their paper). Given this compatibility, they prove that PC algorithm is sound and complete (Theorem 2 and Corollary 1) at discovering the CPDAG of the acyclified graph. Based on this result, we make the following remark.

\begin{remark}[Soundness of \stov with Cycles]\label{rem:S2V_cycles}
    In the setting of \cref{subsub:Scalar_cycles},  
    \stov (\cref{algo:S2V}) with a causal-sufficiency and acyclicity requiring constraint-based CD algorithm $\cA$ that is sound and complete (such as PC), soundly discovers the CPDAG of $\cG$ in the infinite sample limit, with potentially additional orientations. 
\end{remark}
Note that the additional edges introduced in the acyclification of $\cG_{micro}$ are either across vector variables when there is already an edge incoming from one to the other or within a vector variable, therefore coarsening under the canonical partition to the macro graph does not introduce any spurious edges.
In the \cref{app:PC_cycles} we illustrate using an example the soundness of \stov for this setting. Note that the additional orientations are discovered for the reason illustrated in \cref{fig:add_orient}. 

\subsubsection{Faithfulness violations}\label{subsub:Scalar_faithless}

\paranew{Specific Setting} In this setting we note that \cref{ass:CD}, in particular faithfuless of $P(\*X)$ to $\cG$, could be satisfied, but it does not guarantee that faithfulness of $P(\*X)$ w.r.t~micro graph $\cG_{micro}$ will hold too. 
This could happen when internal dynamics violate faithfulness between $P(\*X)$ and $\cG_{micro}$, for instance, due to determinism or a synergistic relation (e.g.~with an XOR gate),  but $P(\*X)$ and $\cG$ remain faithful to each other. 

As the example below illustrates, \stov (\cref{algo:S2V}) is not sound in the setting when faithfulness may be violated internal to the vector variables. An algorithm that treats the vector variables as fundamental causal variables to perform CD over, dubbed with the prefix `vectorized', continues to perform soundly as expected under \cref{ass:CD}. One can construct similar examples with synergistic effects within the vector variable, see \cite{marx_faithfulness}.

    \begin{example}[Determinism]
    Given the following SCM for vector variables $\*X = (X^1,X^2)$ and $\*Y = (Y)$:
        \begin{equation*}
        \begin{split}
            &\*X :=  \begin{pmatrix}\eta_{X_1}  \\  f(\eta_{X_1})\end{pmatrix} \ , \quad 
            \*Y := g(\*X, \eta_Y) \ , 
        \end{split}
        \end{equation*}
        where $f(\cdot)$ is a deterministic function. Then: \\
         (i) \stov (and \stovtwo) with PC yields a completely unconnected macro graph, i.e., no edge between $\*X$ and $\*Y$. \\
        (ii) Vectorized-PC yields sound graph $\*X^1 \doublemarked \*X^2$ due to the inference $(X_1, X_2) \nind Y$. Recall that $(X_1, X_2) \ind Y \Leftrightarrow X_1 \ind Y \text{ and }  X_2 \ind Y|X_1$ (cf.~graphoid axioms). For the SCM above, however, $X_1 \nind Y$, which implies $\*X \nind \*Y$. 
        
    \end{example}


\subsection{Conclusions from the investigation of the component-wise approach}
We now draw conclusions from the investigation of the component-wise approach to different internal dynamics of the vector variables. \stov and \stovtwo (i.e.~\cref{algo:S2V} and \ref{algo:S2Vtwo}) with the PC algorithm are not provably sound in the latent confounding setting. 
In the cyclic setting, \stov and \stovtwo combined with a sound and complete CD algorithm that requires causal sufficiency and maps the independence model to the CPDAG (such as PC), is sound. In the faithfulness violating setting, \stov and \stovtwo are not provably sound. Finally, when the internal dynamics cannot be given a suitable causal interpretation, the soundness of \stov or \stovtwo cannot be proven generally, but needs to be studied on a case-by-case basis. 

\section{Aggregation approach to vector-valued causal discovery}\label{sec:agg_approach}

In \cref{subsub:agg}, we discussed a second approach to answering causal queries among vector variables, namely the aggregation approach. In this section we will investigate the aggregation approach for soundness w.r.t.~the task of vector CD. We advocated this view of task-dependent soundness of aggregation maps for causal discovery in \cref{subsub:motivation_abstraction}. 
The aggregation approach is commonly adopted in settings where the dimensions of the vector variables $\*X^i$ are high 
because, in the finite sample limit, high-dimensional variables pose a computational challenge, especially for CD methods that rely on conditional independence (CI) testing \citep{Shah_Peters}.
Thus, practitioners prefer to learn the causal graph among aggregate causal variables, that is, variables that are defined as an aggregation of observed high-dimensional variables to scalar (or, generally, lower-dimensional) variables. The reduced dimensions of the aggregate variables serves to make CI testing more reliable and allow for a broader class of conditional independence tests to be applicable in case the aggregate variables are scalar. In \cref{app:mult_CI}, we computationally evaluate a few approaches to multivariate CI testing based on combining univariate CI test statistics and empirically show that even for the linear Gaussian case it is difficult to maintain a low type I error and achieve high power for low sample sizes. Given this experimental evidence of the challenges of multivariate CI testing, we take a careful look at the practically desirable aggregation approach and determine the conditions for an aggregation map to be valid for CD. 



\subsection{Aggregation faithfulness and sufficiency properties}

We start by defining an \emph{aggregation map}. 
For the following, recall that $\*X^i, i\in[N],$ denote vector variables of dimension $d_i$, respectively, and $\*X$ signifies the set of all vector variables. 

\begin{definition}[Aggregation Map]
    An aggregation map over $\mathbf{X}$ 
    is defined as a set of maps $g_i(\cdot): \*X^i \mapsto \*Z^i \ , \forall i$, where $g_i (\cdot)$ is a map from the domain of the variable $\*X^i$ to the domain of the variable $\*Z^i$ such that dimension $ \bar{d}_i := dim(\*Z^i) \leq d_i$. In the following, we will refer to the $\*Z^i$'s as aggregate variables. The aggregate variables $\*Z^i, \forall i$ are denoted collectively by $\mathbf{Z}$.
    Note that the aggregate variables can be vector-valued.
\end{definition} 

We now define a central property of the aggregation maps, referred to as `aggregation graph existence', which is required for subsequent results in this section and \cref{sec:agg_consistency_adag}. 
\begin{property}[Existence of a Graph over Aggregate Variables]\label{ass:agg_graph_existence}
    An aggregation map $g(\cdot)$ over $\*X$ is said to satisfy aggregation graph existence, if the aggregate variables $\*Z$ are such that there exists a DAG that is Markovian and faithful to $P(\*Z)$.  
\end{property}
For the connection of the Markov and faithfulness properties to the graphoid axioms, see \cite{verma_pearl_networks, sadeghi_faithfulness}.
We now define certain desirable properties of the aggregation map that will help us define what we mean by the term ``valid aggregation map" in the context of vector CD.
We discuss the intuition behind and possible violations of the properties below.

We follow the following notation: let $g:=(g_1, \ldots, g_N)$ be the aggregation map that maps the tuple $\*X = (\*X^1, \ldots, \*X^N)$ to the aggregate variables $(\*Z^1, \ldots, \*Z^N)$ via $\*Z^i = g_i (\*X^i), \, i \in [N]$. Furthermore, we denote the element-wise aggregation map applied to a set $\cS_X \subset \*X$ by $g(\cS_X)$ and call it $\cS_Z$. Additionally, we continue to impose the general setting of \cref{ass:CD}, in particular, the Markov and faithfulness conditions are assumed for the SCM \cref{eq:SCM}.

\begin{property}[Aggregation Faithfulness]\label{prop:faithfulness}
    Given an aggregation map $g(\cdot)$ s.t.~$g(\*X) = \*Z$, if (conditional) independence between aggregate variables $\*Z$ implies (conditional) independence between corresponding vector variables $\*X$, where the conditioning sets are mapped from the vector to the aggregate level (or vice-versa) using the element-wise aggregation map, i.e., $$ \*Z^i \ind \*Z^j | \ \cS_Z \Rightarrow \*X^i \ind \*X^j | \ \cS_X \ , \ $$ then the map $g(\cdot)$ is said to satisfy the aggregation faithfulness property.
\end{property}
 A consequence of aggregation faithfulness is that aggregation does not lead to accidental cancellation of dependence among the aggregate-version of variables that are dependent at the vector-level.

\begin{property}[Aggregation Sufficiency]\label{prop:sufficiency}
   Given an aggregation map $g(\cdot)$ s.t.~$g(\*X) = \*Z$, if (conditional) independence between vector variables implies (conditional) independence between aggregate variables, where the conditioning sets are mapped from the vector to the aggregate level using the element-wise aggregation map, i.e., $$ \*X^i \ind \*X^j | \ \cS_X \Rightarrow \*Z^i \ind \*Z^j | \ \cS_Z \ , \ $$ then the map $g(\cdot)$ is said to satisfy the aggregation sufficiency property.
\end{property}

In other words, aggregation sufficiency implies that if two aggregate variables $\*Z^1$ and $\*Z^2$ are aggregates of vector variables $\*X^1$ and $\*X^2$ s.t.~$\*X^1$ and $\*X^2$ are not adjacent in $\cG$, then there must exist $\cS_Z \subset \mathbf{Z}\setminus \{\*Z^1, \*Z^2\}$, such that $\*Z^1 \ind \*Z^2 | \cS_Z$. Aggregation faithfulness, on the other hand, requires that the counterpart of $\cS_Z$ at the vector-level $\cS_X$ must be such that $\*X^1 \ind \*X^2 | \cS_X$. 

\paranewspace{Intuition for \cref{prop:faithfulness} and \ref{prop:sufficiency}} 
Aggregation faithfulness excludes the case that independence is discovered between the aggregate variables due to a fine-tuned choice of the aggregation map that disregards information that is responsible for the dependence of the vector variables. 
Aggregation sufficiency implies that the aggregation map is such that if two variables $\*X^1$ and $\*X^2$ are d-separable in $\cG$, then their aggregate counterparts $\*Z^1, \*Z^2$ must be independent given a (possibly empty) subset $\cS_Z$ of the aggregate variables, and it specifies a choice of $\cS_Z$ as the element-wise aggregate set corresponding to the separating set of $\*X^1$ and $\*X^2$ in $\cG$. This is termed as a `sufficiency' assumption because an aggregation map should retain all information from variables $\mathbf{X}$ that is required to establish independence between any pair of them.



We now discuss examples of violations of the properties discussed above, 
and in order to point out how they are pertinent for aggregate CD to be sound. 

\begin{example}[Violation of aggregation faithfulness]\label{ex:violation_agg_faith}
Let $\*X=(\*X^1, \*X^2)$, such that $d_1 = 1, d_2 = 2$, that is, explicitly, $\*X^2 = (X^2_1, X^2_2)$. Let the SCM be given by:
$$ \*X^2 := A \cdot \*X^1 + \mathbf{\eta} \ , $$
where $A = (\alpha, \beta)$ and $\mathbf{\eta} = (\eta_1 , \eta_2)$. 
Let the corresponding aggregate variables $(\*Z^1, \*Z^2)$ be defined such that $\*Z^1 = \*X^1, \*Z^2 = -\beta X^2_1 + \alpha X^2_2$. Then $\*Z^1 \ind \*Z^2$ but $\*X^1 \nind \*X^2$. Therefore, a fine-tuning of the aggregation map $g_2 (\cdot)$ that leads to a `creation' of independence violates aggregation faithfulness.
\end{example}

\begin{example}[Violation of aggregation sufficiency]\label{ex:violation_agg_sufficiency}
This example appears in Section 8.2.1 of  \cite{Wahl_foundations}.
Let $\*X = (\*X^1, \*X^2, \*X^3)$, such that $d_1 = 2, d_2 = 1, d_3 = 1$, i.e. $\*X^1 = (X^1_1, X^1_2)$.  The SCM is given by:
\begin{equation*}
\begin{split}
    X^1_1 &= \eta^1_1  \ , X^1_2 = \eta^1_2 \ ,\\  
     \quad \*X^2 &= X^1_1 + 2 X^1_2 + \eta_2 \ , \\  
     \quad \*X^3 &= X^1_1 + 2 X^1_2 + \eta_3 \ .
\end{split}
\end{equation*}
    
\noindent
Let the aggregate variables be $(\*Z^1, \*Z^2, \*Z^3)$ such that $\*Z^1 = X^1_1 + X^1_2\ ,\  \*Z^2 = X^2 $ and $\*Z^3 = X^3$. Then $\*X^2 \ind \*X^3 | \*X^1$ but $Z^2 \nind Z^3 | Z^1$. Therefore, the aggregation maps fails to `sufficiently' capture the vector variables together with their corresponding independence relations.
\end{example}

Examples \ref{ex:violation_agg_faith} and \ref{ex:violation_agg_sufficiency} serve to illustrate that the aggregation faithfulness and sufficiency properties are required when aggregate CD uses an algorithm that requires the Markov and faithfulness properties for soundness. Since the Markov and faithfulness properties are required to ascertain the presence or absence of causal edges from the independence relations of the observational distribution, we deem aggregation maps invalid for the task of vector CD when they don't soundly capture the independence relations of the vector SCM \cref{eq:SCM}. This observation helps us in defining a \emph{valid} aggregation map for vector CD, that we take up below.

\subsection{Valid aggregation maps for causal discovery}

In this section, we will define the term \emph{valid aggregation map} for the task of sound CD over vectors as motivated in \cref{subsub:motivation_abstraction}.
To define a valid aggregation map, we first need to revisit \emph{independence models}. 
An independence model $\cI(\*X)$ over a set of variables $\*X = \{\*X^1, \ldots \*X^N\}$ is defined as a collection of a triple of sets of indices $\*i, \*j, \*k \subset [N]$ such that, $(\*i, \*j,\*k) \in \cI(\*X) \ \text{if and only if} \ \*X^{\*i} \ind \*X^{\*j} | \*X^{\*k} $, where $\*X^{\*i}$ (resp.~$\*X^{\*j}, \*X^{\*k}$) refers to the set of variables in $\*X$ corresponding to the index set $\*i$ (resp.~$\*j,\*k$). In words, $\cI(\*X)$ is the set of all conditional independence statements that hold in the distribution $P(\*X)$. Given \cref{ass:CD} between $P(\*X)$ and graph $\cG$, an independence model over $\*X$ equivalently contains the set of all d-separation statements over $\cG$, and can, therefore, also be denoted as $\cI(\cG)$. Two independence models $\cI(\*X)$ and $\cI(\*Y)$ are said to be \emph{equivalent} when $(\*i, \*j, \*k) \in \cI(\*X) \Leftrightarrow  (\*i, \*j, \*k) \in \cI(\*Y)$.


Let $g:=(g_1, \ldots, g_N)$ be the aggregation map that maps the tuple of vector variables $\*X = (\*X^1, \ldots, \*X^N)$ to the aggregate variables $(\*Z^1, \ldots, \*Z^N)$ via $\*Z^i = g_i (\*X^i), \, i \in [N]$. 
Note that \cref{prop:faithfulness} implies $\cI(g(\*X)) \subseteq \cI (\*X)$ and \cref{prop:sufficiency} implies $\cI (g(\*X)) \supseteq \cI (\*X)$. Therefore properties \ref{prop:faithfulness} and \ref{prop:sufficiency} for map  $g(\cdot)$ are equivalent to the statement $\cI (g(\*X)) \equiv \cI (\*X). $
This observation helps define the soundness of aggregate CD as a proxy for vector CD. 

    



\begin{definition}[Valid Aggregation Map]\label{def:valid_agg}
    A map $g(\cdot)$ over $\*X$ is defined as a valid aggregation map if it:
    
(i) fulfills \cref{prop:faithfulness} and \cref{prop:sufficiency} or equivalently if, 

(ii) $\cI (g(\*X)) \equiv \cI (\*X)  $ 
\end{definition}
Therefore, in effect we can test the validity \cref{prop:faithfulness} and \ref{prop:sufficiency} by way of testing the equivalence of independence models at the level of vector variables and aggregate variables.

\begin{definition}[Equivalence of Aggregation Maps]\label{def:equiv_class}
    Two aggregation maps $g(\cdot)$ and $h(\cdot)$ over the set of variables $\*X$ are said to be equivalent if the respective independence models $\cI(g(\*X))$ and $\cI(h(\*X))$ are equivalent.
\end{definition} 

It is natural to consider such an equivalence of aggregation maps when determining validity as per \cref{def:valid_agg}, as any aggregation map that yields the correct independent model $\cI(\*X)$ over $\*X$ will qualify as a valid aggregation map. 

\begin{remark}[Invertible maps and equivalence of aggregation maps]\label{rem:bijective}
 Let $g(\cdot)$ and $h(\cdot)$ be aggregation maps over $\*X$, where the domain of $g(\*X)$ (resp.~$h(\*X)$) is denoted by $\cZ_g$ (resp.~$\cZ_h$). Let there exist a map $f(\cdot)$ such that $h_i = f_i \cdot g_i \ , \forall i$ where $f_i : \mathcal{Z}^i_g \to \mathcal{Z}^i_h$ is a measurable invertible map with $\mathcal{Z}^i_g$ denoting the domain of $g_i \in g$.  
 Then $h, g$ are equivalent aggregation maps as defined in \cref{def:equiv_class}. This follows from noting that $\*Z$ is related to random variables $\*Y$ with the invertible maps $f_i, \ \forall i$, as $ \*Y^i = f^i(\*Z^i), \forall i$ and for arbitrary random variables $A,B,C$:

 (i) $A\ind B \ | \ C \Rightarrow u(A)\ind v(B) \ |\ C$, where $u$ and $v$ are arbitrary measurable functions (follows from the graphoid axioms, see C2 in Sec 3.1 in \cite{lauritzen_graphical}),
 
 (ii) For a measurable invertible function $f$, $\sigma (A) = \sigma (f(A))$, where $\sigma(\cdot)$ denotes the sigma-algebra.
\end{remark}

Note that a version of this remark is discussed in Lemma 11 of \cite{schooltink_zennaro} in the context of consistent causal abstractions. In \cref{sec:agg_consistency_adag} we connect this remark to the concept of a \emph{tunable} aggregation map that we define in that section.



\section{Illustrating the pitfalls of classical approaches }\label{sec:experiments}
In this section, we compare the classical component-wise and aggregation approach to vector CD with the vectorized approach, and experimentally illustrate the pitfalls of the former two when the underlying causal model is given by \cref{eq:SCM} that were discussed in the previous sections. 
In \cref{sec:data_gen}, different modeling choices for vector-valued data were discussed within the framework of an SCM among vector-valued variables \cref{eq:SCM}. Here we focus on one data generation mechanism and provide experiments of other data generation mechanisms in \cref{app:further_exps}. 

\paranewspace{Data generation process} Our data generation mechanism assumes $d_{macro}$ number of vector-valued variables $\*X^i$, with $d_{micro}$ number of internal components, respectively. Given a randomly chosen causal order, the DAG between $\*X^i$'s is drawn such that each edge has a probability $1/(d_{macro}-1)$ of occurrence, i.e., the number of edges increases linearly with number of nodes $d_{macro}$. The edge density of cross-variable edges is denoted by $ext_{dens.}$ (for external density), where a maximum of $(d_{micro})^2$ edges among every pair of vector variables are permitted. The internal dynamics of each variable are modeled as a Markov random field (MRF) with an edge density of $int_{dens.}$. The sample size is denoted by $T$, and in case a time series is considered, the maximum lag length is denoted by $\tau_{max}$. $\tau_{max} = 0$ implies the non-time series case.

The mechanism of SCM \cref{eq:SCM} is considered to be linear and is drawn from a coefficient interval of $[c, 0.5]$ where $c \in [-0.5,0]$. That is, for $c = 0$ the coefficient average is positive and for $c = -0.5$, the coefficient average is zero. We name the latter case a `symmetric coefficient interval' and increasing $c>-0.5$ yields an increasingly asymmetric coefficient interval. The additive noises are chosen to be drawn from a multivariate Gaussian distribution.

\paranewspace{Background for experiments}
We consider three different approaches to CD over vector-valued data (i) vector (``vec''), (ii) aggregation, and, (iii) component-wise approach \stov (``component''). In approach (ii) we consider two aggregation maps: (a) averaging (``avg'') and (b) retaining the first principal component of each $\*X^i$ (``pca''). The causal discovery algorithm is chosen to be the PC algorithm \citep{spirtes_causation_1993} for non-time series data and PCMCI+ for time-series data \citep{runge2020discovering} (see \cref{app:further_exps} for experiments in the time series setting). The conditional independence test is partial correlation when the variables are univariate, and pairwise-partial correlation combined with the generalised covariance measure (GCM) and a Gaussian multiplier bootstrap as the aggregation measure of the pairwise univariate statistics as done in \cite{Shah_Peters}. More details on the choice of the multivariate test can be found in the \cref{app:mult_CI}. The results shown are over 100 repetitions and error bars indicate standard errors.

\begin{figure}
    \centering
    \includegraphics[scale=0.3]{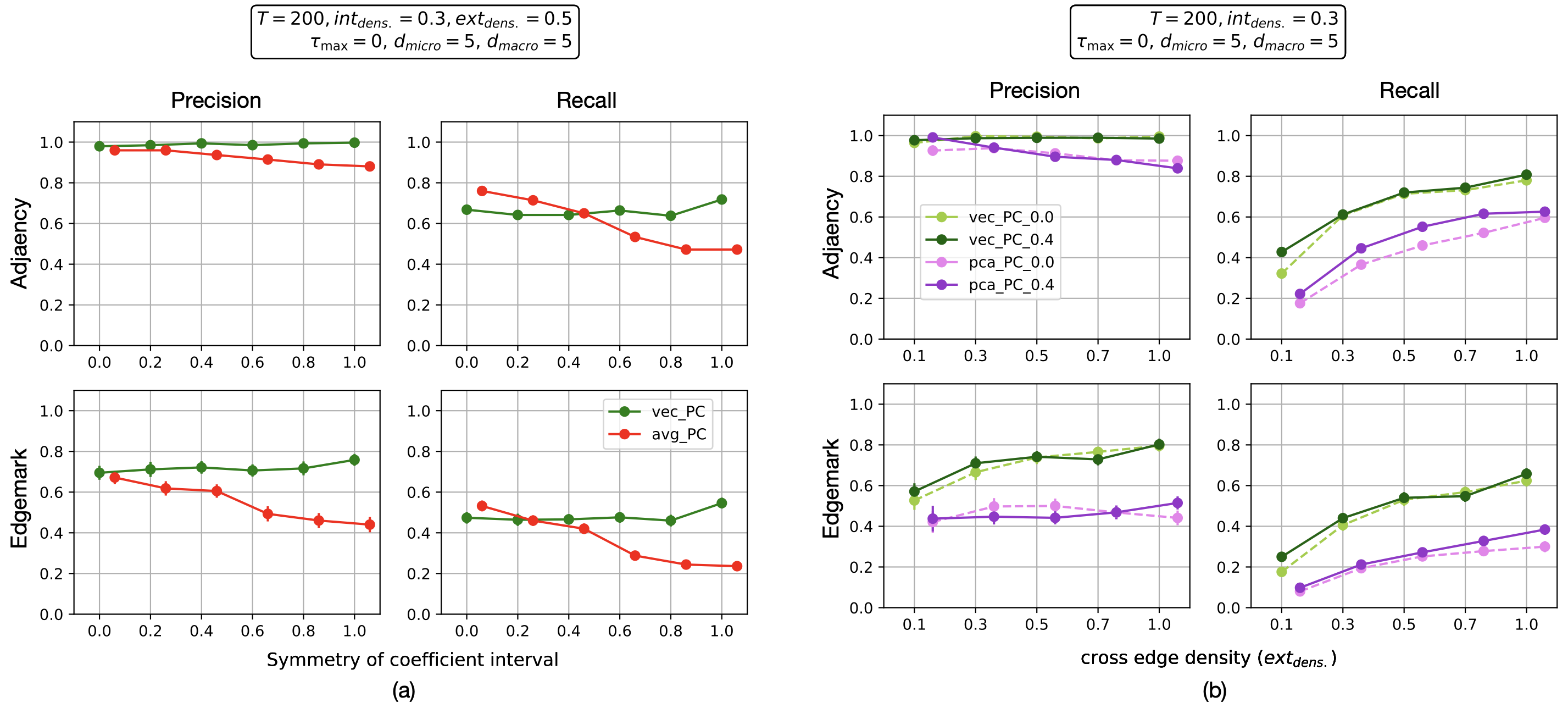}
    \caption{Comparing vector-PC with PC over aggregated variables. Plots of precision and recall of adjacencies and edgemarks, respectively: (a) PC over averaged variables (`$\text{avg}\_\text{PC}$') versus PC over vector-valued variables (`$\text{vec}\_\text{PC}$'), (b) PC over the first principal component of each variable (`$\text{pca}\_\text{PC}$') versus PC over vector-valued variables (`$\text{vec}\_\text{PC}$'). The floats $0.0$ and $0.4$ denote a weighting strategy of the principal components that was used in the data generation, see main text for more details. $T$ denotes sample size, $\tau_{max}$ the maximum lag-length ($=0$ implies non-time series data), $int_{dens.}$, resp.~$ext_{dens.}$, density of edges internal to (resp.~across) the vector variables. The error bars indicate standard errors.}
    \label{fig:avg_pca_vec}
\end{figure}

\begin{figure}
    \centering
    \includegraphics[scale=0.3]{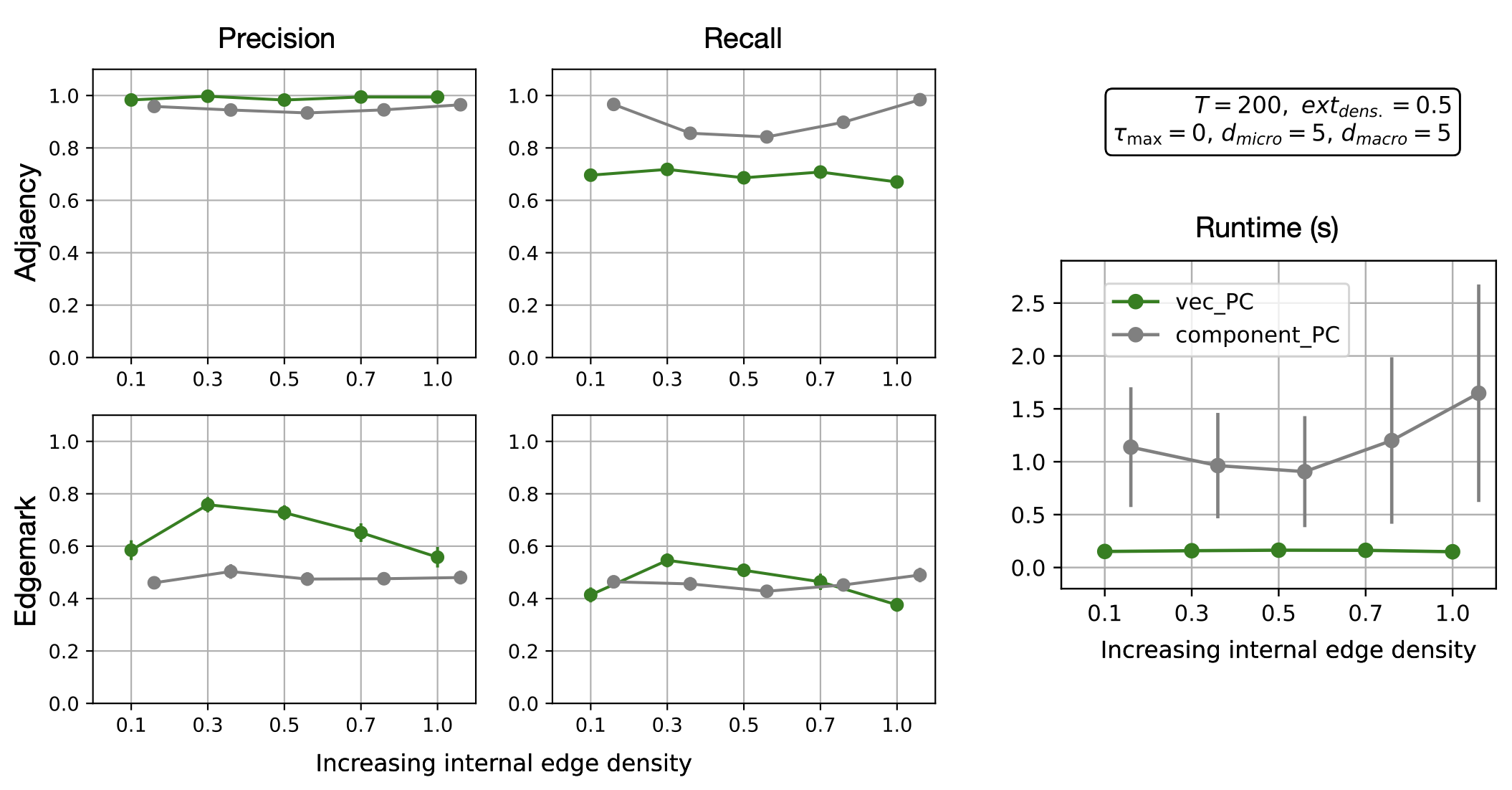}
    \caption{Comparing vector-PC with component-wise PC with the majority edge aggregation strategy. Plots of precision and recall of adjacencies and edgemarks respectively. T denotes sample size, $\tau_{max}$ the maximum lag-length ($=0$ implies non-time series data), $int_{dens.}$, resp.~$ext_{dens.}$, density of edges internal to (resp.~across) the vector variables. The error bars indicate standard errors.}
    \label{fig:component_vec}
\end{figure}

\paranewspace{Experimental configurations}
We consider the precision and recall of the adjacencies and the edgemarks in the output graph as compared to the ground truth graph.  
\begin{itemize}[leftmargin=*]
    \item[(i)] In \cref{fig:avg_pca_vec} (a), we compare the vector and average-aggregation approach with decreasing $c$ from $0$ to $-0.5$ (equivalently increasing symmetry of the coefficient interval from $0$ to $1$). As illustrated in \cref{fig:scalar_agg_pitfalls}, a symmetric coefficient interval leads to a higher probability of effect cancellation when variables are averaged. From the figure we can see that as the symmetry of coefficient interval increases, avg-PC performs worse than vec-PC, in line with the explanation of effect cancellation. The computation times for avg-PC and vec-PC are comparable and therefore not displayed. 
    \item[(ii)] In \cref{fig:avg_pca_vec} (b), we compare the vector and the PCA-aggregation approach for increasing cross-edge density $ext_{prob}$, for two settings: The first principal component of the additive noise corresponding to each variable $\*X^i$ is weighted by a weight of (1) $0.$ and (2) $0.4$ in the data generation process. We do not weigh it higher than $0.4$ because the coefficient strength is capped at an absolute value of $0.5$.     
    Overall, the performance of vec-PC is higher. 
    As the cross-edge density increases, the adjacency recall of both vec-PC and pca-PC increases due to more causal edges among components across variables, however, a lower principal component weight (PC wt.) affects the performance of pca-PC negatively, while leaving vec-PC's performance unchanged. Furthermore, the adjacency precision of pca-PC decreases with increasing $ext_{dens.}$ as it is increasingly unlikely that the principal component chosen corresponds to the actual causally relevant components within the vector variable. The computation times for pca-PC and vec-PC are comparable and therefore not displayed. 
    
    \item[(iii)] In \cref{fig:component_vec}, we compare the vector and the component-wise approach (with \stov) for increasing internal edge density $int_{dens.}$. The lower precision and higher recall of the component-wise approach indicate that it detects a higher number of edges on average, and its lower edgemark precision (and marginally lower edgemark recall) indicate that the edge aggregation strategy for link orientation does not work as well as that of the vectorized approach. Furthermore, the component-wise approach suffers from a much higher computation time due to the increasing number of CI tests it needs to perform as the internal edge density increases, which vector CD is immune to. 
\end{itemize}

\paranewspace{Conclusions} The experiments of this section illustrate that the aggregation and the component-wise approach exhibit modes of failure and inefficiency when their respective assumptions cannot be guaranteed to hold for the SCM between vector-valued variables. Therefore in the lower-dimensional settings (respectively higher sample size setting), the vectorized approach shows better performance due to better conformity with assumptions and reliability of the CI tests in the low dimensional setting. In \cref{app:mult_CI} we illustrate the performance of multivariate CI tests to emphasize that the vectorized approach is only as reliable as the CI tests underlying it. 


\section{Aggregation consistency scores and \adag wrapper}\label{sec:agg_consistency_adag}

\subsection{Testing validity of aggregation maps}

The valid aggregation maps introduced above pose a problem in their usability for the task of causal discovery among vector variables: An aggregation map is considered valid when its independence model coincides with the independence model of the underlying vector variables. However, if we had access to the independence model over the vector variables to verify whether an aggregation map is valid, constructing aggregation maps to simplify the discovery task would become irrelevant. To circumvent this circularity in testing the validity of an aggregation map, we now introduce two scores, an independence and a dependence score for aggregation consistency, that together contribute to the joint \emph{aggregation consistency (AC) score}.

\paranewspace{A note on the usage of `Validity' and `Consistency'} In \cref{def:valid_agg}, we defined a \emph{valid} aggregation map for the task of vector CD. Ideally, we would like to test the validity of the aggregation map based on this definition, but this relies on the availability of \emph{true} independence statements over $\*X$. Since our intention in this section is to establish a test for the validity of the aggregation map in the absence of any ground truth knowledge about $P(\*X)$ and $\cG$, we refer to the scores we establish as aggregation \emph{consistency} scores. We thus intend to highlight that what we can and do test is whether the aggregation map implies testable statements whose fulfillment (or lack thereof) points to an internal (in)consistency. In short, the validity of aggregation is tested by means of consistency in the absence of ground truth knowledge. Recent works have explored quantifying consistency as a technique for CD method evaluation, eg.~\cite{faller_self-compatibility, faltenbacher_incoherency} (see \cref{subsec:related} for further references).

\paranewspace{Notation and Setting} Variables $\*X$ are generated by the SCM \cref{eq:SCM}, and satisfy \cref{ass:CD}. The ground truth DAG over $\*X$ is denoted by $\cG$. The aggregation map $g(\cdot)$ satisfies \cref{ass:agg_graph_existence}, and the graph that is Markovian and faithful to aggregate variables $\*Z$ is denoted by $\cG_Z$. 
The CD algorithm is denoted by $\cA$ and under \cref{ass:CD} outputs the Markov equivalence class (i.e.~CPDAG) of the ground truth DAG. 

\subsubsection{Scores for aggregation consistency}\label{subsub:scores}
Constraint-based CD algorithms are ideally designed such that they can infer the independence model of a set of variables 
from the fewest possible CI tests among the variables. For instance, the PC algorithm improves upon its precursor SGS algorithm in that it restricts the search space for separating sets between variables to the adjacencies of the pair of variables while prefering smaller conditioning sets to make inferences \citep{spirtes_causation_1993}. Note that the information contained in the independence model is equivalent to that contained in the causal graph up to Markov equivalences since Markov equivalent graphs imply the same d-separation statements.  
Given a sound and complete CD algorithm $\cA$ over $\*X$ that assumes Markov and faithfulness property, with correct CI tests, let the set of conditional independence statements discovered during its run be denoted by $\tested (\*X) \subseteq \indmod(\*X)$. Note that $\tested$ consists of tuples of three sets of indices $\*i, \*j, \*k$ s.t.~if $(\*i, \*j,\*k) \in \tested \Rightarrow \*X^\*i \ind \*X^\*j | \*X^\*k$. Here, despite $\tested (\*X)$ being a subset of the full independence model $ \indmod(\*X)$, $ \indmod(\*X)$ is still fully recoverable from $\tested$ because the latter yields the Markov equivalence class of $\cG$ (if all CI tests are correct), which in turn sufficiently determines the $\indmod(\*X)$. 

In the following, let algorithm $\cA$ over aggregate variables $\*Z = g(\*X)$ yield the set of independence statements $\tested (\*Z)$. Note that an aggregation map $g$ is valid if $\indmod(\*Z) \equiv \indmod(\*X)$. Since a sound and complete CD algorithm $\cA$ over $\*X$ yields $\tested (\*X)$ which completely determines $\indmod (\*X)$, our aim in defining the aggregation consistency scores is to test whether $\tested (\*Z)$ is compatible with $\tested (\*X)$, without needing to run $\cA$ on $\*X$. 


\paragraph{Testing independence statements}
 We introduce a first score to test the validity of the aggregation map using $\tested(\*Z)$ that relies on the following simple remark: each conditional independence statement $I_Z \in \tested(\*Z)$ can be tested over the vector variables $\*X$ for soundness. For instance, if $I_Z = (\*Z^1 \ind \*Z^2 | S_Z)$, then the corresponding statement over $\*X$, $I_X = (\*X^1 \ind \*X^2 | S_X)$ with $g(S_X) = S_Z$, can be tested. Each time an agreement is found the score increases, and the final count is normalized by the cardinality of $\tested(\*Z)$. 

Let the set of independence statements over the vector variables $\*X$ that are conducted corresponding to each element in $\tested(\*Z)$ be denoted by $\widehat{\indmod}(\*X)$. 
Because of the order in which the sets $\tested(\*Z)$ and $\widehat{\indmod}(\*X)$ are discovered $\widehat{\indmod}(\*X) \subseteq \tested(\*Z)$. 
Formally, we define the \emph{independence score for aggregation consistency} $c_{ind} (\*Z, \*X, \cA)$ given aggregate (high-level) variables $\*Z$, vector (low-level) variables $\*X$, and a CD algorithm $\cA$ as follows:
\begin{equation}\label{eq:cind}
    c_{ind} (\*Z, \*X, \cA) =  \frac{|\mathfrak{C}^{ind}|}{|\mathfrak{C}^{ind}|+ |\mathfrak{I}^{ind}|} \ ,
\end{equation}  
where $\mathfrak{C}^{ind} := \{I| I \in \tested(\*Z) \cap \widehat{\indmod}(\*X)\}$ denotes the set of consistent independence statements over $\*Z$ that can be verified over $\*X$ and $\mathfrak{I}^{ind} := \{I|I \in \tested(\*Z) \setminus \widehat{\indmod}(\*X)\}$ denotes the set of inconsistent independence statements. In the following, we suppress the arguments of $c_{ind}$ wherever possible. 
The consistency score $c_{ind}$ when the numerator and denominator are zero is set to one by convention. 
This approach for measuring aggregation consistency relies on the correctness of the individual CI tests $I_Z \in \tested(\*Z)$ and $I_X \in \widehat{\indmod}(\*X)$. The score $c_{ind}$ is indicative of independence consistency when $I_Z$ is correct, which is likely because $\*Z$ are lower-dimensional variables. The problem is graver for $I_X$ that is tested over the (possibly high-dimensional) $\*X$, because when the conditional independence $I_X$ is either (i) incorrectly rejected even though independence is true (i.e.~false positive or a type I error occurs) , or (ii) incorrectly not rejected even though dependence is true (i.e.~false negative or a type II error occurs), then the consistency score $c_{ind}$ will be falsely deflated (resp.~inflated). However, a strategy that uses individual independence tests over (possibly high-dimensional variables) $\*X$ rather than inferring the causal graph over $\*X$ using $\tested(\*Z)$ is still more reliable because the latter is prone to cascading errors if the CI statements in earlier stages of the algorithm are incorrect. The proofs for all the results in this section are presented in \cref{app:proofs_agg}.


\begin{lemma}\label{lem:c_ind_faithfulness}
    Let variables $\*X$ satisfy \cref{eq:SCM} and \cref{ass:CD}.
    Additionally, the aggregation map $g(\cdot)$ that yields aggregate variables $\*Z$ from variables $\*X$ satisfies \cref{ass:agg_graph_existence}.  
    Assuming all conditional independence tests are correct, if $g(\cdot)$ satisfies aggregation faithfulness (\cref{prop:faithfulness}) then the independence score $c_{ind}$ is maximal (i.e.~it equals 1). 
\end{lemma}

In order to prove the other direction, namely that a maximal independence score implies aggregation faithfulness, we require the CD algorithm to fulfill certain properties that we term causal input completeness.
\begin{definition}[Causal Input Completeness]\label{def:input_completeness}
    Let the ground truth causal DAG over a set of variables $\*Y$ be denoted by $\cG_\*Y$. Let a sound and complete CD algorithm $\cA$ over $\*Y$ output a CPDAG $\widehat{\cG}_\cC$,  and assume correct conditional independence tests. Let $\cG_\cD$ be a DAG compatible with $\widehat{\cG}_\cC$ and let $\cT$ denote a valid topological ordering of $\cG_{\cD}$.    
    Then $\cA$ is said to be \emph{causal input complete} w.r.t.~$\cG_\*Y$ if the set of conditional independence statements $I_{\cA} (\*Y)$ found during the run of $\cA$ are such that $(\cT, I_{\cA} (\*Y))$ form a causal input list as defined in \cite{verma_pearl_networks} (the term `stratified protocol' is also used often) for all valid $\cT$. $\cA$ is said to be \emph{causal input complete} if it is causal input complete with respect to all ground truth DAGs $\cG_\*Y$.  
\end{definition}

In words, an algorithm $\cA$ is said to be  \emph{causal input complete w.r.t.~a causal graph $\cG_\*Y$}, if all CI statements corresponding to the local Markov property, namely `every variable is independent of its non-descendants given its parents', was checked during the run of $\cA$. In \cref{app_subsec:causal_input} we provide details on this definition and an example of a ground truth graph w.r.t.~which the PC algorithm is causal input complete.

\begin{corollary}[Abridged corollary 1 from \cite{verma_pearl_networks}]\label{cor:verma_result_graphoid}
    A triplet is d-separated in the DAG if and only if the conditional independence statement corresponding to the triplet can be derived from the causal input list using the semi-graphoid axioms. 
\end{corollary}
Here, a triplet refers to a tuple of index sets $(\*i, \*j, \*k)$ and a triplet is said to be d-separated in a DAG if the corresponding sets of nodes $(\*X^\*i, \*X^\*j, \*X^\*k)$ are such that $\*X^\*i\ind_d \*X^\*j | \*X^\*k$ holds in the DAG. We are now ready to prove the opposite direction of \cref{lem:c_ind_faithfulness}.

\begin{lemma}\label{lem:c_ind_faithfulness_complete}
    Let the assumptions of \cref{lem:c_ind_faithfulness} be satisfied and assume in addition that the 
    CD algorithm $\cA$ is causal input complete, then the independence score $c_{ind}$ is maximal (i.e.~it equals 1) if and only if $g(\cdot)$ satisfies aggregation faithfulness (\cref{prop:faithfulness}).     
\end{lemma}

\begin{remark}[Augmented CD algorithm]\label{rem:augmented_alg}
    Any sound and complete CD algorithm $\cA$ w.r.t.~a ground truth graph $\cG_\*Y$ over variables $\*Y$ can be augmented to be causal input complete in the following way: Construct a candidate DAG $\cG_\cD$ of the CPDAG $\widehat{\cG}_\cC$ output by $\cA$. Test all CI statements that are a consequence of the local Markov property on $\cG_\cD$, that were not already found during the run of $\cA$, and append them to the set $I_{\cA} (\*Y)$ of independence statements found during the run of $\cA$. 
\end{remark}



\begin{figure}
    \centering
    \includegraphics[scale=0.35]{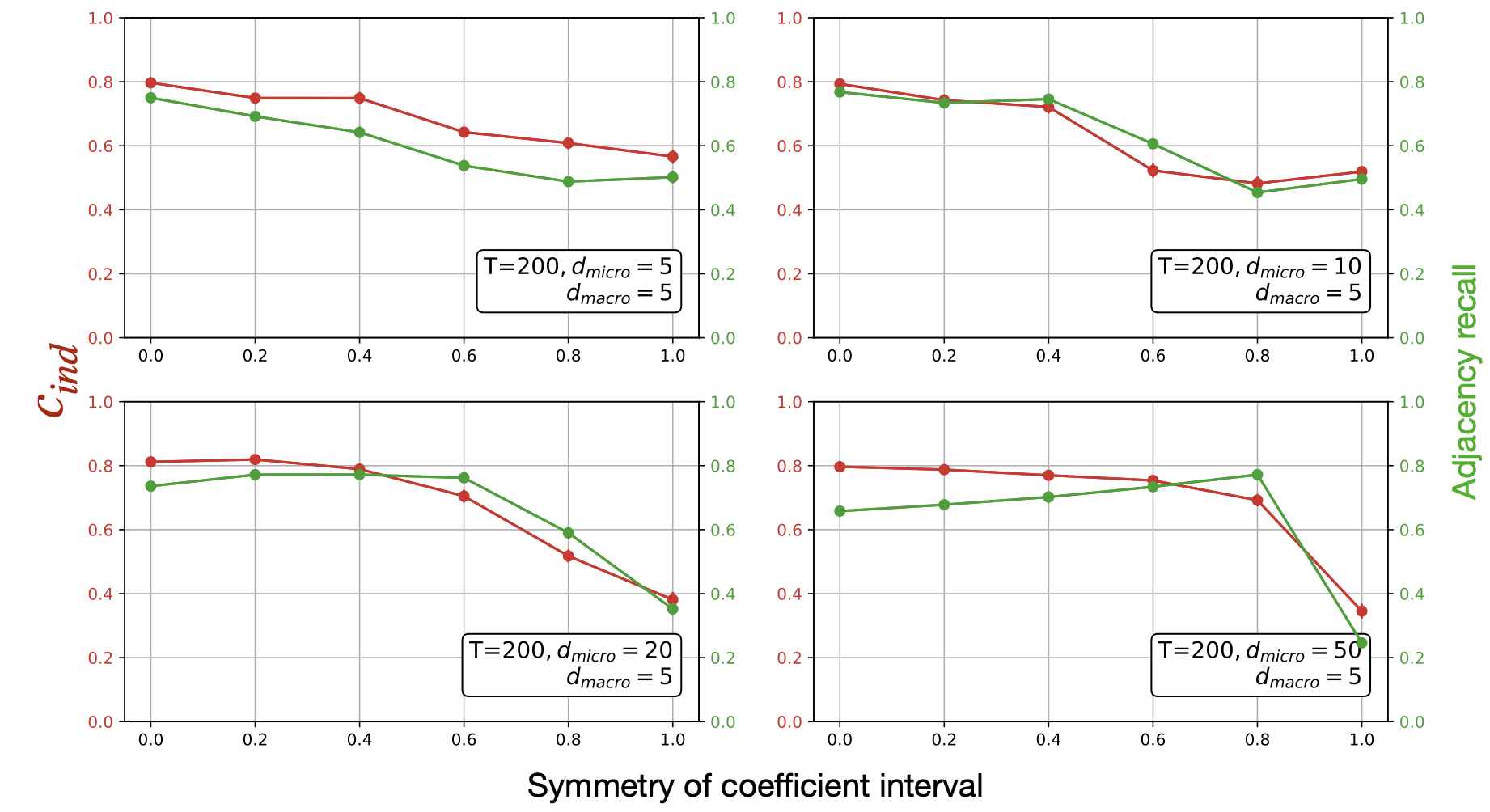}
    \caption{$c_{ind}$ faithfully captures aggregation faithfulness violations: On the x-axis is the degree of symmetry around zero of the interval from which causal coefficients are chosen. This is similar to the data generation process described in \cref{sec:experiments}, where 1 signifies perfect symmetry (namely positive and negative coefficients are equally likely), whereas 0 signifies only positive coefficients. On the left y-axis is the $c_{ind}$ score for the averaging aggregation map. On the right y-axis is the recall of adjacencies of the graph resulting from applying PC over averaged data. $c_{ind}$ faithfully captures the trend of recall without ground truth knowledge, since a decrease in recall, namely increase in false negative rate, is matched by a reduced independence score, i.e., CI tests at the aggregate level do not hold true at the vector level. 
    $T$ stands for sample size, $d_{macro}$ for the number of vector variables and $d_{micro}$ for the number of scalar components within every vector variable.}
    \label{fig:plot_cind_recall}
\end{figure}

    
In \cref{fig:plot_cind_recall}, we plot the performance of $c_{ind} (\*Z, \*X, \cA)$, where the $\*X$ are generated as per \cref{eq:SCM} for a linear mechanism and multivariate Gaussian additive noises, the aggregation map $\*Z = g(\*X)$ is given by the element-wise average of $\*X^i \in \*X, \ \forall i$, and the algorithm $\cA$ is the PC algorithm. On the x-axis is the average of the range from which the coefficients for the SCM \ref{eq:SCM} are chosen, on the y-axis to the left is $c_{ind}$ and to the right is adjacency recall of the graph discovered by $\cA(\*Z)$ compared to the true graph $\cG$ over $\*X$. For a range of $d_i = |\*X^i|$ one sees that the independence $c_{ind}$ score faithfully represents the closeness of the aggregate graph to the ground truth graph, even though the PC algorithm was not augmented to be causally complete for each graph. We plot the comparison of $c_{ind}$ to the structural hamming distance (SHD) to the ground truth graph in \cref{app:further_exps_agg}.

\begin{corollary}\label{cor:agg_super_graph_1}
    Let the assumptions of \cref{lem:c_ind_faithfulness_complete} hold.
    Then, if the independence score for aggregation consistency $c_{ind} = 1, $ 
    then the graph $\cG_Z$ which is Markov and faithful to $P(\*Z)$ is s.t. $sk(\cG) \subseteq sk(\cG_Z)$, where $sk(\cdot)$ denotes the skeleton of the graph in the argument. 
    
    
\end{corollary}

Note that $c_{ind}$ is akin to a precision measure for the independence model over $\*X$, which measures the fraction of independence statements over $\*Z$ that are verifiable over $\*X$, i.e.~at the vector level, out of all the independence statements found over $\*Z$.

\paragraph{Testing dependence statements}
 In order to test the correctness of the dependence statements implicit in $\tested(\*Z)$, we need to define a strategy to verify dependence statements over $\*X$ that are found during the course of running the CD algorithm $\cA$ over $\*Z$. The brute-force way to do this would be to test every dependence statement that holds over $\*Z$ and testing the counterpart in $\*X$, however, this is an unfeasible task that practically requires us to rely on multiple CI tests on $\*X$ that we set out trying to avoid. We term such a strategy to compute the \emph{dependence score for aggregation consistency} (termed $c_{dep} (\*Z, \*X, \cA)$) a \emph{complete} strategy. Formally the dependence score for aggregation consistency is given by,
\begin{equation}\label{eq:cdep}
    c_{dep}(\*Z, \*X, \cA) =  \frac{|\mathfrak{C}^{dep}|}{|\mathfrak{C}^{dep}|+ |\mathfrak{I}^{dep}|} \ ,
\end{equation}
where $\mathfrak{C}^{dep} := \{L | L\in \cD^{dep}(\*Z) \cap \widehat{\cD}^{dep}(\*X)\}$ denotes those statements in the complete set $\cD^{dep}(\*Z)$ of dependence statements over $\*Z$ that also hold over $\*X$, captured in $\widehat{\cD}^{dep}(\*X)$.   Similarly,  $\mathfrak{I}^{dep} := \{L | L\in \cD^{dep}(\*Z) \setminus \widehat{\cD}^{dep}(\*X)\}$ denotes the set of inconsistent dependence statements, i.e.~those dependencies in $\*Z$ that cannot be verified at the level of $\*X$.

\begin{lemma}\label{lem:c_dep_sufficiency}
    Let variables $\*X$ satisfy \cref{eq:SCM} and \cref{ass:CD} and assume that conditional independence tests are correct.
    Additionally, the aggregation map $g(\cdot)$ that yields aggregate variables $\*Z$ from variables $\*X$ satisfies \cref{ass:agg_graph_existence}.  If the strategy to compute $c_{dep} (\*Z, \*X, \cA)$ is complete then $c_{dep}$ is maximal (i.e.~it equals 1) if and only if $g(\cdot)$ satisfies aggregation sufficiency as defined in \cref{prop:sufficiency}. 
\end{lemma}

In practice, as we discussed, it is unfeasible to employ a complete strategy to compute $c_{dep}$, and therefore completely capturing violations of aggregation sufficiency is not possible due to computational constraints. 
For practical use, we define the following strategy to test dependence statements. 
First, we need to assume that given a dependence statement $\*Z^i \nind \*Z^j | \cS_Z$, if the counterpart $\*X^i \nind \*X^j | \cS_X$ (where $\cS_Z = g(\cS_X)$) holds too, then $\*X^i \in \adj(\*X^j, \cG)$. In \cref{algo:cdep_eff}, we propose an \emph{incomplete} strategy which employs p-values of the CI tests to choose the most reliable dependence statement and compute the \emph{effective} dependence score $\bar{c}_{dep} (\*Z, \*X, \cA)$ for aggregation consistency. In \cref{app:complete_strategy_dep}, we provide a general \cref{algo:meta_cdep} that allows for opting various types of complete and incomplete strategies to test dependence statements at the aggregate level. 
\begin{algorithm}
\caption{Algorithm for effective dependence score $\bar{c}_{dep} (\*Z, \*X, \cA)$ for aggregation consistency}\label{algo:cdep_eff}
\begin{algorithmic}[1]
\State \textbf{Input:} CD algorithm $\cA$, data samples for vector-variables $\*X$ and aggregate variables $\*Z$. \vspace{0.2cm}
\State \textbf{Step 0:} Initialize $\bar{c}_{dep}(\*Z, \*X, \cA) = 0$.\vspace{0.2cm}
\State \textbf{Step 1:} List all unordered pairs of adjacent nodes $(\*Z^i, \*Z^j)$ in the (skeleton) graph $\widehat{\cG}$ discovered from $\tested (\*Z)$, and denote the list as $\cD^{adj} (\*Z)$. \vspace{0.2cm}
\State \textbf{Step 2:} For each $L_{ij} \in \cD^{adj} (\*Z)$ such that $\*Z^i \in \adj(\*Z^j, \widehat{\cG} )$, sort all (conditional) independence statements tested among $\*Z^i$ and $\*Z^j$ in $\cA$ in increasing order of p-values, and denote the sorted list for each pair $(i,j)$ as $\cD^{adj}_{ij} (\*Z)$. Specifically, an element $L \in \cD^{adj}_{ij}(\*Z)$ is a set of indices  $\*k \subset [N]$ such that $\*Z^i \nind \*Z^j | \*Z^{\*k}$, where $\*Z^{\*k}$ denotes the set of variables in $\*Z$ corresponding to the index set $\*k$. Note that all p-values considered for $\cD^{adj}_{ij}(\*Z)$ are smaller than the significance level $\alpha$. \vspace{0.2cm}
\State \textbf{Step 3:}  Repeat this step for all elements $L_{ij}\in \cD^{adj}(\*Z)$: For the first element $L_1 \in \cD^{adj}_{ij}(\*Z)$ (i.e.~with the smallest p-value
), test whether $\*X^i \ind \*X^j | L_1^X$, where $L_1^X$ corresponds to the set of variables in $\*X$ with the index set $L_1$. If dependence is found, increase $\bar{c}_{dep}$ by one and append the pair $(i,j)$ to the list $\widehat{\cD}^{adj} (\*X)$ of adjacencies in $\*X$.  \vspace{0.2cm}
\State \textbf{Step 4:} Normalize $\bar{c}_{dep}(\*Z, \*X, \cA)$ by the cardinality of $\cD^{adj}(\*Z)$.    \vspace{0.2cm}  
\State \Return $\bar{c}_{dep}(\*Z, \*X, \cA)$
\end{algorithmic}
\end{algorithm}

Formally the effective dependence score for aggregation consistency is given by,
\begin{equation}\label{eq:cdep_eff}
    \bar{c}_{dep}(\*Z, \*X, \cA) =  \frac{|\mathfrak{C}^{adj}|}{|\mathfrak{C}^{adj}|+ |\mathfrak{I}^{adj}|} \ ,
\end{equation}
where $\mathfrak{C}^{adj} := \{L | L\in \cD^{adj}(\*Z) \cap \widehat{\cD}^{adj}(\*X)\}$ denotes the set of adjacencies that are consistent between $\*Z$ and $\*X$ (namely that an adjacency in $\*Z$ implies an adjacency in $\*X$), and $\mathfrak{I}^{adj} := \{L | L\in \cD^{adj}(\*Z) \setminus \widehat{\cD}^{adj}(\*X)\}$ denotes the set of inconsistent adjacencies. 
As with the score $c_{ind}$ in \cref{eq:cind}, for the dependence score, too, we suppress the arguments wherever possible and set it to one by convention when both numerator and denominator are zero. 

Note that the asymmetry in the computation of $c_{ind}$ (\cref{eq:cind}) and $\bar{c}_{dep}$(\cref{eq:cdep_eff}) arises from the fact that for a pair of variables, typically a CD algorithm requires only one CI statement to hold for edge-deletion to take place, whereas several conditional dependence statements among a pair of variables are found in the course of the algorithm. This is not exactly true for order-dependent or conservative versions of CD algorithms, but more dependence statements per pair of variables on an average justifies the strategy to adapt p-value sorting for the effective dependence score. Note also that other types of sorting strategies, including testing more than one dependence statement per pair of variables are also possible. Recent works have proposed checks for the Markov property that can be adapted to dependence scores for aggregation consistency \citep{ramsey_markov_checker, jeong_ejaz_testingcausalmodelshidden}. 
 
\paragraph{Joint aggregation consistency score} We now combine the independence and the dependence score for aggregation consistency into a joint score. 
\begin{definition}\label{ac}
    The joint aggregation consistency ($AC$) score is defined as the arithmetic mean of the independence score $c_{ind}$ \cref{eq:cind} and the dependence score $c_{dep}$ \cref{eq:cdep}, i.e.,
    \begin{equation}\label{eq:ac_score}
        AC (\*Z, \*X, \cA) := \frac{c_{ind}(\*Z, \*X, \cA) + c_{dep}(\*Z, \*X, \cA)}{2} \ .
    \end{equation}
    Replacing $c_{dep}(\*Z, \*X, \cA)$ with $\bar{c}_{dep}(\*Z, \*X, \cA)$ (\cref{eq:cdep_eff}) yields the effective (joint) aggregation consistency score $\overline{AC}$.
\end{definition}

\begin{theorem}\label{thm:acs_consistency}
    Let the assumptions of \cref{lem:c_ind_faithfulness_complete} be satisfied and let the strategy adopted to measure $c_{dep}$ be complete. 
    Then, the joint aggregation consistency score is maximal (i.e.~it equals 1), if and only if $g(\cdot)$ is a valid aggregation map as defined in \cref{def:valid_agg}. 
\end{theorem}


\begin{figure}
    \centering
    \includegraphics[scale=0.25]{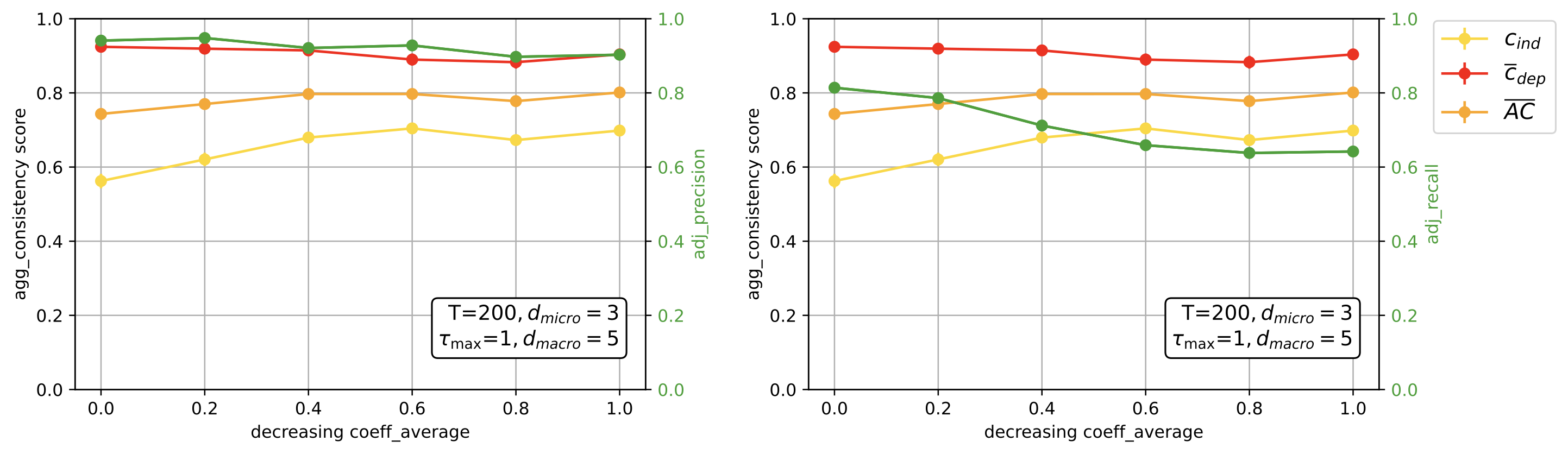}
    \caption{Illustrating the relevance of both independence and (effective) dependence (low dimensional time series setting). Even though $\bar{c}_{dep}$ is high overall, its trend for both adjacency precision and recall as the coefficient average decreases (namely the coefficient interval becomes more symmetrically distributed around zero), matches the trend of the adjacency recall (curve in green) better than $c_{ind}$'s alone does. $\overline{AC}$ stands for the effective aggregation consistency score (\cref{eq:ac_score}), $T$ stands for sample size, $\tau_{max}$ for the maximum lag length in the time series, $d_{macro}$ for the number of vector variables and $d_{micro}$ for the number of scalar components within every vector variable.}
    \label{fig:plot_all_scores}
\end{figure}

In \cref{fig:plot_all_scores} we plot the performance the independence, dependence and the joint score and compare it to adjacency precision and recall to the true graph.

\subsection{Consistency-guided aggregation for causal discovery: \adag wrapper}

Motivated by the results that allow us to check the validity of the aggregation map from the aggregation consistency scores introduced above, in this section we propose a wrapper \adag (for adaptive aggregation) for consistency-guided aggregation for the task of vector CD, in the absence of ground truth knowledge.
\adag works with any of the three scores (\cref{eq:cind}), \cref{eq:cdep}, and \cref{eq:ac_score}) that we defined in the previous section, a tunable aggregation map (defined below), and a sound and complete constraint-based CD algorithm $
\cA$ (that is additionally causal input complete if the score used contains the independence score \cref{eq:cind}). The idea behind \adag is to start with a desired value of the chosen aggregation consistency score, and tune the aggregation map until this desired score is achieved. We present the pseudocode for \adag in \cref{algo:adag}. 

\adag is adaptable to those aggregation maps for which there exists a tuning parameter that can increase the dimension of the aggregate variables in order to capture more information from the low-level variables $\*X$. For instance, if the aggregation map is such that it selects the first $n$ principal components of each $\*X^i \in \*X$, such that for each $i, \ |\*Z^i| = n$, then increasing $n$ makes the aggregate variables higher-dimensional and thus more expressive. We define an aggregation map with such a tuning parameter a \emph{tunable aggregation map}, and denote it by $g^{\*m}(\cdot) := (g^{m_1}_1, \ldots, g^{m_N}_N)$ where $\*m  = (m_1, \ldots, m_N) \in \mathbb{N}^N$ is the tuning parameter, and $m_i$ denotes the dimension of the aggregate variable $\*Z^i$. The maximum value of $\*m$ is given by  $\*m = (d_1, d_2, \ldots d_N)$.
The averaging aggregation map can similarly be interpreted as a tunable aggregation map $g^\*m (\cdot)$, such that $m_i=1 \Rightarrow \*Z^i = \Sigma (\*X^i)$ where $\Sigma (\cdot)$ denotes the averaging operation. For $m_i>1$, the aggregation map can be defined as a weighted sum of the arguments where each weight $\*w^i_{m_i} \in (0,1)^{d_i}$, is a $d_i$ dimensional array that takes values in the interval $(0,1)$, such that the aggregation map $g^{m_i}_i$ yields the aggregate variable $\*Z^i  = \*w^i_{m_i} \cdot \*X^i$. Naturally, no two weight vectors $\overline{\*w}_{m_a} := (\*w^1_{m_a}, \ldots, \*w^N_{m_a})$ and $\overline{\*w}_{m_b} :=(\*w^1_{m_b}, \ldots, \*w^N_{m_b})$, for $m_a, m_b \in \*m$ must be linearly dependent. That is, the matrix $\*W = (\*W_1, \ldots, \*W_N)$ where $\*W_i = (\overline{\*w}_1, \ldots, \overline{\*w}_{d_i})^T$ must be full rank, such that increasing $m_i$'s results in more informative aggregate variables. 

\paranewspace{Notation} We denote the set of the three scores by $\aggscore$, and an element of $\aggscore$ by $q$ which we refer to as $q$-score. That is, the $q$-score could be $
c_{ind}$ (\cref{eq:cind}), $c_{dep}$ (\cref{eq:cdep}) or $AC$ (\cref{eq:ac_score}). The value of the $q$-score is denoted by $\alpha_q$, with $0 \leq \alpha_q \leq 1$ (denoted $\widehat{\alpha}_q$ when the value is estimated). That is, \adag has arguments $(q, \alpha_q, \cA, g^{\*m})$ that we will suppress when possible. The maximum value of $\*m$ is denoted by $\operatorname{max}(\*m)$. The update statement $\*m \pluseq [1]^N$ implies that all components of $\*m$ are increased by one, up to their respective maximum value, i.e.~$m_i \leq d_i$ always holds. Therefore, $\*m < \operatorname{max}(\*m)$ implies $\exists \ i \in [N]$ s.t.~$m_i < d_i$. We remark here other types of updates of $\*m$, for instance, increasing the tuning parameter for only a subset of all variables or continuing to increase the tuning parameter for a set of variables despite other variables having reached maximum tuning, is also possible.

\begin{algorithm}
\caption{Wrapper \adag$(q, \alpha_q, \cA, g^{\*m})$ for consistency guided aggregation for vector CD}
\begin{algorithmic}[1]
\State \textbf{Input:} Data samples for variables $\*X$, sound and complete (and possibly causal input complete) CD algorithm $\cA$, Tunable aggregation map $g^{\*m}(\cdot)$, $q$-score $q \in \aggscore$ with desired value $0 \leq \alpha_q \leq 1$
\State \textbf{Output:} Causal Graph ${\cG}_Z$ over aggregate variables $\*Z^i$ with q-score $\alpha_q$

\State Initialize $\*m=[0]^N$ and $\widehat{\alpha}_q = 0$ 
\While{$\widehat{\alpha}_q < \alpha_q$ and $\*m < \operatorname{max}(\*m)$}
\State Define aggregate variables $\*Z_\*m = g^\*m (\*X)$
\State Run CD algorithm $\cA$ over $\*Z_\*m$ to yield graph $\widehat{\cG}$ and compute value of q-score $\widehat{\alpha}_q$
\If {$\widehat{\alpha}_q < \alpha_q$ and $\*m < \operatorname{max}(\*m)$}
        \State $\*m \pluseq [1]^N$
\Else
    \State{break}
\EndIf
\EndWhile
\State \Return ${\cG}_Z$
\end{algorithmic}
\label{algo:adag}
\end{algorithm}


\paranewspace{Termination condition of \adag}
\adag has two termination conditions, and will terminate when either of them is satisfied:
(i) The condition $\widehat{\alpha}_q \geq \alpha_q$ is achieved, namely the empirical q-score exceeds the desired value of the q-score, and 
(ii) the tuning parameter $\*m$ of the tunable aggregation map $g^\*m(\cdot)$ cannot be increased any further for any of the vector variables $\*X^i \in \*X$. Note that according to \cref{rem:bijective}, $\*m = \operatorname{max}(\*m)$ implies that $\widehat{\alpha}_q = 1$, however, this is only true in the infinite sample limit and thus this practical condition of termination is required in the finite sample setting. 

\begin{theorem}\label{thm:adag_soundness}
    Let variables $\*X$ be generated by \cref{eq:SCM} and satisfy \cref{ass:CD}.  
    In the infinite sample limit, \adag$(q, \alpha_q, \cA, g^{m})$ together with a causal input complete CD algorithm $\cA$, a tunable aggregation map $g^m (\cdot)$ that satisfies \cref{ass:agg_graph_existence}, and the $q$-score given by 
    
    (i) the joint aggregation consistency score (\cref{eq:ac_score}) with $\alpha_q = 1$, outputs a CPDAG ${\cG}_Z$ that is equivalent to the CPDAG of the ground-truth graph $\cG$ over variables $\*X$.  
    
    (ii) the independence score for aggregation consistency $c_{ind}$ (\cref{eq:cind}) with $\alpha_q = 1$ outputs a CPDAG ${\cG}_Z$ s.t. $sk(\cG) \subseteq sk(\cG_Z)$, where $sk(\cdot)$ denotes the skeleton of the graph in the argument.
\end{theorem}

\begin{figure}
    \centering
    \includegraphics[scale=0.35]{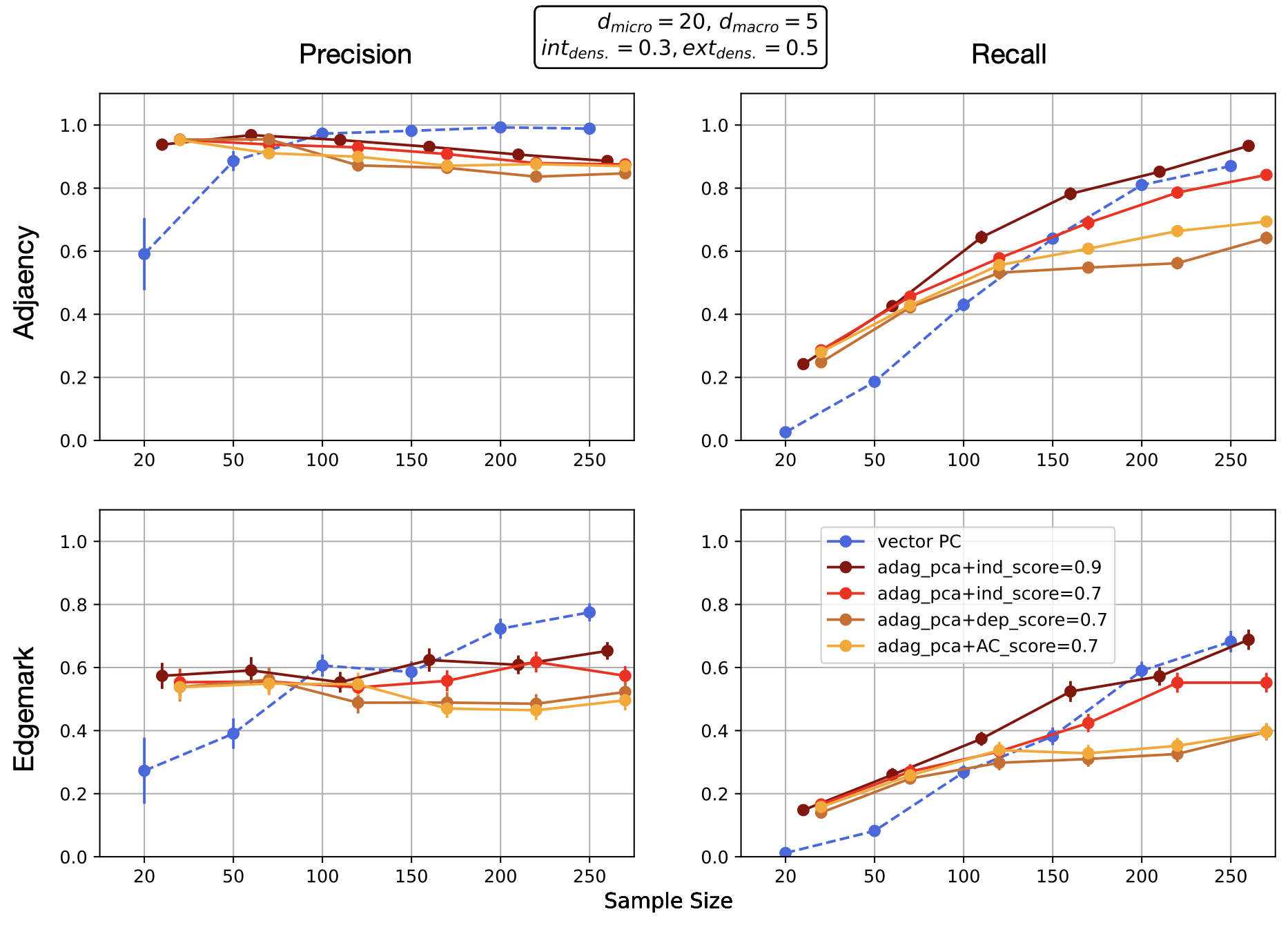}
    \caption{\adag performance comparison with vector CD: Adjacency (respectively edgemark) precision and recall for vector CD (denoted 'vec') and \adag withindependence score $c_{ind}$ (\cref{eq:cind}), effective dependence score $\bar{c}_{dep}$ (\cref{eq:cdep_eff}), and, effective $AC$ score (\cref{eq:ac_score}), as sample size increases. For lower sample sizes \adag with either of the three scores performs better. For higher sample sizes vector CD performs better than \adag for a lower score ($=0.7$), whereas \adag gains in the recall when the score is increased to $0.9$. 
    Other parameters of the data-generating process are shown in the box in the top centre. $d_{macro}$ is the number of vector variables and $d_{micro}$ is the number of scalar components within every vector variable. $int_{dens.}$ is the internal edge density, and $ext_{dens.}$ is the edge density across vector variables.}
    \label{fig:adag_vs_vec}
\end{figure}

In \cref{fig:adag_vs_vec}, the performance of \adag is compared with vector CD, using the PC algorithm. The tunable aggregation map $g^{\*m} (\cdot)$ is given by extracting the first $\*m$ principal components of the vector variables $\*X$ element-wise. The results shown are for 100 repetitions of experiments with different linear Gaussian SCMs corresponding to the parameters shown in the figure on the top center. In \cref{app:further_exps_agg}, we show experiments for vector variables with other internal dimensions. The experiments help illustrate two central points: (i) \adag outperforms vector CD for lower sample sizes, but as sample size increases, the value of the $q$-score, $\alpha_q$, has to be adjusted accordingly to keep \adag competitive; (ii) \adag combined with the independence score typically performs better than when combined with dependence or joint AC-score, which may be traced back to the particular incomplete strategy used to compute the independence score outlined in \cref{algo:cdep_eff}. In \cref{app:complete_strategy_dep} we provide an overview of strategies to compute the dependence score and do not exclude the possibility that other user-designed dependence scores better fitting to the problem may work better with \adag. However, we emphasize that already optimizing for the independence score yields a much better CD result than aggregate CD that does not check for consistency, as it ensures fewer wrongful edge deletions, which is a typical problem of constraint-based CD \citep{malinsky_cautious}. 

\section{Application of \adag to spatio-temporal causal discovery} \label{sec:savar}

\begin{figure}
    \centering
    \includegraphics[width=\linewidth]{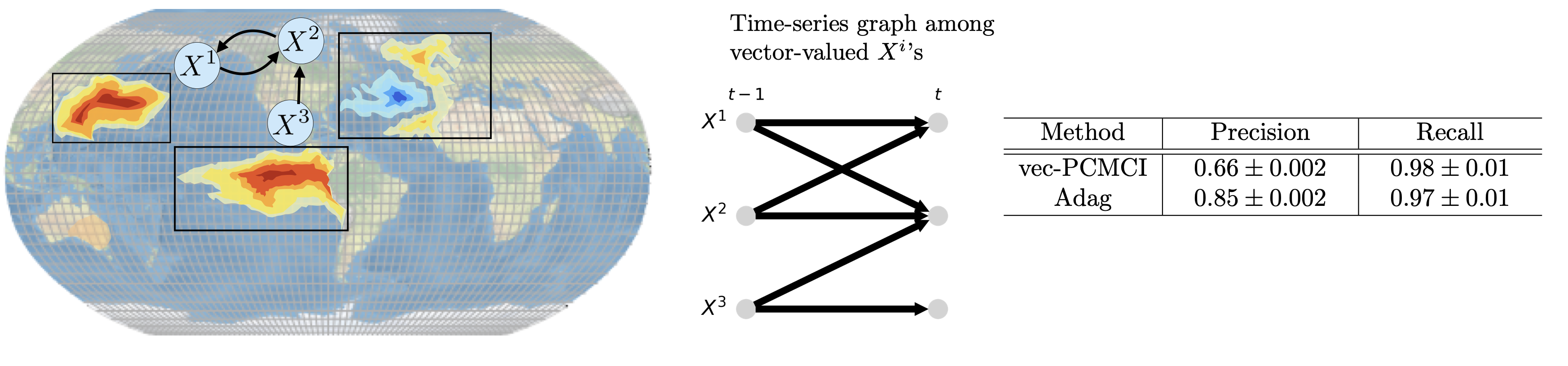}
    \caption{Application of \adag to spatiotemporal stochastic climate model data: On the \textbf{left} is a schematic illustration of the SAVAR model (figure adapted from \cite{tibau_savar} where it is licensed under the CC BY 4.0 license: \url{https://creativecommons.org/licenses/by/4.0/}). The high-dimensional grid-level variables $X^i$'s (the boxes around shaded regions) are defined by experts. The summary graph shown is a toy-model, with all edges at lag 1, and no real teleconnections are shown. 
    In the \textbf{center} is the ground truth time-series graph corresponding to the SAVAR model on the left with $d_{macro} = 3, \ d_{micro} = 49, \ T= 200, \ \tau_{max} = 1$. Note that all edges are at lag 1.  On the \textbf{right} is the table of precision and recall for vec(torized)-PCMCI and \adag-PCA-PCMCI methods (PCMCI is a version of PCMCI+ that assumes only lagged links). \adag boasts a higher gain in precision while not sacrificing severely in recall compared to the vectorized version. }
\label{fig:savar}
\end{figure}


Spatio-temporal data is found commonly in domain sciences where vector-valued variables are sensitive to spatial location (e.g.~indices of scalar components correspond to distinct spatial grid points) and the recorded data as a time series, such as in climate science and neuroscience. Within climate science, one sub-field of research is to study teleconnections that connect climatic phenomena at widely separated spatial locations \cite{Kretschmer_Teleconnections}. To supplement the analysis of real-world climate teleconnections, and help evaluate methods that draw inferences from spatio-temporal datasets with teleconnections, a synthetic benchmark model named SAVAR (for Spatially-Aggregated VAR model) was introduced in \cite{tibau_savar}. The SAVAR model distinguishes between the recorded observations on the spatial grid (`grid-level') from the modes of climate variability (`mode-level') that are viewed as lower dimensional maps of the grid-level observations. The SAVAR model is then defined as a VAR model at the mode level, with a corresponding representation at the grid level, using the pseudo-inverse of the grid-to-mode map. We refer the reader to the original paper for further details. In \cref{fig:savar} (left), the SAVAR model is illustrated on the Earth grid (figure adapted from \cite{tibau_savar}). 
The vector-valued grid-level data produced from the SAVAR model can be used to benchmark vector-valued causal discovery methods since the ground truth causal graph between the climate modes follows from the underlying VAR model. 

\paranewspace{Experimental configuration and results} We choose the PCMCI algorithm \citep{Runge17Science} as the CD algorithm for time-series data and compare vectorized CD with consistency-guided aggregate CD using the \adag wrapper. We choose retention of the first $\*m$ principal components as the tunable aggregation map and opt for a value of $\alpha_q = 0.8$ for the independence consistency score (\cref{eq:cind}) due to its efficiency. The number of vector variables $d_{macro}$ equals $3$, and the components therein $d_{micro} $ equals $ 49$ ($d_{micro}$ is set to a perfect square for ease since each vector variable is internally modeled as a $\sqrt{d_{micro}} \times \sqrt{d_{micro}}$-dimensional grid). We generate a time-series of 200 points from the displayed time-series graph (\cref{fig:savar} center) using the SAVAR model package (\url{https://github.com/xtibau/savar}) and compare the average adjacency precision and recall of edges of vectorized-PCMCI and \adag-PCA-PCMCI for 100 repetitions. The results are shown in \cref{fig:savar}.

\section{Conclusions and outlook}\label{sec:conclusions}

In this work, we have taken an exhaustive look at vector-valued CD by first studying traditional component-wise and aggregation approaches and identifying their failure modes in various settings. 
We showed that the vectorized approach overcomes common pitfalls of the currently prevalent component-wise and aggregation-based strategies. This opens up the use of vectorized constraint-based CD methods together with suitable multivariate CI tests to the problem settings where an underlying vectorized causal model may be assumed, from non-time series data in economics to spatio-temporal data in Earth sciences.
Secondly and crucially, for the setting when the macro variables are high-dimensional, or the sample sizes are low, and the aggregate approach serves as a tool for efficiency and robustness, we proposed a novel concept of a valid aggregation map for the task of vector CD. Furthermore, in the absence of ground truth knowledge to test the validity of an aggregation map, we introduced three types of aggregation consistency scores that motivate consistency-guided aggregation. Operationally, we introduced the \adag wrapper for adaptive aggregation that, given a target aggregation consistency score, a tunable aggregation map, and a causal discovery algorithm, tunes the aggregation map to achieve the desired value of the consistency score. 

Several key issues did not receive significant attention in our work to contain the scope, that we deem relevant to a holistic discussion on the topic of vector-valued CD. Our work builds on the premise that the vector-valued data is modeled by a causal model among the vector-valued variables, and that key CD assumptions such as the Markov property and faithfulness hold at the vector level. We did not, however, justify the strength of these enabling assumptions at the vector-level as opposed to the micro-level, or exhaustively study how likely they are to hold in realistic distributions. For the case that the internal dynamics correspond to a Bayesian network, \cite{ParKas17} empirically studied the extension of faithfulness to the macro-level, whereas for the internal dynamics corresponding to DAGs and DMGs, \cite{Wahl_foundations} have provided sufficient conditions for macro-level Markov and faithfulness properties. However, many other internal dynamics that are discussed in our work have not enjoyed similar systematic analyses, nor is it immediately clear which model of internal dynamics fits best to which real-world application. Depending on such considerations, the order of suitability of approaches to vector CD may change. 

We presented an argument in this work in favor of task-dependent aggregations that need not require establishing the existence of a high-level SCM, as required by the consistent abstraction framework, to be practically useful.
Our adaptive approach to achieving a sufficiently valid aggregation map using the pre-defined aggregation consistency scores hinges on the correctness of CI tests, which are the bedrock of constraint-based CD methods. However, CI queries in the vector-valued setting are strictly harder than in the scalar setting, therefore for very high-dimensional vectors, the independence score may be high due to the reduced power of multivariate CI tests, and similarly, the dependence score may be low. That is, the reliability of the individual aggregation consistency scores is weakened in the low-power or high type I error regime of CI tests. This challenge can be circumnavigated in part by using the total aggregation consistency score that weighs independence and dependence consistency scores equally. There are, however, immitigable challenges here too since computing the complete dependence score requires several CI tests. Novel tests for aggregation consistency that do not rely entirely on high-dimensional CI testing constitute interesting future research directions. 

Even though the adaptive approach to aggregation focuses squarely on constraint-based CD methods in our work, we note here that the general philosophy is easily extendable to other classes of CD algorithms (and causally insufficient settings). This is because the graph output by any algorithm implies a set of conditional (in)dependence statements that can be tested in turn at the macro level. The ambiguity arises from prescribing which separation statements to prioritize or weigh highly, and constraint-based algorithms provide an easy remedy to this ambiguity by saving a set of CI tests that are used to learn the output graph. Furthermore, the \adag wrapper can be used in conjunction with user-defined aggregation consistency scores that weigh testable (in)dependence statements differently and are more aptly fit to particular problems. 





\section{Acknowledgments}
We thank Simon Bing for sharing code to generate Markov random field data, Rebecca J.~Herman for comments on an early draft of this paper, and Tom Hochsprung for discussions on multivariate conditional independence testing. U.N. and J.R.~received funding from the European Research Council (ERC) Starting Grant CausalEarth under the European Union's Horizon 2020 research and innovation program (Grant Agreement No. 948112). J.W.~received support from the German Federal Ministry of Education and Research (BMBF) as part of the project MAC-MERLin (Grant Agreement No. 01IW24007). This work used resources of the Deutsches Klimarechenzentrum (DKRZ) granted by its Scientific Steering Committee (WLA) under project ID 1083.


\bibliography{main.bib}
\bibliographystyle{plainnat}

\appendix

\section{Remarks on the component-wise approach to vector CD}\label{app:FCI}

\subsection{\stovtwo}

We now define the meta-algorithm \stovtwo (\cref{algo:S2Vtwo}), corresponding to the second type of strategy that may be adopted in the component-wise approach, see first paragraph in \cref{sec:scalar_approach} for an overview of the strategies. It differs from \stov (\cref{algo:S2V}) because it only implements the skeleton phase of algorithm $\cA$ and replaces the edge-aggregation step with the orientation phase at the vector-level directly. The implementation of orientation rules on the vector level implies that \stovtwo does not require an edge-aggregation strategy as input, however, this also implies that the certain orientations that could be found by applying orientation rules at the component-level (see \cref{fig:add_orient}), are not found in the \stovtwo approach. 
\begin{algorithm}
\caption{Meta-algorithm \stovtwo for vector CD using the component-wise approach}\label{algo:S2Vtwo}
\begin{algorithmic}[1]
\State \textbf{Input:} CD algorithm $\cA$, data samples for variables $X^i_j$ \vspace{0.2cm}
\State \textbf{Step 0:} Run the skeleton phase of algorithm $\cA$ on the scalar variables $X^i_j$ to obtain the skeleton graph $|\widehat{\cG}_{micro}|$, i.e.~graph with undirected edges signifying adjacencies. \vspace{0.2cm}
\State \textbf{Step 1:} Coarsen $|\widehat{\cG}_{micro}|$ among scalar variables $X^i_j, \forall i\neq j$, to the undirected graph $\widehat{\cG}$ among vector variables $\*X^i, \forall i$ s.t. $\*X^i - \*X^j$ in $\widehat{\cG}$ \emph{iff} there exists $a,b$, s.t. $X^i_a - X^j_b$ in $|\widehat{\cG}_{micro}|$. \vspace{0.2cm}
\State \textbf{Step 2:} Let an unshielded triple of the form $\*X^i - \*X^j - \*X^k$ in $\widehat{\cG}$ be denoted by the tuple $(i,j,k)$ of indices. Then, let $\widehat{\cU}$ denote the list of all tuples of unshielded triples. 
Further, denote the list of (conditional) independence statements found in the skeleton phase of $\cA$ as $\widehat{\cI}$. For each $(i,j,k) \in \widehat{\cU}$, define the following two sets: 
\textbf{(i)} $S^{(i,j,k)} := \{(a,b) \ |\ a \in [d_i], b \in [d_k]\}$, and, 
\textbf{(ii)} $T^{(i,j,k)} := \{\cS_{(a,b)} \ |\  X^i_a \ind X^k_b | \cS_{(a,b)} \in \widehat{\cI} \text{ for } \cS_{(a,b)} \subset \*X\setminus \{X^i_a, X^k_b\}\}$. 
\vspace{0.2cm}
\State \textbf{Step 3:} For each $(i,j,k) \in \widehat{\cU}$, define $\overline{S}^{(i,j,k)} := \{(a,b) \ |\  (a,b) \in S^{(i,j,k)} \text{ and } X^i_a \ind X^k_b | \cS_{(a,b)}\setminus \*X^j \ ,  \forall \cS_{(a,b)} \in T^{(i,j,k)} \}$. That is, $\overline{S}^{(i,j,k)}$ is defined by all components within $\*X^i$ and $\*X^k$ that remain independent when the conditioning set that rendered them independent no longer contains any component of the middle node $\*X^j$.
If $S^{(i,j,k)} = \overline{S}^{(i,j,k)}$, then mark the triple $(i,j,k)$ as $\*X^i \to \*X^j \leftarrow \*X^k$ in $\widehat{\cG}$.


\vspace{0.2cm}
\State \textbf{Step 4:} Apply the Meek orientation rules to the resultant graph $\widehat{\cG}$.  \vspace{0.2cm}
\State \Return $\widehat{\cG}$
\end{algorithmic}
\end{algorithm}

The weakness of \stovtwo as opposed to \stov is that it suffers further in the finite sample limit due to the additional CI tests that need to be performed in Step 3. A majority or conservative strategy to mark the unshielded triples as colliders may be adopted similar to the conservative PC algorithm (that deals, unconnectedly, with the orientation faithfulness violating setting), see \citep{ramsey_adjacency-faithfulness_2006} for details. Furthermore, similar to \stov, the soundness of \stovtwo cannot be generally proven for all internal dynamics of the vector variables.

\subsection{Latent confounding}
In this section, we recall the key differences between the PC and the FCI algorithm for causal graph discovery \citep{spirtes_causation_1993}. In doing so, we aim to clarify the implications of applying the PC algorithm over all the scalar variables in the setting of \cref{subsub:Scalar_latent}, i.e., to a problem with latent confounding, and study the soundness of the discovery of the causal graph $\cG$ over vector variables $\*X$ with this approach. A complete discussion of causal discovery without causal sufficiency can be found in Chapter 6 of \cite{spirtes_causation_1993} and \cite{zhang_completeness}.

The PC algorithm consists of the (i) skeleton phase, (ii) collider orientation phase and, (iii) Meek rules-based orientation phase. The violation of causal sufficiency has ramifications for the correctness of each of these phases. First, we will study the skeleton phase.

\subsubsection{Sources of spurious edges in the PC skeleton}
The edge-deletion of the PC algorithm is based on CI statements that are found using the strategy of searching for d-separators among the adjacency sets of a pair of variables. Due to faithfulness (specifically adjacency faithfulness), the edge-deletion remains sound, however, not complete. That is, all edges absent in the PC skeleton are also absent in the ground truth graph, but not vice-versa. As noted in \cite{spirtes_causation_1993}, the sources of the spurious edges are: (i) the presence of \emph{inducing paths} and that (ii) the adjacency sets of a pair of nodes are in general not sufficient to block all paths between the respective (inadjacent) nodes, which leads to the formulation of sets that are sufficient to d-separate two nodes in a graph with latent confounding, termed \emph{possible d-separating sets}, denoted by $\poss$. We will now define these terms and illustrate how they result in spurious edges in the skeleton graph.

For the following, let there be a graph $G$ with vertices $V$, and a set $O \subseteq V$ be the observed variables and $L \equiv V \setminus O$ be the unobserved variables. We assume the absence of selection bias.

\begin{definition}[Inducing path \citep{zhang_completeness}]
    Let $X, Y \in O$ be two distinct observed vertices in $G$. A path $p$ between $X$ and $Y$ is defined as an inducing path relative to $L$ if every non-endpoint vertex on $p$ is either in $L$ or a collider, and every collider on $p$ is an ancestor of $X$ or $Y$.  
\end{definition}

The presence of an inducing path between $X, Y \in O$ implies that there exists no $S \subset O \setminus \{X,Y\}$ that can d-separate\footnote{d-separation in the case of mixed graphs, i.e., graphs with directed and bidirected edges, is often called m-separation \citep{Richardson_Ancestral}.} $X$ and $Y$. Therefore, if an inducing path is present between a pair of vertices in the underlying mixed graph, then the PC, or for that matter any conditional independence based CD algorithm, will lead to spurious edges in the skeleton between the respective pair of vertices. This issue is alleviated in the FCI algorithm by modifying the causal interpretation of edges accordingly.  

\begin{figure}
    \centering
    \includegraphics[scale=0.3]{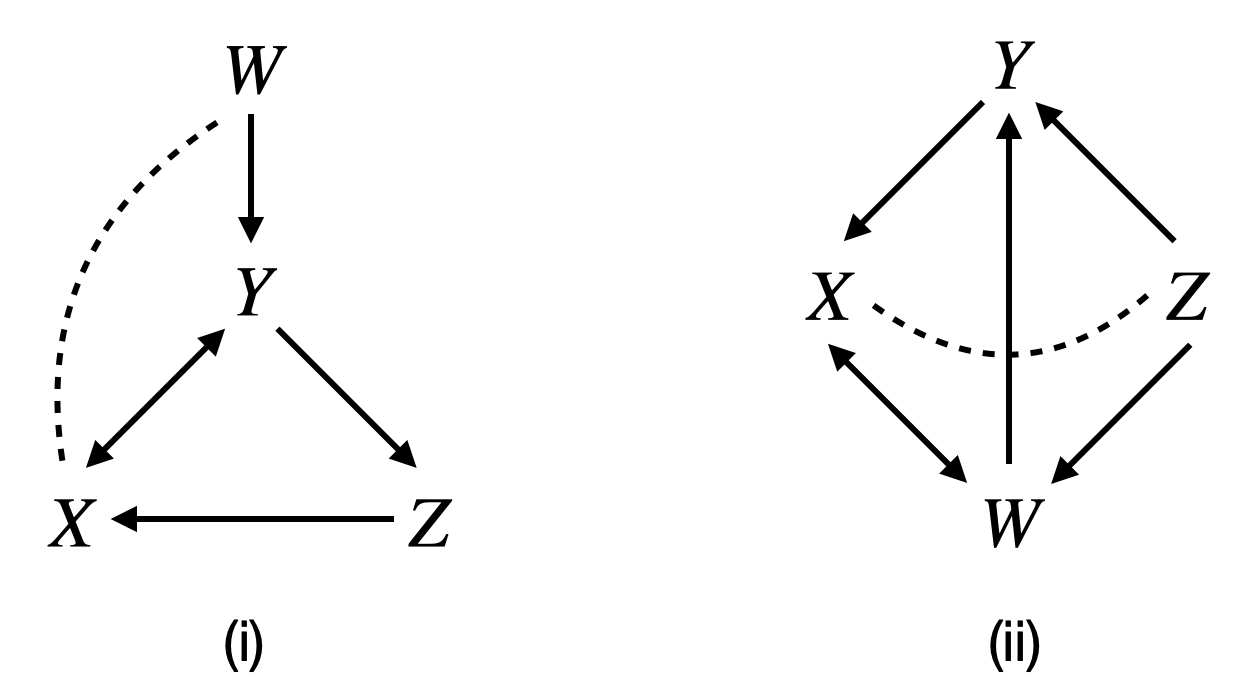}
    \caption{Examples of graphs with inducing paths. The vertices with an inducing path between them are connected by a dashed edge. (i) $X \leftrightarrow Y \leftarrow W$ is an inducing path because  the collider $Y$ is an ancestor of at least one endpoint vertex, here $X$. (ii) $X \leftrightarrow W \leftarrow Z$ is an inducing path because  the collider $W$ is an ancestor of $X$. }
    \label{fig:inducing_paths}
\end{figure}

In \cref{fig:inducing_paths}, we illustrate two examples of mixed graph where inducing paths are present. Interestingly, note that in both the examples shown, had the skeleton and colliders been detected correctly, the Meek rules would have been unsound. In (i), the edge $X \leftarrow Z$ would have been oriented incorrectly using rule 2, after rule 1 was correctly applied on $Y \to Z$. In (ii), $W \to Y$ would have been oriented incorrectly due to rule 3. 
It is for this purpose, namely to have graphs in the latent confounding setting with a suitable Markov property, that \emph{inducing path graphs} and \emph{maximal ancestral graphs} were defined, see \cite{zhang_causal_2008} for a definition and difference between the two. In short, these two types of graphs attach a different interpreation to what it means for two vertices to be adjacent and have a different orientation strategy for the `spurious' edges that are created due to inducing paths. 

\begin{figure}
    \centering
    \includegraphics[scale=0.3]{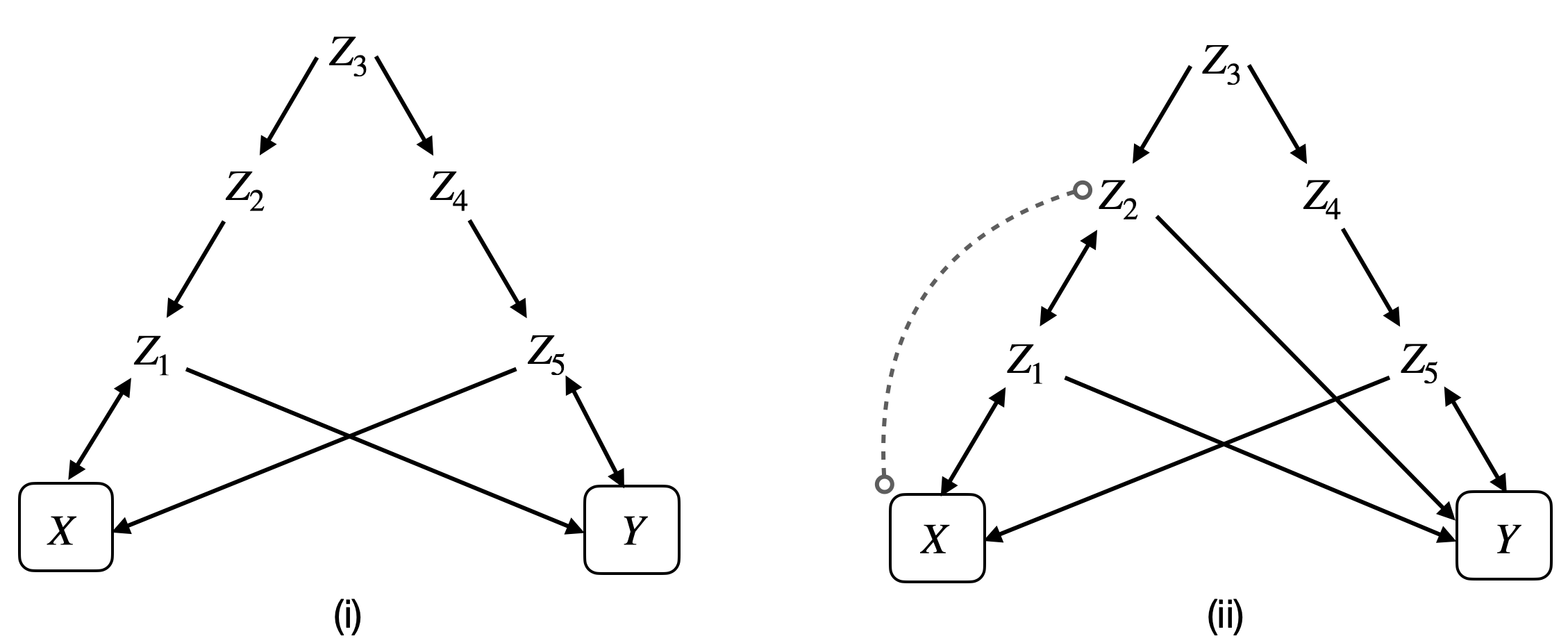}
    \caption{Examples of graphs where $\poss(X,Y) \cup \poss(Y,X) \not\subseteq \adj(X,G) \cup \adj(Y,G)$. In (i) and (ii), $X \nind Y | S$ for all $S$ such that $S \subseteq \adj(X,G) \cup \adj(Y,G) = \{Z_1, Z_5\}$. 
    \textbf{(i)} $\poss(X,Y) = \{Z_1, Z_2, Z_5\}$ where $Z_2$ is included because the path $X \leftrightarrow Z_1 \leftarrow Z_2$ is such that the middle vertex is a collider. Similarly $\poss(Y,X) = \{Z_1, Z_4, Z_5\}$. Note that  $X \ind Y | S$, where $S = \{Z_1, Z_2, Z_5, Z_4\} \subseteq \poss(X,Y) \cup \poss(Y,X)$. \textbf{(ii)} $\poss(X,Y) = \{Z_1, Z_2, Z_3, Z_5\}$, where $Z_2$ is included because the only sub-path $X  \leftrightarrow Z_1 \leftarrow Z_2$ is such that middle vertex is a collider, and would still be included if $X$ and $Z_2$ were adjacent (denoted by the dashed-edge with circle edge-marks) even though the collider at $Z_1$ would not be detectable due to the non-definite noncollider condition in \cref{def:possible_d-sep}. $Z_3$ is included because $Z_2$ is a collider and $Z_1$ is either a collider or a non-definite noncollider with $X,Z_1,Z_2$ forming a triangle (i.e.~a shielded triple). See \cite{spirtes_causation_1993} for further explanation on the possible d-separating sets. }
    \label{fig:possible_d-sep}
\end{figure}

In the latent confounding setting, the adjacency sets of a pair of nodes $X,Y$ are not sufficient to d-separate $X$ and $Y$, as illustrated in \cref{fig:possible_d-sep}. This requires the PC algorithm to be modified such that after executing the standard PC skeleton phase and orienting colliders, each pair of vertices $(X,Y)$ that remain adjacent must again be subject to edge-removal given a subset of the $\poss(X,Y)$, followed by the final edge-orientation phase \citep{spirtes_causation_1993}.

\begin{definition}[$\poss$]\label{def:possible_d-sep}
    For a given graph $G$, a node $Y$ is in $\poss(X,Z)$ if $X \neq Y$ and there is an undirected path $U$ between $X$ and $Y$ in $G$ such that for every sub-path $\langle A,B,C \rangle$ of $U$ either $B$ is a collider on the subpath, or $B$ is not a definite noncollider and $A,B$ and $C$ form a triangle in $G$.
\end{definition}
A definite noncollider on a path is a vertex that is known to not be a collider on this path. Note that $\poss(X,Z)$ may not be equal to $\poss(Z,X)$. 
In \cref{fig:possible_d-sep}, two examples of graphs with latent confounding are shown where $\poss(X,Y)$ (and $\poss(Y,X)$) is not a subset of $\adj(X, G) \cup \adj(Y,G)$, therefore a sound edge-removal phase in the most general latent confounding setting requires the possible separating sets to be taken into account. 

In the context of studying the soundness of the PC algorithm in the latent confounding setting, we   state the following lemma.

\begin{lemma}[Exclusion of certain inducing paths]\label{lem:no_inducing}
    Assume the general setting of \cref{sec:scalar_approach} and specific setting of \cref{subsub:Scalar_latent} to hold. There exists no $\cG_{micro}$ consistent with this setting such that in $\cG_{micro}$ there exists an inducing path between $X^i_a \in \*X^i$ and $X^j_b \in \*X^j$, with $i \neq j$ and $\*X^i \notin \adj(\*X^j, \cG)$. 
\end{lemma}

\emph{Proof: } Let there be $i \neq j$ and $a,b$ such that there exists a path $\pi$ between $X^i_a$ and $X^j_b$, which can be denoted explicitly as $\pi(X^i_a,X^j_b)$, that is an inducing path over $\cG_{micro}$.
Then, $\pi$ can only contain a unidirectional edge (i.e. "$\to$" or "$\leftarrow$") where one end of the edge is at an endpoint vertex, because otherwise all non-enpoint vertices cannot be colliders, specifically the non-endpoint vertex at the tail-end of the unidirectional edge. And, in addition, if there are these undirectional edges then they need to point away from the respective endpoint vertex.
In case there is only one non-endpoint vertex, this leads to a cycle due to the directed path from the non-endpoint vertex to the endpoint vertex, that needs to exist because by definition of inducing path all its colliders must be ancestors of at least one of the endpoint vertices, but this is is not possible because $\cG_{micro}$ is an ADMG.
Let $\pi_1 \subset \pi$ be such that $\pi_1= \pi \setminus \{X^i_a, X^j_b \}$. Then $\pi_1 = A_1 \leftrightarrow A_2 \leftrightarrow \ldots \leftrightarrow A_n$, where $A_1, \ldots A_n$ are all the non-endpoint vertices. Then, there exists $k \in [N], \ k \neq i,j$, such that $A_s \in \*X^k, \ \forall s$, because the macro-graph $\cG$ is a DAG. W.l.o.g., let $A_s$ be an ancestor of $X^i_a$, then this leads to a cycle in the coarse graph between $\*X^k$ and $\*X^i$. This is because the edge from $X^i_a$ points away from $X^i_a$ because the following vertex must be a collider and the edge cannot be a bidirected because the edge is between different vector variables. This contradicts the acyclicity assumption on $\cG$. (Note that if $k \neq i,j$ had not been imposed, namely $\*X^i \notin \adj(\*X^j, \cG)$ had not been imposed in the lemma statement, then an inducing path between $X^i_a$ and $X^j_b$ could be constructed by noting that if $k=i$ w.l.o.g., then $A_s, \ \forall s$ are allowed to be ancestors of $X^i_a$ without violating any adjacencies. But the case $\*X^i \in \adj(\*X^j, \cG)$ is not of interest to this lemma since it states the absence of spurious edges in the macro graph, and the inducing paths among already adjacent macro nodes will not introduce unsound edges when the micro graph is coarsened to the macro graph.)
$\square$

\subsubsection{PC orientation rules and latent confounding}

\begin{figure*}[!ht] 
    \centering
    \begin{minipage}{0.32\textwidth}
        \centering
        \includegraphics[scale=0.35]{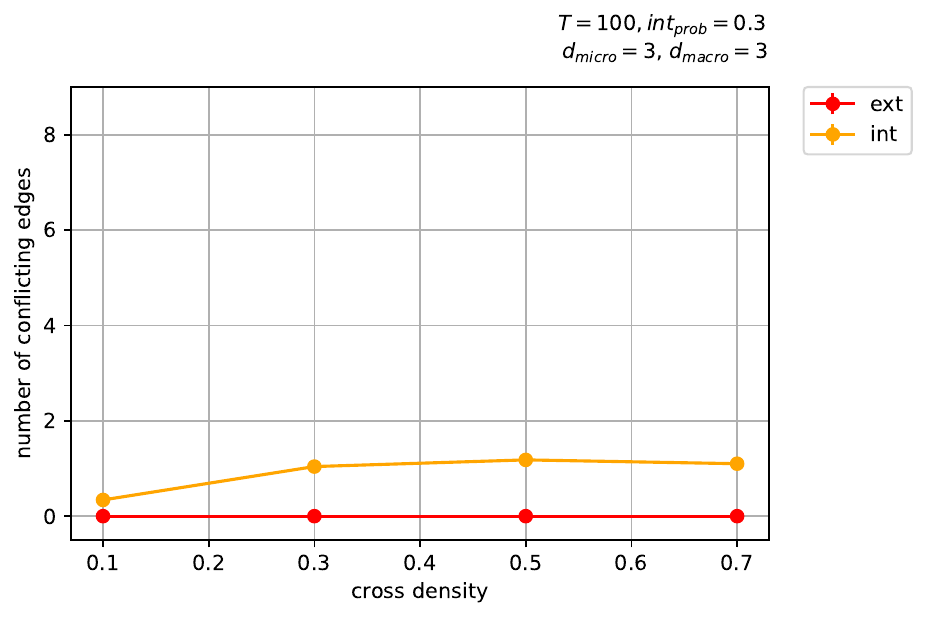}
    \end{minipage}%
    ~ 
    \begin{minipage}{0.32\textwidth}
        \centering
        \includegraphics[scale=0.35]{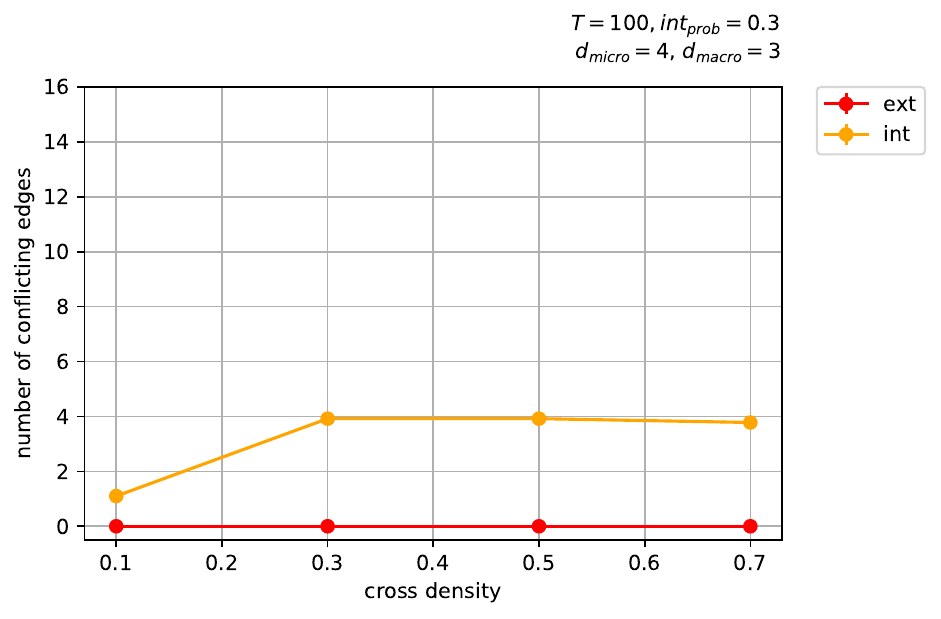}
    \end{minipage}
~
    \begin{minipage}{0.32\textwidth}
        \centering
        \includegraphics[scale=0.35]{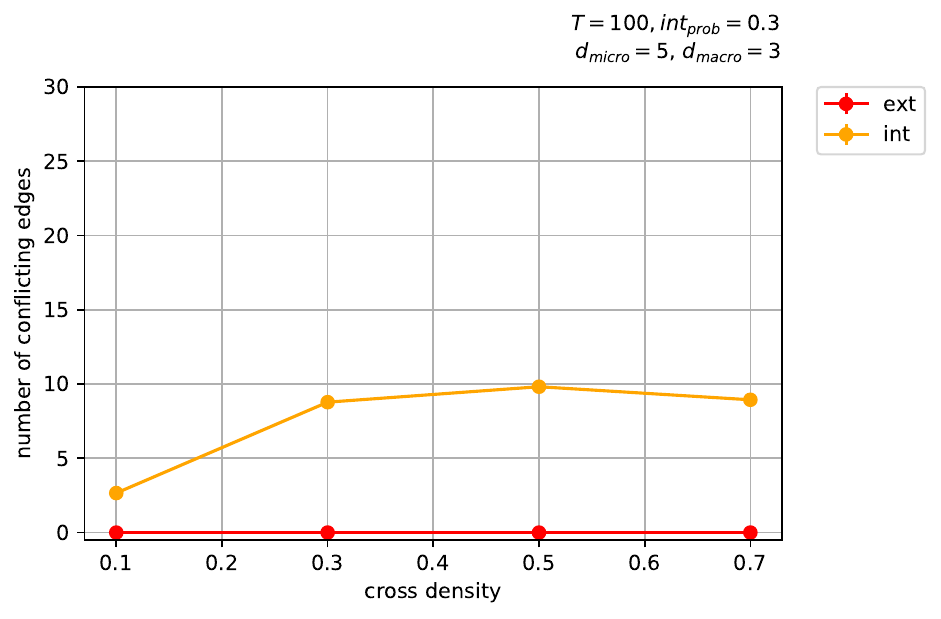}
    \end{minipage}
    \caption{Experiments to illustrate the number conflicting edges found \textbf{internal (int)} to the vector variable versus \textbf{external (ext)}, i.e.~across vector variables, in the micro graph when the PC algorithm is applied component-wise to the latent confounding setting of \cref{subsub:Scalar_latent} as the cross density, i.e.~the density of edges across the vector variables, increases. $T$ stands for sample size, $int_{prob}$ for the probability of edges internal to the vector variables, $d_{micro }$ are the number of scalar components inside the vector variable and $d_{macro}$ the number of vector variables.} \label{fig:PC_conflicts}
\end{figure*}

In \cref{fig:PC_conflicts}, we illustrate empirically that the orientation phase in \stov leads to practically no orientation conflicts  in edges across vector variables. In the interest of efficiency and finite sample considerations, we recommend the usage of \stov as compared to \stovtwo (\cref{algo:S2Vtwo}) within the component-wise approach for the latent confounding setting


\subsection{Cycles: Example to illustrate Remark \ref{rem:S2V_cycles}}\label{app:PC_cycles}
\begin{figure}
    \centering
    \includegraphics[scale=0.25]{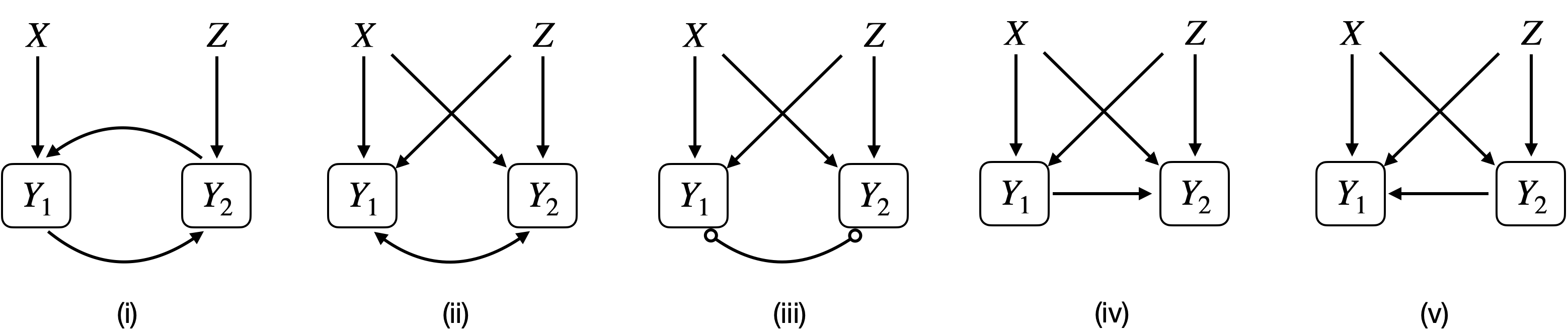}
    \caption{Soundness of causally sufficient CD algorithms under cyclic relationships: \textbf{(i)} Cyclic graph corresponding to SCM \cref{eq:bongers_cyclic}, \textbf{(ii)} The acyclified graph with a bidirected edge due to the presence of $\eta_{Y_2}$ in the equations of $Y_1$ and $Y_2$, \textbf{(iii)} The Markov equivalence class (MEC) of the acyclified graph, \textbf{(iv)} and \textbf{(v)} Two DAGs that are compatible with the MEC in (iii). }
    \label{fig:pc_cycles}
\end{figure}

In \cref{subsub:Scalar_cycles}, we discussed the setting where the internal dynamics of the vector variables contain cyclic relationships, but the graph $\cG$ is acyclic.
In the light of \cref{rem:S2V_cycles},
we will illustrate the soundness of the PC algorithm on graphs with cycles using the graph and the discussion on cyclic SCMs and the acyclification operation in Example A.8 from \cite{Bongers_foundations}. See \cref{fig:pc_cycles} for further details on this example. 
Let a cyclic SCM over variables $X, Y_1, Y_2, Z$ be given by:
\begin{equation}\label{eq:bongers_cyclic}
    \begin{split}
        X &:= \eta_X \ , \\
        Z &:= \eta_Z \ , \\
        Y_1 &:= X\cdot Y_2 + \eta_{Y_1} \ , \\
        Y_2 &:= Z \cdot Y_1 + \eta_{Y_2} \ .       
    \end{split}
\end{equation}
This SCM can be `solved' (see \cite{Bongers_foundations} for definitions of solvability and solutions of SCMs with cycles and latent confounding) in terms of all exogenous and endogenous variables by substituting the R.H.S.~of $Y_2$ into the structural equation for $Y_1$, and substituting this solution of $Y_1$ back into the structural equation for $Y_2$:
\begin{equation}\label{eq:bongers_acyclified}
    \begin{split}
        X &:= \eta_X \ , \\
        Z &:= \eta_Z \ , \\
        Y_1 &:= \frac{X \cdot \eta_{Y_2} + \eta_{Y_1}}{1-X\cdot Z} \ , \\
        Y_2 &:= \frac{Z \cdot \eta_{Y_1} + \eta_{Y_2}}{1-X\cdot Z} \ .       
    \end{split}
\end{equation}
This yields an acyclified SCM, the graph corresponding to which can be seen in part (ii) of \cref{fig:pc_cycles}. This graph is an ADMG, the PAG (partial ancestral graph, namely the output of a constraint-based CD algorithm that account for latent confounding, such as FCI) corresponding to which is shown in part (iii) of the same figure. 
Notice that if the PC algorithm were run for the SCM \cref{eq:bongers_acyclified}, then the set of conditional independencies, namely $X \ind Z$, would have yielded a CPDAG that is equivalent to the PAG shown in (iii). Parts (iv) and (v) display the DAGs that are compatible with the CPDAG in (iii). For our purpose, namely to illustrate \cref{rem:S2V_cycles}, note that if $Y_1$ and $Y_2$ belonged to a vector variable $\*Y$, then the setting of \cref{subsub:Scalar_cycles} would be satisfied and \stov would yield the sound CPDAG $X \to \*Y \leftarrow Z$, thus illustrating \cref{rem:S2V_cycles}. 

\section{General strategies to compute the dependence score for aggregation consistency}\label{app:complete_strategy_dep}

In \cref{subsub:scores}, we discussed complete and incomplete strategies to compute the dependence score for aggregation consistency. Since there are many possible choices involved when computing such a score, of which we presented one in \cref{algo:cdep_eff}, in \cref{algo:meta_cdep} we present a meta-algorithm that provides an overview of the choices involved when designing a dependence score for aggregation consistency. For this, we will require (i) a dependence strategy $\mathfrak{D}$ which can either be `adjacency-based' (denoted by $\mathfrak{D}_{adj}$) or d-connection-based (denoted by $\mathfrak{D}_{con}$) and refers to whether it is all adjacencies or all possible d-connections in the aggregate graph that are to be tested at the vector level, (ii) conditioning strategy $\mathfrak{C}$ which can either be `tested' (i.e.~only considers the dependence statements that were found during the run of $\cA$ corresponding to a pair of variables) or `all' (i.e.~considers all possible dependence statements corresponding to a pair of variables), and last (iii) a sorting strategy $\mathfrak{S}$ that signifies how many of the dependence statements given by the conditioning strategy are to be tested at the vector level, which, when left to `None' corresponds to considering all dependence statements given by the conditioning strategy. A combination of the $\mathfrak{D}_{con}$ dependence strategy, together with an `all' conditioning strategy and a `None' sorting strategy yields a \emph{complete} strategy for computing the dependence score $c_{dep} (\*Z, \*X, \cA)$, whereas all other combinations yield an incomplete strategy and hence an effective dependence score $\bar{c}_{dep} (\*Z, \*X, \cA)$. Therefore, in \cref{algo:meta_cdep}, we refer to the (potentially effective) dependence score as $\tilde c_{dep} (\*Z, \*X, \cA)$. Note that a combination of the $\mathfrak{D}_{adj}$ dependence strategy, together with a `tested' conditioning strategy and a p-value sorting strategy yields the \emph{incomplete} strategy to compute the effective dependence score that was presented in \cref{algo:cdep_eff}.

\begin{algorithm}
\caption{Meta-algorithm for dependence score $\tilde c_{dep} (\*Z, \*X, \cA)$ for aggregation consistency}\label{algo:meta_cdep}
\begin{algorithmic}[1]
\State \textbf{Input:} CD algorithm $\cA$, data samples for vector-variables $\*X$ and aggregate variables $\*Z$, dependence strategy $\mathfrak{D}$, conditioning strategy $\mathfrak{C}$, sorting strategy $\mathfrak{S}$ \vspace{0.2cm}
\State \textbf{Step 0:} Initialize $\tilde c_{dep}(\*Z, \*X, \cA) = 0$.\vspace{0.2cm}
\State \textbf{Step 1:} List all unordered pairs of nodes $(\*Z^i, \*Z^j)$ to be tested given by the dependence strategy $\mathfrak{D}$, and denote the list as $\cD^{\mathfrak{D}} (\*Z)$. \vspace{0.2cm}
\State \textbf{Step 2:} For each $L_{ij} \in \cD^{\mathfrak{D}}  (\*Z)$ 
list all triples of sets of indices resulting from the conditioning strategy $\mathfrak{C}$, and denote it as $\cL^{\mathfrak{D},\mathfrak{C}}$. \vspace{0.2cm}
\State \textbf{Step 3:}  Repeat this step for all elements $L_{ij}\in \cD^{\mathfrak{D}} (\*Z)$: Construct  $ \cL^{\mathfrak{D},\mathfrak{C}}_{\mathfrak{S}} \subseteq \cL^{\mathfrak{D},\mathfrak{C}}$ based on the sorting strategy $\mathfrak{S}$. Test whether the dependence statement given by $\cL^{\mathfrak{D},\mathfrak{C}}_{\mathfrak{S}}$ holds in $\*X$. If dependence is found, increase $\tilde{c}_{dep}$ by one and append the pair $(i,j)$ to the list $\widehat{\cD}^{\mathfrak{D}} (\*X)$ of dependencies in $\*X$.
\vspace{0.2cm}
\State \textbf{Step 4:} Normalize $\tilde{c}_{dep}(\*Z, \*X, \cA)$ by the cardinality of $\cD^{\mathfrak{D}}(\*Z)$.    \vspace{0.2cm}  
\State \Return $\tilde{c}_{dep}(\*Z, \*X, \cA)$
\end{algorithmic}
\end{algorithm}

\section{Proofs of results in section \ref{sec:agg_consistency_adag}} \label{app:proofs_agg}

\subsection{Details on causal input completeness}\label{app_subsec:causal_input} 

\cite{verma_pearl_networks} defined a \emph{causal input list} as follows:

\begin{definition}[Causal Input List \citep{verma_pearl_networks}]\label{def:causal_input_list_Verma} 
    Given an ordering  $\cT$ of variables $\*X$, a causal input list corresponding to an independence model over $\*X$ is an ordered list of triples $I(x, B_x, R)$ for each variable $x \in \*X$ such that $x \ind R | B_x $. Here $B_x$ denotes the tail boundary of $x$ and $R$ the set of all variables that appear before $x$ in $\cT$ excluding $B_x$. A tail boundary of a variable $x$ is any subset of the set of all variables that appear before $x$ in $\cT$ that renders $x$ independent of $R$. 
\end{definition}

\begin{example}[Causal Input Completeness (\cref{def:input_completeness})]
    Let a DAG $\cG_{\*Y}$ over $\*Y = (Y_1, Y_2, Y_3)$ be given by $Y_1 \leftarrow Y_2 \to Y_3$. Then the PC algorithm over $\*Y$ is causal input complete because it outputs the CPDAG $\widehat{\cG}_\cC: Y_1 \doublemarked Y_2 \doublemarked Y_3$ and finds a singleton of CI test statements $I_{\cA} (\*Y) = \{(Y_1 \ind Y_3 | Y_2)\}$, such that any topological ordering $\cT$ compatible with (any DAG the represents) $\widehat{\cG}_\cC$ together with $I_{\cA} (\*Y)$ is sufficient to construct a causal input list.  For instance, let $\cT = (Y_1, Y_2, Y_3)$, then $B_{Y_3} = \{Y_1\}$ and $B_{Y_1} = B_{Y_2} = \emptyset$.
\end{example}

Note that the PC algorithm cannot be proven to be causal input complete with respect to \emph{any} input graph. For instance,  for the ground truth graph $Y_1 \leftarrow Y_2 \to Y_3 \to Y_4$, PC will not test $Y_4 \ind (Y_1, Y_2) | Y_3$. 


\subsection{Proofs}

\paranewspace{Lemma \ref{lem:c_ind_faithfulness}}
    \emph{Let variables $\*X$ satisfy \cref{eq:SCM} and \cref{ass:CD}.
    Additionally, the aggregation map $g(\cdot)$ that yields aggregate variables $\*Z$ from variables $\*X$ satisfies \cref{ass:agg_graph_existence}.  
    Assuming all conditional independence tests are sound, if $g(\cdot)$ satisfies aggregation faithfulness (\cref{prop:faithfulness}) then the independence score $c_{ind}$ is maximal (i.e.~it equals 1). }
\begin{proof}
    Note that aggregation faithfulness is equivalent to $\cI(\*Z) \subseteq \cI(\*X) $. If $\cI(\*Z) \subseteq \cI(\*X) $, then  $\mathfrak{I}^{ind} = \emptyset$ because there is no CI statement that holds true on $\*Z$, but not on the corresponding variables in $\*X$, under the assumption that CI tests are sound. Therefore $c_{ind} = 1$. 
\end{proof}

\paranewspace{Lemma \ref{lem:c_ind_faithfulness_complete}}
    \emph{Let the assumptions of \cref{lem:c_ind_faithfulness} be satisfied and assume in addition that the 
    CD algorithm $\cA$ is causal input complete, then the independence score $c_{ind}$ is maximal (i.e.~it equals 1) if and only if $g(\cdot)$ satisfies aggregation faithfulness (\cref{prop:faithfulness}).}

\begin{proof}
    In the following, let the CPDAG discovered over $\*Z$ using algorithm $\cA$ be denoted by $\widehat{\cG}_Z$, and the true causal graph over $\*X$ be denoted by $\cG$. Note that the two graphs can be compared due to the direct correspondence between $\*Z$ and $\*X$.

The direction ($g$ satisfies \cref{prop:faithfulness}) $\Rightarrow (c_{ind} = 1)$ has been proven in \cref{lem:c_ind_faithfulness}.

For the direction $(c_{ind} =1) \Rightarrow (g$ satisfies \cref{prop:faithfulness}), note that  if $c_{ind} = 1$, then: 
    \begin{itemize}
        \item[] Case 1: $|\mathfrak{C}^{ind}| = |\mathfrak{I}^{ind}| = 0 \Rightarrow \tested (\*Z) = \widehat{\cI}(\*X) =  \emptyset$.
        \item[] Case 2: $|\mathfrak{C}^{ind}| \neq 0$ and $|\mathfrak{I}^{ind}| = 0 \Rightarrow \tested(\*Z) = \widehat{\cI}(\*X)$.  
    \end{itemize}
    Due to \cref{ass:agg_graph_existence}, $\tested (\*Z)$ is Markov and faithful to a graph whose CPDAG is $\widehat{\cG}_Z$ (under correct CI statements and soundness and completeness of $\cA$). Note that a DAG $\cG_\cD$ compatible with $\widehat{\cG}_Z$ yields a topological ordering, which together with the set of CI statements $\tested (\*Z)$ found during the run of $\cA$ on $\*Z$ form a causal input list (\cref{def:causal_input_list_Verma} ). 
    According to \cref{cor:verma_result_graphoid}, 
    the causal input list in turn yields the entire independence model $\cI(\*Z)$. Furthermore, the same CI tests which yield the independence model of $\*Z$ from its causal input list and graphoid axioms can be performed on $\*X$ since $\tested(\*Z) = \widehat{\cI}(\*X)$ to yield a superset $\widetilde{\cI}(\*X) \supset \widehat{\cI}(\*X)$. This superset is by definition a subset of the independence model over $\*X$, namely, $\widetilde{\cI}(\*X) \subseteq \cI (\*X)$.
    Hence: 
    \begin{itemize}
        \item if $\tested (\*Z) = \emptyset \Rightarrow \cI(\*Z) = \emptyset  \subseteq \cI(\*X)$ and, 
        \item if $\tested (\*Z) \neq \emptyset$ then: let $\mathfrak{G} (\cdot)$ denote the semi-graphoid closure given the causal input list found by $\cA$ on $\*Z$. Then, 
        \begin{equation}
            \mathfrak{G}(\tested (\*Z)) = \cI(\*Z) = \mathfrak{G}(\widehat{\cI}(\*X)) = \widetilde{\cI}(\*X) \subseteq \cI (\*X).
        \end{equation}
    \end{itemize}

    Thus, both Case 1 and 2 imply $\cI(\*Z) \subseteq \cI(\*X) $, i.e., $g$ satisfies \cref{prop:faithfulness}.
 \end{proof}

\paranewspace{Corollary \ref{cor:agg_super_graph_1}}
    \emph{Let variables $\*X$ satisfy \cref{eq:SCM} and \cref{ass:CD} w.r.t.~a graph $\cG$.
    Additionally, the aggregation map $g(\cdot)$ that yields aggregate variables $\*Z$ from variables $\*X$ satisfies \cref{ass:agg_graph_existence}. Assuming sound conditional independence tests and the CD algorithm to be causal input complete, if the independence score for aggregation consistency $c_{ind} = 1, $ 
    then the graph $\cG_Z$ which is Markov and faithful to $P(\*Z)$ is s.t. $sk(\cG) \subseteq sk(\cG_Z)$, where $sk(\cdot)$ denotes the skeleton of the graph in the argument. }
    
    

\begin{proof} From \cref{lem:c_ind_faithfulness_complete}, $c_{ind} = 1$ if and only if aggregation faithfulness condition is satisfied.
Recall, adjacency faithfulness is defined as the relationship between the distribution and the graph over a set of variables according to which if two variables are (conditionally) independent given any subset of the remaining variables, they must be nonadjacent in the graph; see \cite{ramsey_adjacency-faithfulness_2006} for a formal definition. Since the faithfulness condition implies adjacency faithfulness, for every $I_Z \in \tested (\*Z)$, which yields $I_X \in \widehat{\cI}(\*X)$, the corresponding variables are nonadjacent in $\cG_Z$ and $\cG$, due to \cref{ass:CD} and property \ref{ass:agg_graph_existence}. Since $\widehat{\cI} (\*X) = \tested (\*Z) $ for $c_{ind} = 1$ and $\widehat{\cI} (\*X) \subseteq \cI(\*X)$, this implies $ sk(\cG) \subseteq sk(\cG_Z)$.  
\end{proof}

 \paranewspace{Lemma \ref{lem:c_dep_sufficiency}}
   \emph{ Let variables $\*X$ satisfy \cref{eq:SCM} and \cref{ass:CD} and assume that conditional independence tests are sound.
    Additionally, the aggregation map $g(\cdot)$ that yields aggregate variables $\*Z$ from variables $\*X$ satisfies \cref{ass:agg_graph_existence}.  If the strategy to compute $c_{dep} (\*Z, \*X, \cA)$ is complete then $c_{dep}$ is maximal (i.e.~it equals 1) if and only if $g(\cdot)$ satisfies aggregation sufficiency as defined in \cref{prop:sufficiency}. }
\begin{proof}
    Contrapositive of \cref{prop:sufficiency} implies that for every $i\neq j \in [N]$ and any $\cS_Z \subset \*Z\setminus\{\*Z^i, \*Z^j\}$ such that $\*Z^i \nind \*Z^j | \cS_Z \Rightarrow \*X^i \nind \*X^j | \cS_X$ where $\cS_Z = g(\cS_X)$ which is exactly what a complete strategy checks. 
\end{proof}

 \paranewspace{Theorem \ref{thm:acs_consistency}}
    \emph{Let the assumptions of \cref{lem:c_ind_faithfulness_complete} be satisfied and let the strategy adopted to measure $c_{dep}$ be complete. 
    Then, the joint aggregation consistency score is maximal (i.e.~it equals 1), if and only if $g(\cdot)$ is a valid aggregation map as defined in \cref{def:valid_agg}.}

\begin{proof}
    Note that $AC$ is maximal when its composite parts $c_{ind}$ and $c_{dep}$ are respectively maximal (i.e.~equal one).
From the two lemmas, \ref{lem:c_ind_faithfulness_complete} and  \ref{lem:c_dep_sufficiency},  we conclude $\indmod (\*Z) \equiv \indmod (\*X)$, thus the aggregation map is valid as per \cref{def:valid_agg}. 
\end{proof} 

\paranewspace{Theorem \ref{thm:adag_soundness}}
    \emph{Let variables $\*X$ be generated by \cref{eq:SCM} and satisfy \cref{ass:CD}.  
    In the infinite sample limit, \adag$(q, \alpha_q, \cA, g^{m})$ together with a causal input complete CD algorithm $\cA$, a tunable aggregation map $g^m (\cdot)$ that satisfies \cref{ass:agg_graph_existence}, and the $q$-score given by     
    (i) the aggregation consistency score (\cref{eq:ac_score}) with $\alpha_q = 1$, outputs a CPDAG ${\cG}_Z$ that is equivalent to the CPDAG of the ground-truth graph $\cG$ over variables $\*X$.    
    (ii) the independence score for aggregation consistency $c_{ind}$ (\cref{eq:cind}) with $\alpha_q = 1$, outputs a CPDAG ${\cG}_Z$ s.t. $sk(\cG) \subseteq sk(\cG_Z)$, where $sk(\cdot)$ denotes the skeleton of the graph in the argument.}
\begin{proof} (i) follows from \cref{thm:acs_consistency} and \cref{def:valid_agg} and (ii) follows from \cref{cor:agg_super_graph_1} 
\end{proof}

\section{Further experiments}\label{app:further_exps}

\subsection{Evaluation of multivariate conditional independence tests}\label{app:mult_CI}
In this section, we present experimental evaluations of three multivariate conditional independence tests. The data generation process is explained below. The SCM generating the data is assumed to be linear with additive Gaussian noises. We use ordinary least squares regression to generate pairwise residuals $(r_{X,Z}^{j}, r_{Y,Z}^{k}) , \forall (i,j),$ obtained by regressing  each component $X^j \in \*Y$ and $Y^k \in \*Y$ on $\*Z$, respectively. Given the residual pairs $(r_{X,Z}^{j}, r_{Y,Z}^{k})$, we consider four ways of testing the null hypothesis $\*X \ind \*Y | \*Z$ with a pre-defined type 1 error rate (namely the likelihood of rejecting the null hypothesis when it is true) of  $\alpha$:
\begin{enumerate}
    \item \textbf{Max-Corr}: We consider a test of multivariate conditional independence, which we name `Max-Corr', which computes the pairwise Pearson correlation coefficients $\rho^{j,k}$ of the residual pairs $(r_{X,Z}^{j}, r_{Y,Z}^{k})$, and computes the Bonferroni corrected analytic p-value for the test statistic $\rho  = \operatorname{max}_{i,j}(\rho^{j,k})$. This approach is equivalent to computing the correlation matrix among the residual vectors $\*r_{X,Z}$ and $\*r_{Y,Z}$, (where $\*r_{X,Z} = (r_{X,Z}^{1}\ldots,r_{X,Z}^{dim_X})$ and similarly for $\*r_{Y,Z}$), and for each individual entry of the correlation matrix, performing a standard t-test using the Bonferroni correction and rejecting the null if at least on of the tests rejects it. 
    \item \textbf{Generalized Covariance Measure (GCM) \citep{Shah_Peters}}: GCM computes a covariance-like measure between the residuals $(\*r_{X,Z}, \*r_{Y,Z})$, where $\*r_{X,Z} = (r_{X,Z}^{1}\ldots,r_{X,Z}^{dim_X})$ (and similarly for $\*r_{Y,Z}$), and computes an aggregate test statistic by taking the maximum value of this covariance-like matrix and computes the corresponding p-value using the Gaussian multiplier bootstrap \citep{GMB_Chernozhukov}. 
    \item \textbf{Hilbert-Schmidt Independence Criterion (HSIC) \citep{Gretton_HSIC_Kernel, fukumizu_HSIC_Kernel, Zhang_KCIT}}: This methods defines the test statistic as the Hilbert-Schmidt norm of the conditional cross covariance operator, which can be obtained from the residual pairs $(\*r_{X,Z}, \*r_{Y,Z})$. It computes the p-value from the estimated empirical null distribution.   
    \item \textbf{Box-M-Test} The Box-M test is used to test the equality of covariance matrices \citep{box_M_test}. We adapt the Box-M-test for multivariate CI testing as follows: Regress $\*Y$ on $(\*X,\*Z)$ and once only on $\*X$, and test for the equality of the two covariance matrices of residuals arising from the two regressions using the Box-M-test.  
\end{enumerate}
Empirically, we found the Box-M-Test to be highly inefficient for high-dimensional regressors together with a rather high type 1 error rate, therefore in the evaluations we do not present the results for this case.

\paranewspace{Data Generation Process}
The data generation model to evaluate the multivariate CI tests listed above is generated by the SCM among vector-valued variables $\*X, \*Y, \*Z$ given by:
\begin{equation}
    \begin{split}
        \*Z &:= \*\eta_{\*Z} \\
        \*X &:= c \cdot \*Z + \*\eta_{\*X} \\
        \*Y &:= c \cdot \*Z + \*\eta_{\*Y} \ .
    \end{split}
\end{equation}
Here, $c$ is scalar value signifying the causal effect strength and additive noises $\eta_i$'s drawn from a multivariate normal distribution with a unit mean and identity covariance matrix. Further parameters such as the internal connectivity of $\*Z$ and the fraction of components of $\*Z$ that act as confounders can be adjusted in the data generation process. The code to  generate the data is shown in the Python code below.  

\vspace{1cm}

\begin{python}
def multivariate_confounder(samples, dimxy, dimz, coef, z_int=None, n_conf = 'high', random_state = None):
'''
Generates data corresponding to confounder X <- Z -> Y, with multivariate X, Y, Z

Parameters:
----------
samples: int, Sample size
dimxy: int, dimension of X and Y (dim_X = dim_Y)
dimz: int, dimension of Z
coef: float, coefficient strength of the edge Z->X and Z->Y,
             should be set to zero for testing type 1 error rate
z_int: str or float, internal density of Z
n_conf: str, fraction of components in Z that act as confounders
random_state: random number generator 

Returns:
-------
data: numpy array of shape (sample, 2*dimxy + dimz)
'''
    
    if random_state is None:
        seed = None
        random_state = np.random.default_rng(seed=seed)
    
    data = random_state.standard_normal((samples, 2*dimxy + dimz))
    confounder = random_state.standard_normal((samples, 1))
    
    if z_int: #z_int denotes how dense z is internally connected (None implies unconnected)        
        if type(z_int) == str:        
            if z_int == 'high': # z is fully connected 
                data[:,int(2*dimxy):] += confounder
            elif z_int == 'low': # num_int components of z are connected
                num_int = int(dimz/3)    
                data[:,int(2*dimxy):int(2*dimxy)+num_int] += confounder            
        elif type(z_int) == int: # z_int components of z are connected
            data[:,int(2*dimxy):int(2*dimxy)+z_int] += confounder
        
    if n_conf == 'high':
        n_conf = dimz  # all components of z act as confounders
    elif n_conf == 'low':
        if dimz>2:
            n_conf = int(dimz/3) # "n_conf" components of z act as confounders
        elif dimz>0:
            n_conf = 1
        else:
            raise ValueError('dim_z must be greater than zero for confounder data')
    else:
        raise ValueError('n_conf should be in ["high", "low"]')
    
    data[:,0:dimxy] += data[:,int(2*dimxy):int(2*dimxy+n_conf)].mean(axis=1).reshape(samples, 1) # Z causes X
    data[:,dimxy:int(2*dimxy)] += data[:,int(2*dimxy):int(2*dimxy)+n_conf].mean(axis=1).reshape(samples, 1) # Z causes Y
    data[:,dimxy:int(2*dimxy)] += coef*data[:,0:dimxy] # X causes Y
    
    return data

\end{python}

\begin{figure}
    \centering
    \includegraphics[scale=0.45]{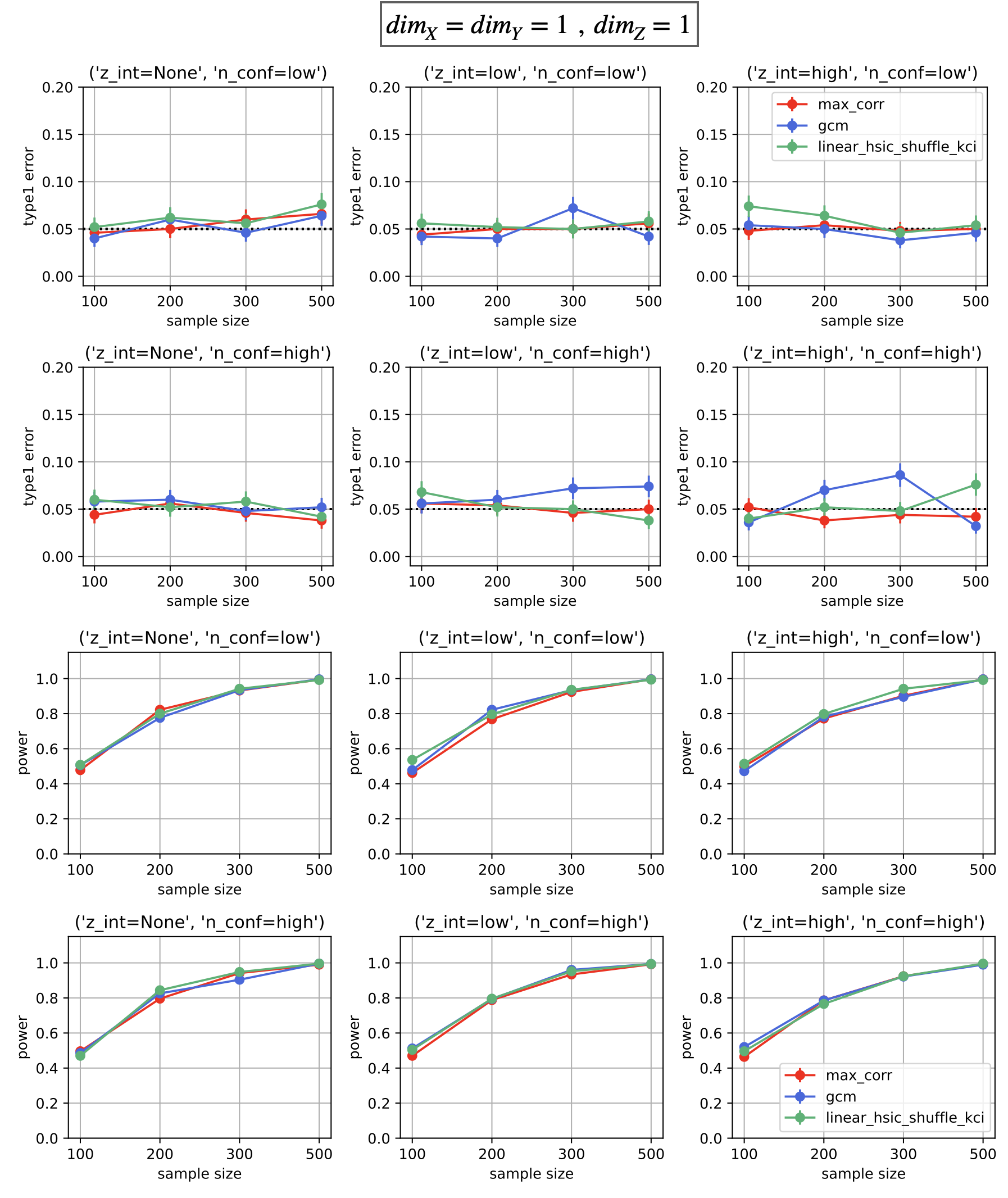}
    \caption{For univariate $\*X, \*Y, \*Z$, there is not a significant difference in the performance of max-corr, GCM and HSIC, except for a GCM suffering from a higher type 1 error rate in the setting $(z_{int} = high, n_{conf} = high)$, namely $\*Z$ is densely connected internally and a high fraction of the components of $\*Z$ act as confounders for $\*X$ and $\*Y$. }
    \label{fig:mult_CI_1_1}
\end{figure}

\begin{figure}
    \centering
    \includegraphics[scale=0.45]{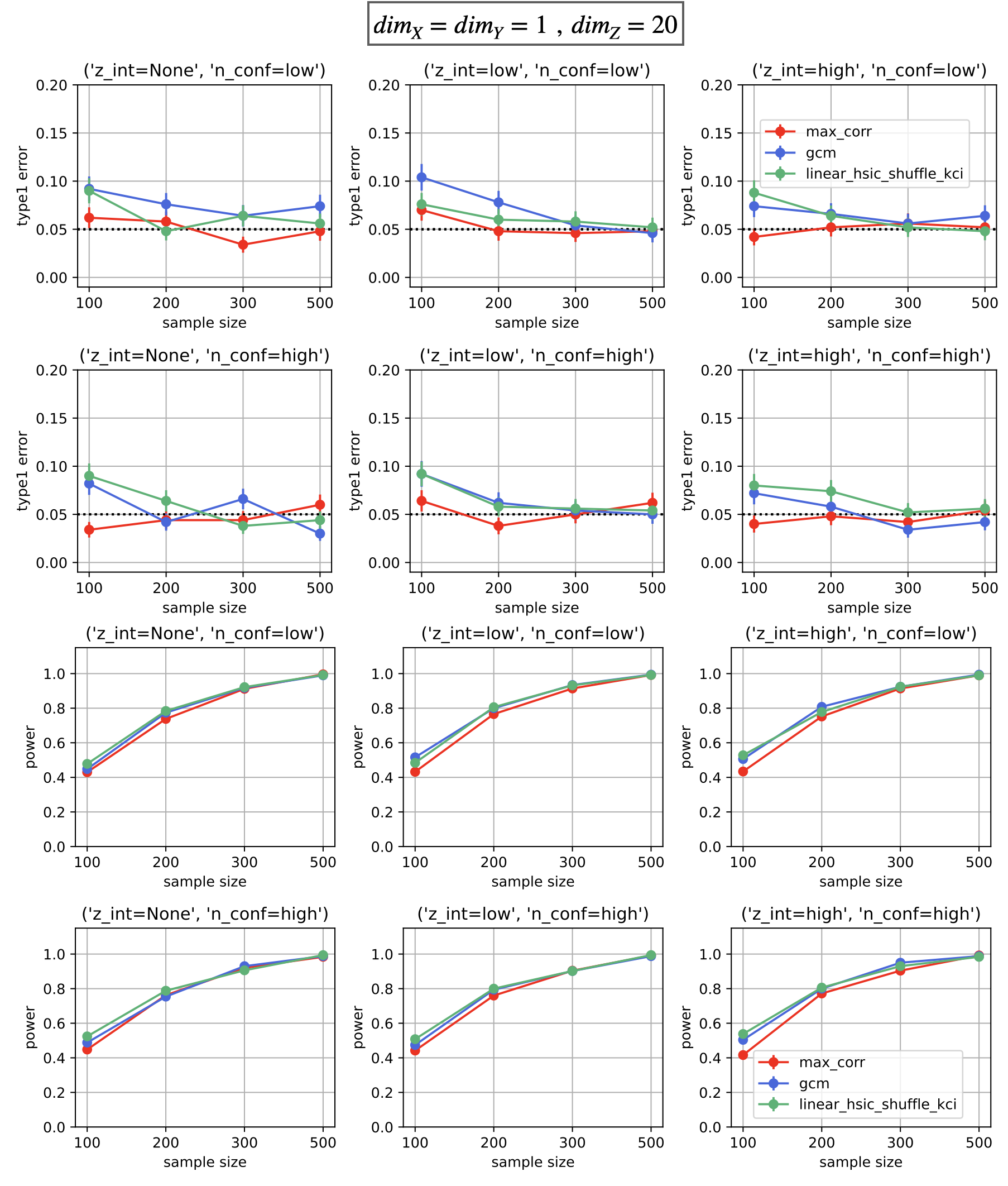}
    \caption{For univariate $\*X, \*Y$ and multivariate $\*Z$, the type 1 error rate for GCM and HSIC is higher for lower sample sizes compared to Max-Corr, while the power for all three methods remains comparable, with Max-Corr seeing a marginal decrease in power in the setting $(z_{int} = high, n_{conf} = low/high)$, namely $\*Z$ is densely connected internally and both low and high fraction of the components of $\*Z$ act as confounders for $\*X$ and $\*Y$.}
    \label{fig:mult_CI_1_20}
\end{figure}

\begin{figure}
    \centering
    \includegraphics[scale=0.45]{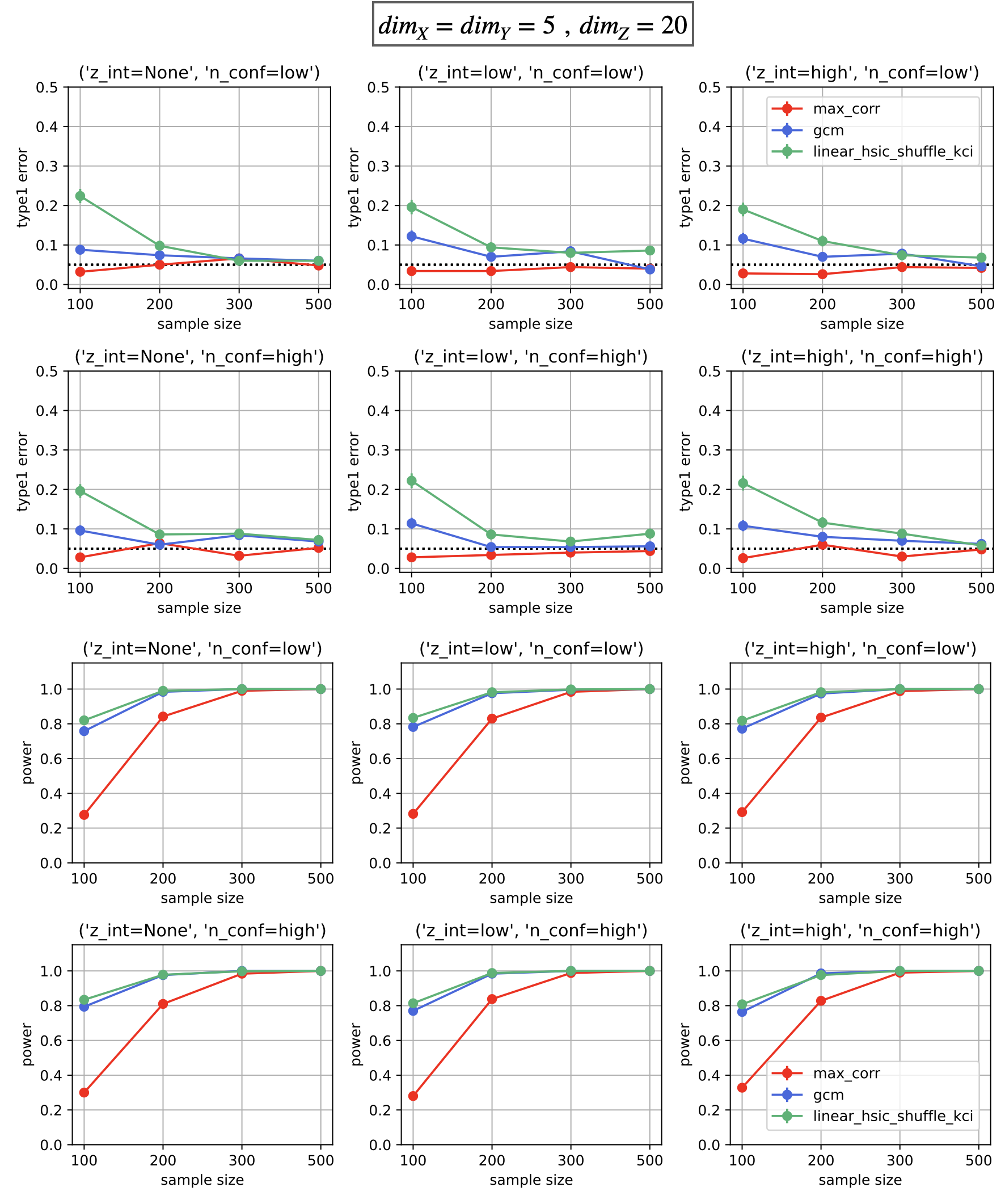}
    \caption{For multivariate $\*X, \*Y,\*Z$, the type 1 error rate of HSIC is highest, and Max-Corr is lowest. However, in this setting, the power of Max-Corr starts to suffer significantly in the low sample regime. Furthermore, the distinction between various internal density settings as well as fraction of active confounders within $\*Z$ has become minimal.}
    \label{fig:mult_CI_5_20}
\end{figure}

\begin{figure}
    \centering
    \includegraphics[scale=0.45]{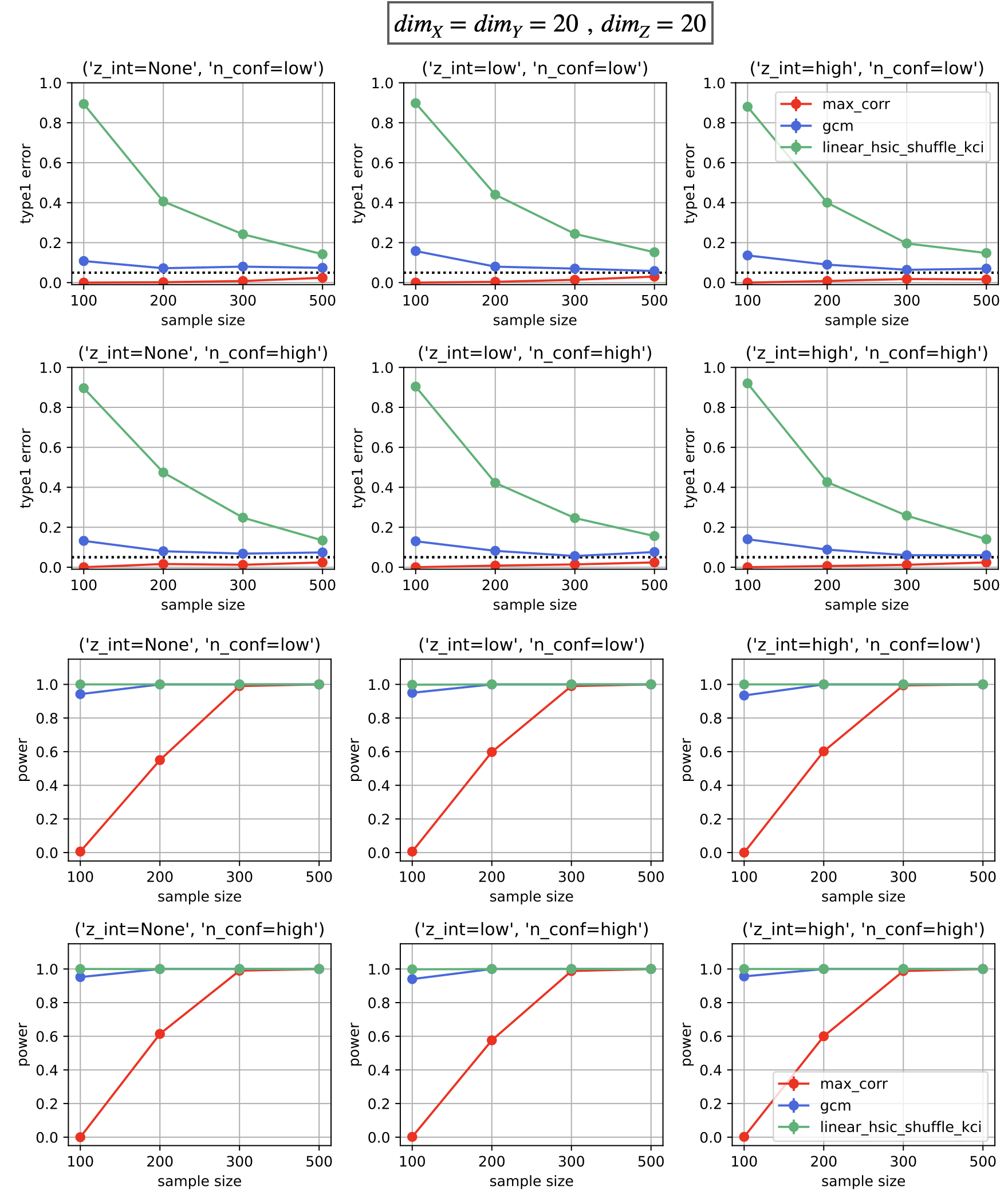}
    \caption{For multivariate $\*X, \*Y,\*Z$, the type 1 error rate of HSIC is highest, and Max-Corr is lowest. The power of Max-Corr starts to suffer yet more severely compared to lower dimensional $\*X,\*Y$ in the low sample regime. Furthermore, the distinction between various internal density settings as well as fraction of active confounders within $\*Z$ has become minimal. GCM maintains a relatively low type 1 error for high power, therefore we have opted OLS together with GCM as the means of obtaining the aggregate statistic as our multivariate CI test of choice for the experiments in this paper.}
    \label{fig:mult_CI_20_20}
\end{figure}

\paranewspace{Experiments} In figures \ref{fig:mult_CI_1_1}, \ref{fig:mult_CI_1_20}, \ref{fig:mult_CI_5_20} and \ref{fig:mult_CI_20_20} we present results on the type 1 error rate and power (i.e., 1-type 2 error rate) of the setting where the dimensions of $\*X,\*Y $ span $ \{1,5,20\}$ and the dimension of $\*Z$ is 1 or 20. Increasing dimensions of $\*X$ (or $\*Y$) and $\*Z$ have varying effects on the performance of the multivariate CI tests discussed above. 

\subsection{Experiments pertaining to Section \ref{sec:experiments}}\label{app:further_exps_vecCI}
\begin{figure}
    \centering
    \includegraphics[scale=0.3]{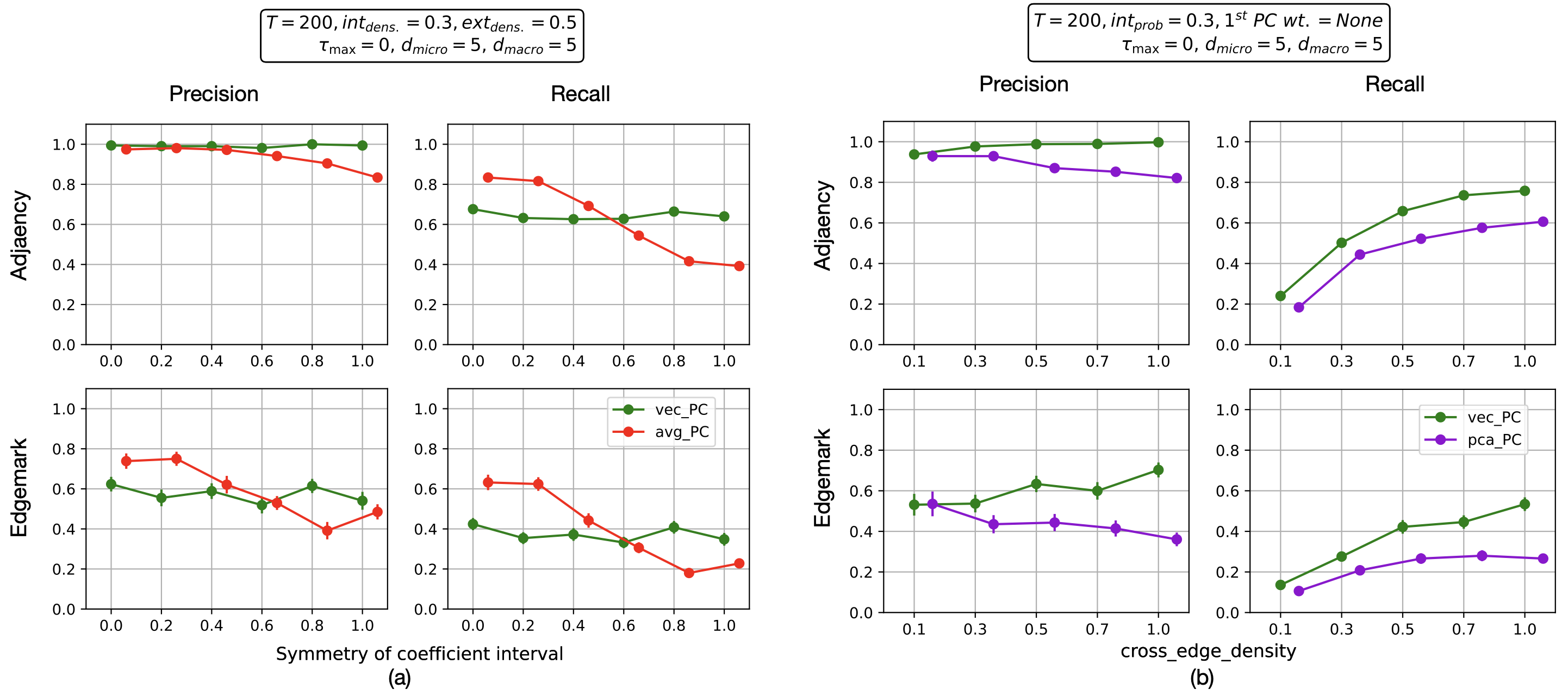}
    \caption{Comparing vector-PC with PC over aggregated variables when $\cG_{micro}$ is a DAG. Plots of precision and recall of adjacencies and edgemarks respectively: (a) PC over averaged variables (`$\text{avg}\_\text{PC}$') versus PC over vector-valued variables (`$\text{vec}\_\text{PC}$'), (b) PC over first principal component of each variable (`$\text{pca}\_\text{PC}$') versus PC over vector-valued variables (`$\text{vec}\_\text{PC}$'). T denotes sample size, $\tau_{max}$ the maximum lag-length ($=0$ implies non-time series data), $int_{dens./prob.}$, resp.~$ext_{dens.}$ (or cross\_edge\_density), density of edges internal to (resp.~across) the vector variables. For an explanation on the `$1^{st} $ PC wt.', see \cref{sec:experiments}.}
    \label{fig:coarse_avg_pca_vec}
\end{figure}

\begin{figure}
    \centering
    \includegraphics[scale=0.35]{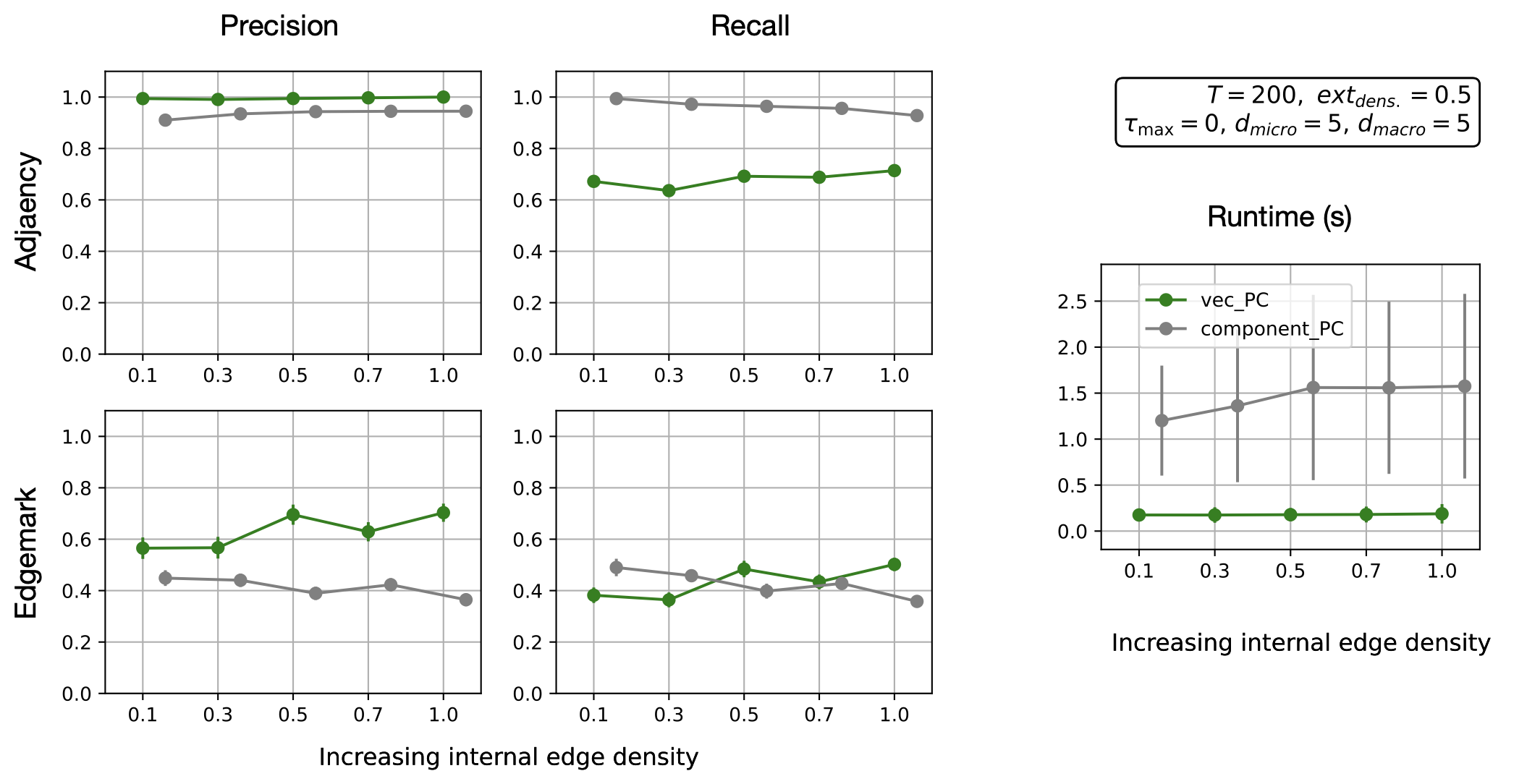}
    \caption{Comparing vector-PC with component-wise PC with the majority edge aggregation strategy when $\cG_{micro}$ is a DAG. Plots of precision and recall of adjacencies and edgemarks respectively. T denotes sample size, $\tau_{max}$ the maximum lag-length ($=0$ implies non-time series data), $int_{dens.}$, resp.~$ext_{dens.}$, density of edges internal to (resp.~across) the vector variables.}
    \label{fig:coarse_component_vec}
\end{figure}

\begin{figure}
    \centering
    \includegraphics[scale=0.3]{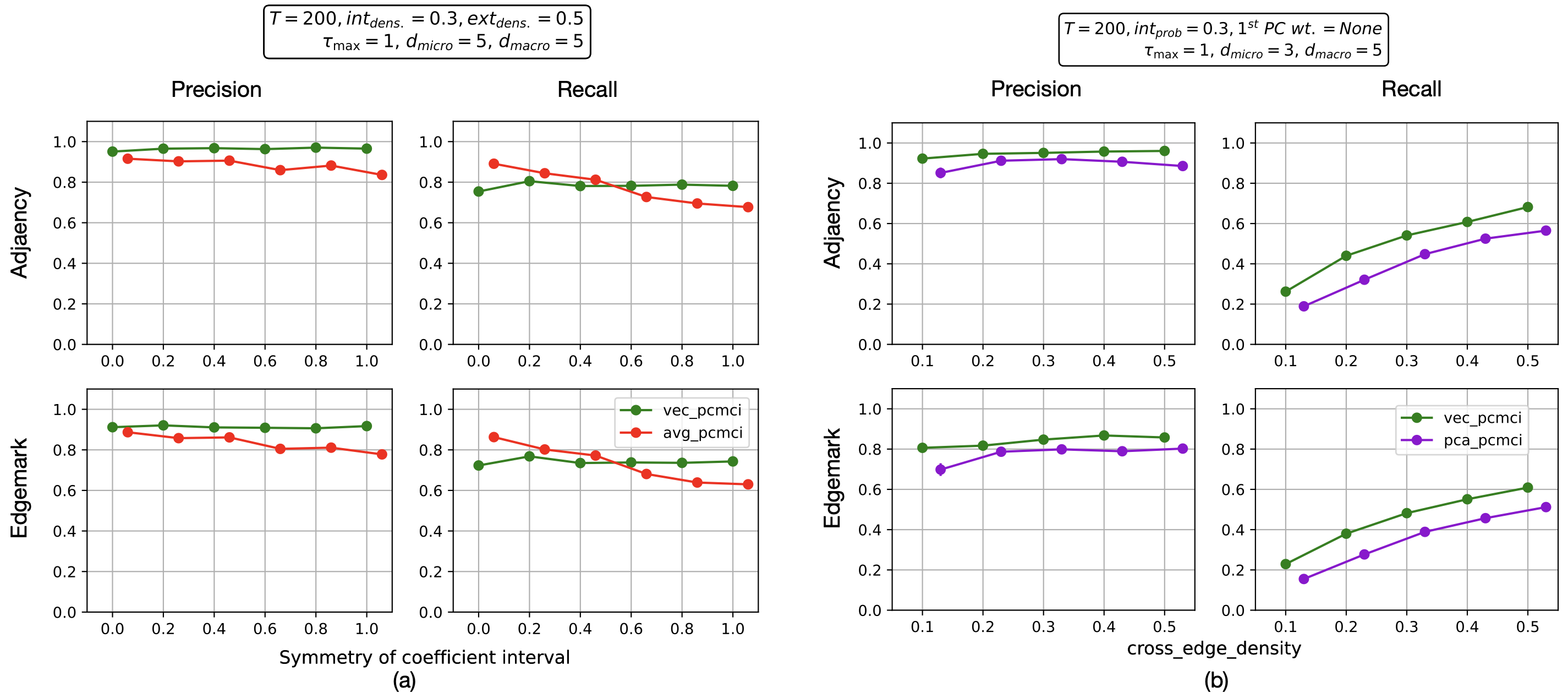}
    \caption{Comparing vector-PCMCI+ with PCMCI+ over aggregated variables for time series data. Plots of precision and recall of adjacencies and edgemarks respectively: (a) PC over averaged variables (`$\text{avg}\_\text{PC}$') versus PC over vector-valued variables (`$\text{vec}\_\text{PC}$'), (b) PC over first principal component of each variable (`$\text{pca}\_\text{PC}$') versus PC over vector-valued variables (`$\text{vec}\_\text{PC}$'). T denotes sample size, $\tau_{max}$ the maximum lag-length, $int_{dens./prob.}$, resp.~$ext_{dens.}$ (or cross\_edge\_density), density of edges internal to (resp.~across) the vector variables. For an explanation on the `$1^{st} $ PC wt.', see \cref{sec:experiments}.}
    \label{fig:TS_avg_pca_vec}
\end{figure}

\begin{figure}
    \centering
    \includegraphics[scale=0.35]{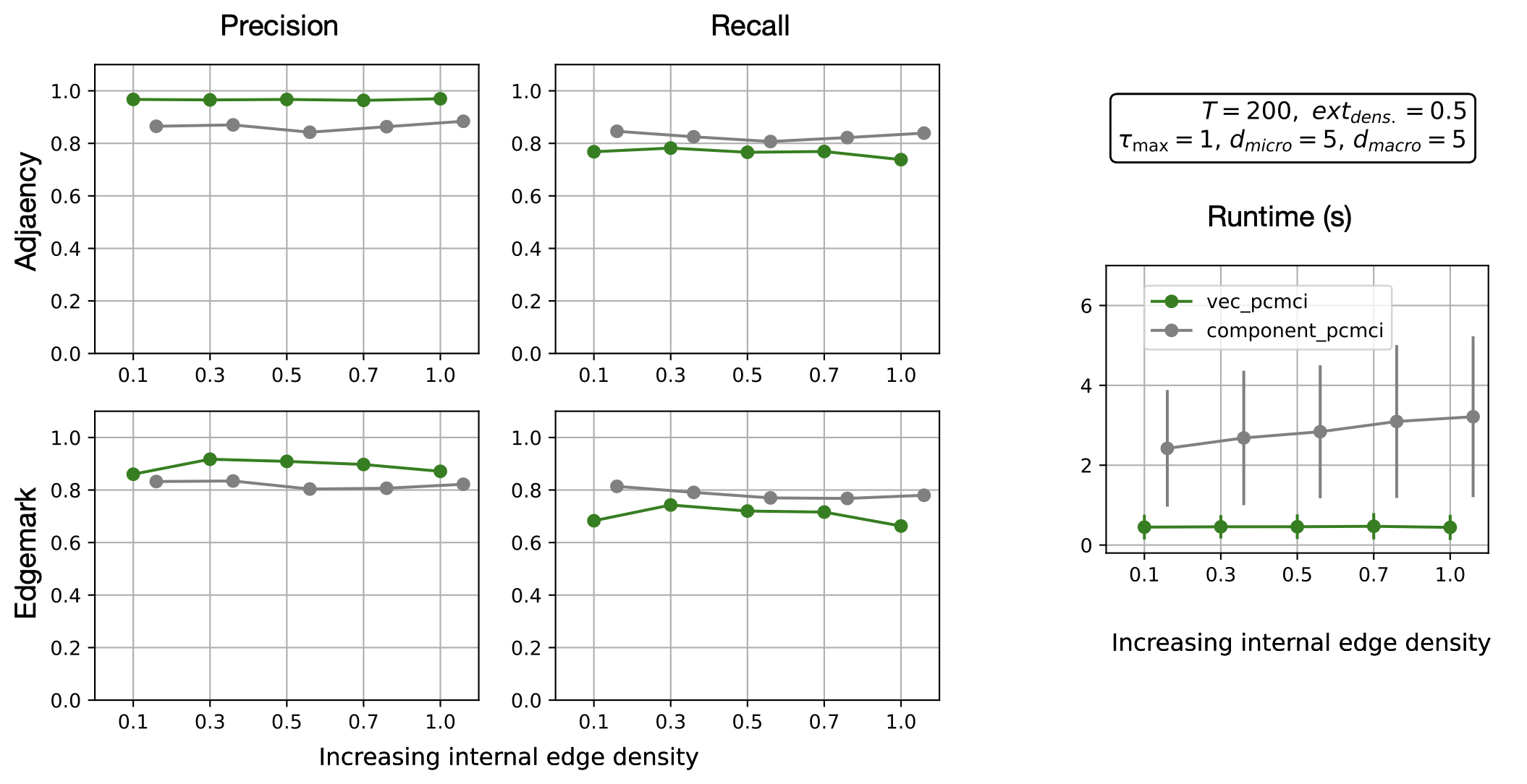}
    \caption{Comparing vector-PCMCI+ with component-wise PCMCI+ with the majority edge aggregation strategy for time series data. Plots of precision and recall of adjacencies and edgemarks respectively. T denotes sample size, $\tau_{max}$ the maximum lag-length, $int_{dens.}$, resp.~$ext_{dens.}$, density of edges internal to (resp.~across) the vector variables.}
    \label{fig:TS_component_vec}
\end{figure}

In this section, we provide further experiments (\cref{fig:coarse_avg_pca_vec}, \cref{fig:coarse_component_vec}, \cref{fig:TS_avg_pca_vec}, \cref{fig:TS_component_vec}) to compare component-wise, aggregate (averaging and retaining the first principal component of each) and vectorized CD. In particular we present experiments for:
\begin{itemize}
    \item[(i)] The data generation process where the internal dynamics can be modeled as a DAG, namely $\cG_{micro}$ is a DAG (\cref{fig:coarse_avg_pca_vec}, \cref{fig:coarse_component_vec}). 
    \item[(ii)] The same data generation process as in \cref{sec:experiments} but on time series data, namely there are contemporaneous as well as lagged links. The CD algorithm chosen is PCMCI+ \citep{runge2020discovering} (\cref{fig:TS_avg_pca_vec}, \cref{fig:TS_component_vec}).
\end{itemize}

\subsection{Experiments pertaining to Section \ref{sec:agg_consistency_adag}}\label{app:further_exps_agg}
\begin{figure}
    \centering
    \includegraphics[scale=0.35]{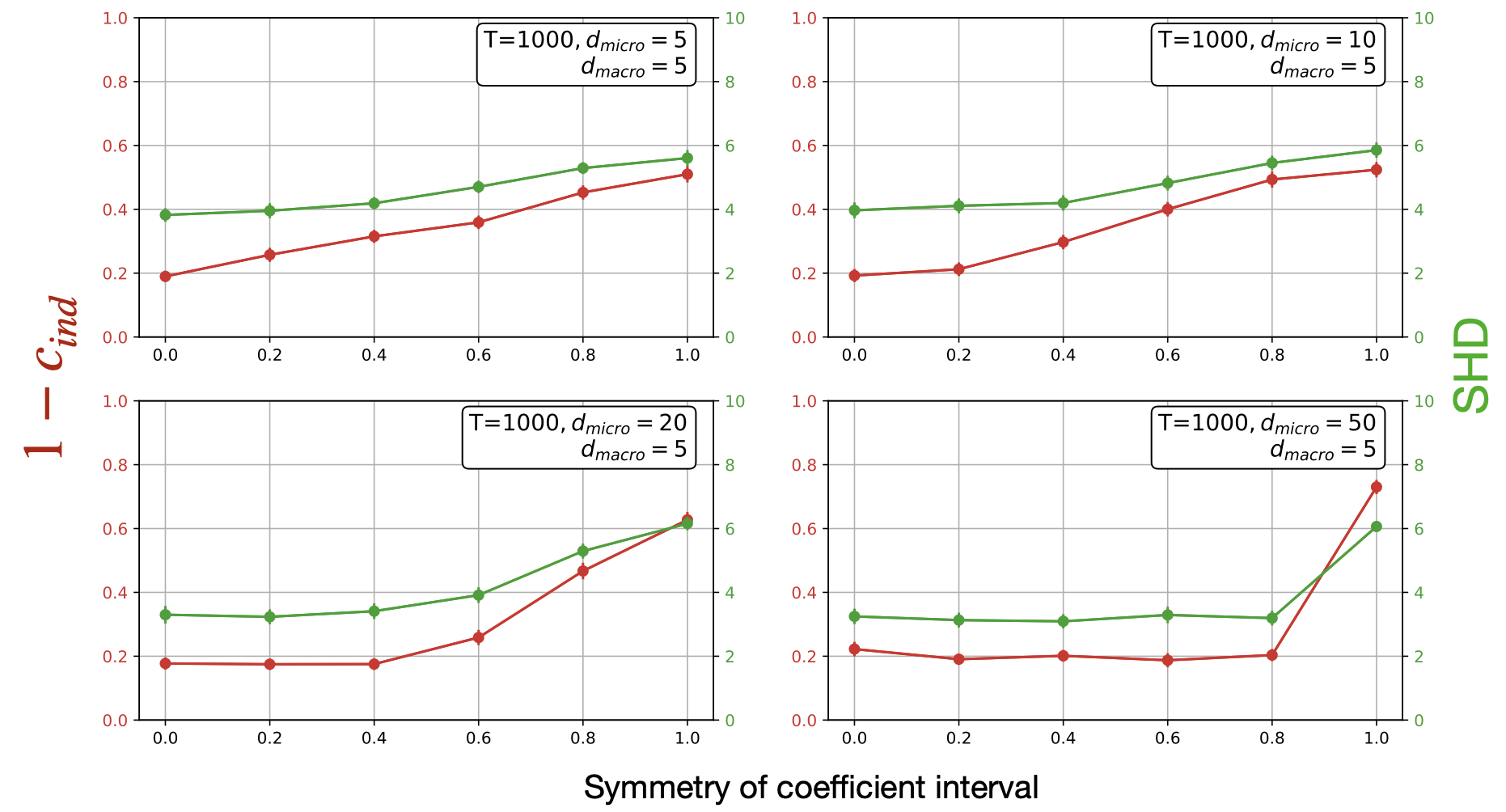}
    \caption{Performance of independence consistency score $c_{ind}$ as the structural hamming distance (SHD) to the ground truth graph increases, namely, aggregate CD becomes less sound: On the x-axis is the degree of symmetry around zero of the interval from which causal coefficients are chosen. This is the same type of plot as in \cref{fig:plot_cind_recall}. On the left y-axis is the $1-c_{ind}$ score for the averaging aggregation map. On the right y-axis is the recall of adjacencies of the graph resulting from PC over averaged data.  $T$ stands for sample size, $d_{macro}$ to the number of vector variables and $d_{micro}$ to the number of scalar components within every vector variable.}
    \label{fig:plot_cind_shd}
\end{figure}

\begin{figure}
    \centering
    \includegraphics[scale=0.3]{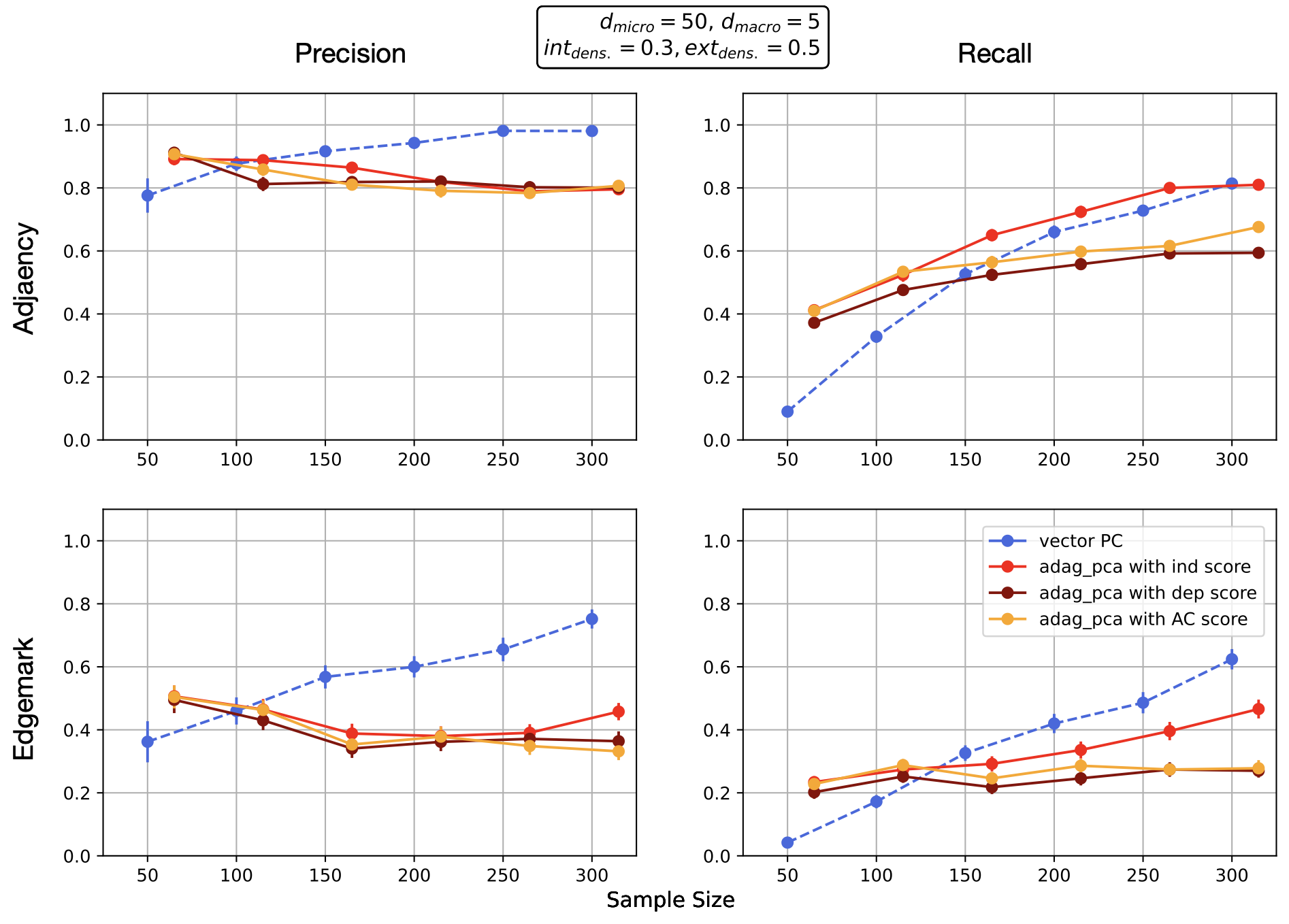}
    \caption{\adag performance comparison with vector CD: Adjacency (respectively edgemark) precision and recall for vector CD (denoted 'vec') and \adag (with tunable aggregation map $g^{\*m} (\cdot)$ given by extracting the first $\*m$ principal components of the vector variables $\*X$ element-wise) with \textbf{(a)} independence score $c_{ind}$ (\cref{eq:cind}), \textbf{(b)} effective dependence score $\bar{c}_{dep}$ (\cref{eq:cdep_eff}), and,  \textbf{(c)} effective $AC$ score (\cref{eq:ac_score}), as sample size increases. For lower samples sizes \adag with either score (and therefore also the effective aggregation consistency score \cref{eq:ac_score}) performs better, whereas for higher sample sizes the performance of vector CD gains as the CI tests become more reliable. Other parameters of the data generating process are shown in the box in the top centre. $int_{dens.}$ is the edge density internal to a vector variable, and $ext_{dens.}$ is the edge density across vector variables.}
    \label{fig:adag_vs_vec_50}
\end{figure}

In this section we present the performance of the independence consistency score (\cref{fig:plot_cind_shd}) and the 
\adag wrapper (\cref{fig:adag_vs_vec_50}) for different sample sizes and internal dimensions of the vector variables. 


\end{document}